\pdfoutput=1
\documentclass[format=acmlarge, review=false, screen=true]{acmart}

\usepackage{booktabs} 

\usepackage{enumitem}
\usepackage{lscape}
\usepackage{tikz}
\usepackage{tikz-qtree}
\usetikzlibrary{trees}
\usepackage{soul}
\usepackage{lineno}
\usepackage{amsmath}
\usepackage{amsfonts}
\usepackage{commath}
\usepackage{float}
\usepackage{multicol}
\usepackage{graphicx}
\usepackage{subcaption}
\usepackage{mathtools}
\usepackage{filecontents}
\usepackage{balance}
\usepackage{multirow}
\usepackage{xcolor}
\usepackage{rotating}
\modulolinenumbers[5]

\newcommand{\specialcell}[2][l]{%
	\begin{tabular}[#1]{@{}l@{}}#2\end{tabular}}

\begin{document}
\title{Multiple Workflows Scheduling in Multi-tenant Distributed Systems: A Taxonomy and Future Directions}

\author{Muhammad H. Hilman}
\orcid{0000-0003-2772-9216}
\email{hilmanm@student.unimelb.edu.au}

\author{Maria A. Rodriguez}
\orcid{0000-0002-2831-8526}
\email{maria.rodriguez@unimelb.edu.au}

\author{Rajkumar Buyya}
\orcid{0000-0001-9754-6496}
\email{rbuyya@unimelb.edu.au}

\affiliation{%
	\institution{\\Cloud Computing and Distributed Systems (CLOUDS) Laboratory, \\School of Computing and Information Systems, The University of Melbourne}
	\streetaddress{Parkville}
	\city{Melbourne}
	\state{VIC}
	\postcode{3000}
	\country{Australia}}

\renewcommand\shortauthors{Hilman, M.H. et al}

\begin{abstract}
The workflow is a general notion representing the automated processes along with the flow of data. The automation ensures the processes being executed in the order. Therefore, this feature attracts users from various background to build the workflow. However, the computational requirements are enormous and investing for a dedicated infrastructure for these workflows is not always feasible. To cater to the broader needs, multi-tenant platforms for executing workflows were began to be built. In this paper, we identify the problems and challenges in the multiple workflows scheduling that adhere to the platforms. We present a detailed taxonomy from the existing solutions on scheduling and resource provisioning aspects followed by the survey of relevant works in this area. We open up the problems and challenges to shove up the research on multiple workflows scheduling in multi-tenant distributed systems.
\end{abstract}

%
%


%
%


\maketitle

\section{Introduction}

The workflow concept has emerged from the manufacturing and office process terminology to a broader notion representing a structured process flow design. Workflow has been used to automate and manage the execution of a complex problem that involves the flow of data with different applications in each process. Many problems adopt this model to tackle the limitation of an individual application to process the vast increasing volume of data. Examples of the workflows include the Scientific Workflows \cite{Taylor:2014:WES:2655383} that consists of HPC applications for e-Science and the MapReduce Workflows \cite{6427527} that is used to process Big Data analytics. Such workflows are large-scale applications and require extensive computational infrastructure to get a reasonable amount of processing time. Therefore, workflows are commonly deployed on distributed systems, in particular, cluster, grid, and cloud computing environments.

Utilizing available distributed systems for deploying particular workflow applications is an affordable option than investing for a dedicated supercomputer system. The luxury for accessing a massive computational power has expanded to broader users with the increasing trend of the community grids infrastructure and the emerging market of cloud environments. From the users' perspective, the presence of these infrastructures is a great benefit to run the more productional process in the industrial area, the research community, and day-to-day governmental business. However, from the providers' point-of-view, this emergence is a new challenge. They must be able to manage at least two complexity, the complication from managing the workflow applications and the complexity of handling multi-tenants with various requirements.

\begin{table*}[!t]
	\begin{center}
		\caption{Resume on Existing Taxonomy of Workflow Scheduling Algorithms}
		\label{table:relatedwork}
		\resizebox{\textwidth}{!}{\begin{tabular}{@{\extracolsep{4pt}} r c c c c c c l@{}}
				\hline \noalign{\vskip 1mm}
				\multicolumn{1}{c}{\multirow{2}{*}{\textbf{Existing Works}}} &\multicolumn{2}{c}{\textbf{Review Types}}&\multicolumn{2}{c}{\textbf{Environments}} &
				\multicolumn{2}{c}{\textbf{Workloads}}&\multicolumn{1}{c}{\multirow{2}{*}{\textbf{Focus of Discussion}}}  \\ \cline{2-3} \cline{4-5} \cline{6-7} \noalign{\vskip 1mm}
				
				&\textbf{Survey}&\textbf{Systematic}&\textbf{Grids}&\textbf{Clouds} &
				\textbf{Single}&\textbf{Multiple}&  \\
				
				\hline \noalign{\vskip 1mm}
				\multirow{1}{*}{\cite{Yu2005}} &\checkmark&-&\multirow{1}{*}{\checkmark}&\multirow{1}{*}{-}& \multirow{1}{*}{\checkmark}&-&Workflow management system architecture\\
				\multirow{1}{*}{\cite{Wieczorek2008}}&\checkmark&-&\multirow{1}{*}{\checkmark} & \multirow{1}{*}{-} &\multirow{1}{*}{\checkmark}&-&Multiple criteria on scheduling workflow \\
				\multirow{1}{*}{\cite{ALKHANAK20153}}&\multirow{1}{*}{-}  &\multirow{1}{*}{\checkmark} &\multirow{1}{*}{-}&\checkmark&\checkmark&limited&Cost-aware scheduling in clouds \\
				\multirow{1}{*}{\cite{SMANCHAT20151}}&\multirow{1}{*}{\checkmark} &\multirow{1}{*}{-} & \multirow{1}{*}{-}&\checkmark&\checkmark&-&Scheduling workflow in cloud environments \\
				\multirow{1}{*}{\cite{Wu2015}} &\multirow{1}{*}{\checkmark}&\multirow{1}{*}{-} & \multirow{1}{*}{-}&\checkmark&\checkmark&limited&Cloud workflow scheduling strategy\\
				\multirow{1}{*}{\cite{Singh2016}}&\multirow{1}{*}{-} &\multirow{1}{*}{\checkmark} &\multirow{1}{*}{-}&\checkmark&\checkmark&limited&Resource scheduling strategy in clouds \\
				\multirow{1}{*}{\cite{CPE:CPE4041}}&\multirow{1}{*}{\checkmark} &\multirow{1}{*}{-} & \multirow{1}{*}{-}&\checkmark&\checkmark&limited&Scheduling workflow in cloud environments\\
				\multirow{1}{*}{Our current work}&\multirow{1}{*}{\checkmark} &\multirow{1}{*}{-} & \multirow{1}{*}{\checkmark}&\checkmark&-&\checkmark&Multi-tenancy in multiple workflows scheduling\\
				\hline \noalign{\vskip 1mm}
		\end{tabular}}
	\end{center}
\end{table*}

Accommodating multi-tenants with different requirements creates a high complexity management system. The very first problem lies on how such a system handles the various workflow applications. A variety of applications involve different software libraries, dependencies, and hardware requirements. The users should be able to customize the specific configurations along with their defined Quality of Service (QoS) when submitting the workflows. Furthermore, multi-tenant systems must have a general scheduling approach to handle different types of computational requirements from different workflows. Another consideration related to multi-tenancy is the strategy to maintain fairness between multiple users that should be achieved through clear prioritization in the scheduling and the automatic scaling of the resources. The last aspect that should be noticed in multi-tenant platforms is the performance variability in computational resources as virtualization-based infrastructures like clouds may encounter performance degradation due to the multi-tenancy, virtualization overhead, geographical location, and temporal aspects \cite{leitner2016performance}.

The contribution of this work is the study of the multiple workflows scheduling problems in multi-tenant distributed systems. The structure of the rest of this paper is as follows. Section 2 discusses the multiple workflows scheduling problems while Section 3 addresses its relevancy in multi-tenant platforms for scientific applications. The proposed taxonomy is presented in Section 4, and the review of existing solutions is covered in Section 5 along with their classification into the taxonomy. Section 6 offers future directions, and Section 7 summarizes the paper. The larger the shape, the higher the LID and the more outlying the query will be relative to others.

\section{Scheduling Multiple Workflows in Multi-tenant nvironments}

Workflow scheduling was studied and surveyed extensively during the cluster, and grid computing era \cite{Yu2005} \cite{Wieczorek2008}. Subsequently, when the cloud computing technology emerged as a new paradigm with market-oriented focus,  workflows community got a promising deployment platform offering multiple benefits. However, it also brought forth additional challenges. Solutions for cloud workflow scheduling have been extensively researched, and a variety of algorithms have been developed \cite{Wu2015} \cite{Singh2016}. Furthermore, various existing taxonomies of workflow scheduling in clouds focus on describing the particular scheduling problem as well as its unique challenges and ways of dealing with them \cite{ALKHANAK20153} \cite{SMANCHAT20151}  \cite{CPE:CPE4041}. The resume of these works is presented in Table \ref{table:relatedwork}.

\begin{figure*}[!t]
	\begin{subfigure}[b]{0.28\textwidth}
		\includegraphics[width=\textwidth]{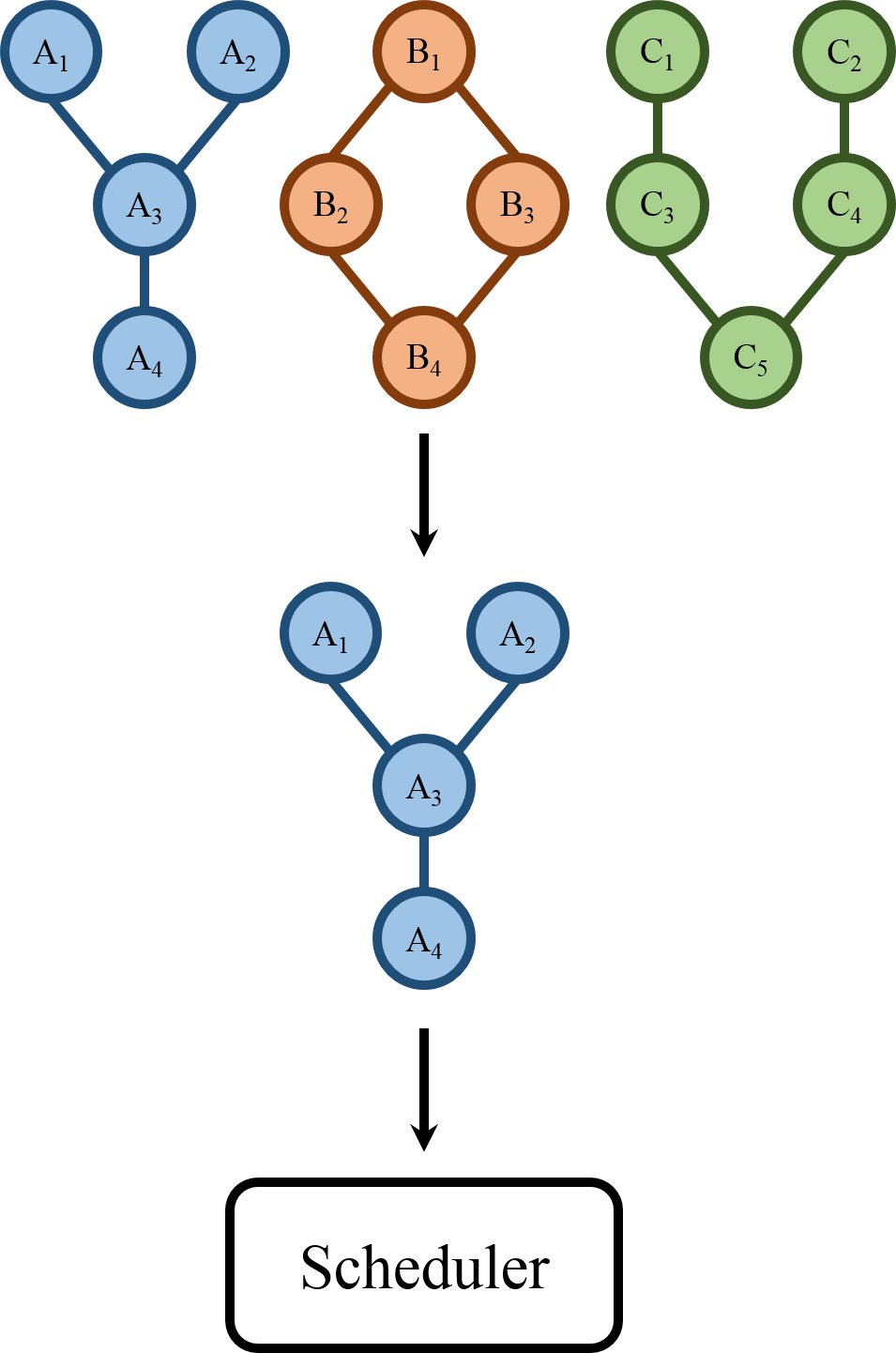}
		\caption{Independent scheduling}    
		\label{fig:indepenttask}
	\end{subfigure}    
	\quad
	\begin{subfigure}[b]{0.275\textwidth}
		\includegraphics[width=\textwidth]{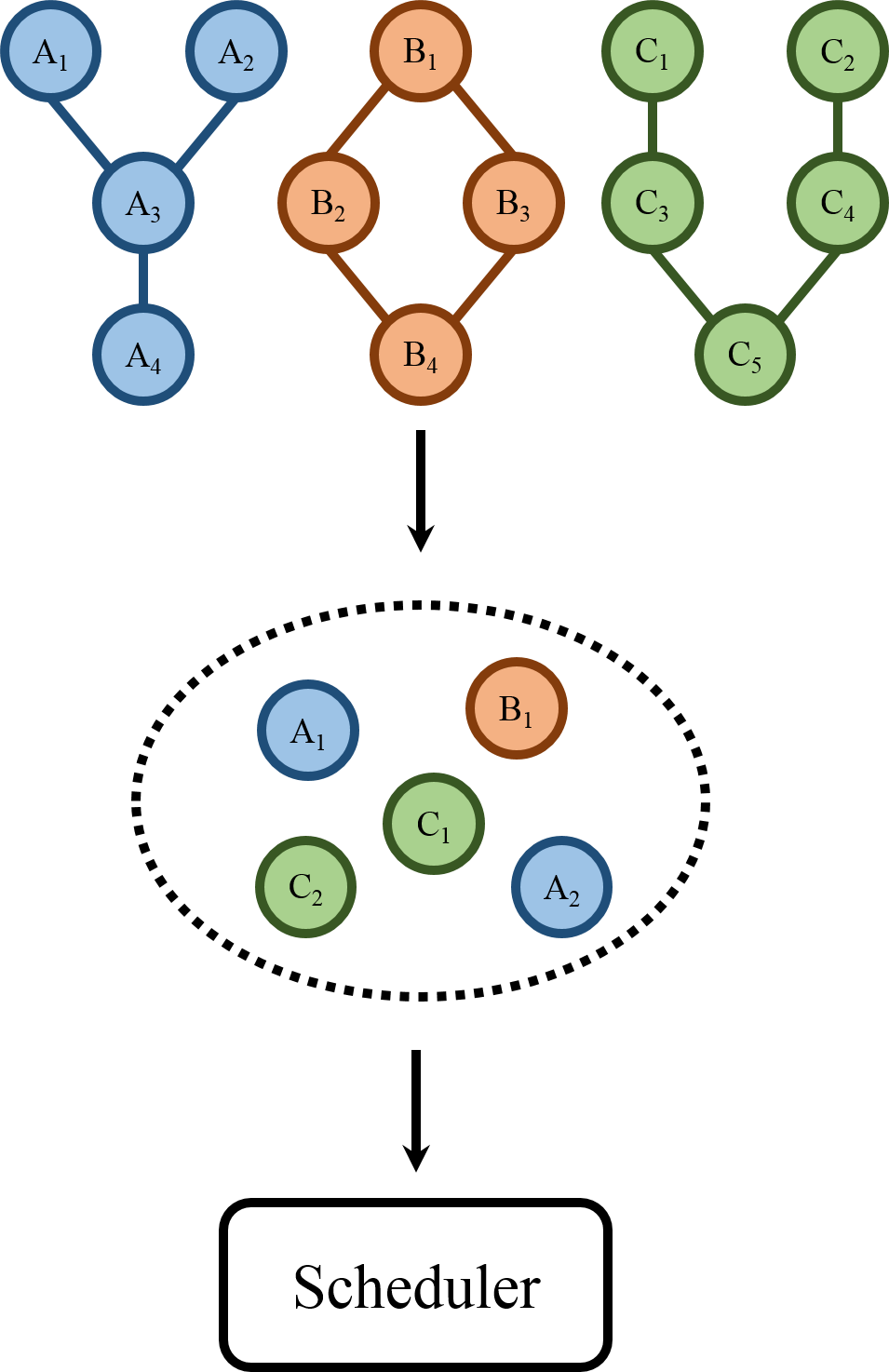}
		\caption{Simultaneous scheduling}
		\label{fig:intertask}
	\end{subfigure}
	
	\caption{Scheduling Multiple Workflows}
\end{figure*}

Contrary to these previous studies which focus mostly on a single workflow scheduling, this study addresses the scheduling problem from a higher level view; it considers the scheduling of multiple workflows that arrive continuously into the multi-tenant platforms. The advent of multi-tenant environments like clouds and the shifting trend from the traditional on-premises to the utility computing era has led to the emergence of platforms that serve multiple workflows processing as a service. These platforms, in theoretical, continuously receive many requests for workflow executions from different users and their various QoS requirements. The provider must then be able to schedule these workflows in a way that each of their requirements is fulfilled. A simple way to achieve this is by allocating a set of dedicated resources to execute each workflow. However, the inter-dependent tasks produce unavoidable idle gaps in the schedule. Hence, dedicating a set of resources for each user can be considered inefficient in environments where multiple workflows are involved as it leads to resources being underutilized. This approach, in turn, may cause a significant loss for the providers that generate revenue from the utilization of resources. Consequently, the strategies implemented in such platforms should aim to improve resource utilization while still complying with the unique requirements of different users.

The scope of this work is limited to the theoretical scheduling algorithms for multiple workflows that are modeled into directed acyclic graphs (DAGs) where a workflow $W$ consists of a set of tasks $T = (t_1, t_2, \dots, t_n)$ and a set of directed edges $E = (e_{12}, e_{13}, . . ., e_{mn})$ in which an edge $e_{ij}$ represents a data dependency between task  $t_i$ (parent task) and task $t_j$ (child task). Hence, $t_j$ will only be ready for execution after $t_i$ has completed. In this way, the purpose of DAG scheduling is to allocate the tasks to computational resources in such fashion that the precedence constraints among the tasks are preserved. Within this context, the workflow scheduling problem is defined by the application model of multi-tenant platforms. Specifically, the workflows submitted to multi-tenant platforms belong to different users and are not necessarily related to each other. As a result, heterogeneity becomes a defining characteristic of the workload that covers various aspects of workflows including the type of applications, the size of the workflows, and the user-defined QoS.

Furthermore, even though the information of workflows (i.e., topological structure, computational requirement, size, input) is available when these workflows arrive into the platforms, planning the schedule by exploiting this information as being implemented in static workflow scheduling is not plausible. For example, a strategy of partitioning tasks before runtime in a static workflow scheduling to minimize the data transfer is proven to be efficient for data-intensive workflows \cite{7034777}. This strategy actually can be done in multi-tenant platforms, but then, it becomes an inevitable bottleneck since the time required for partitioning a workflow may delay the next queue of arriving workflows for scheduling. The waiting time may increase significantly if the planning involves a metaheuristic optimization technique known for its intensiveness in computing. As the size of the workflow increases, this pre-processing time may become longer and produce more massive queue with significant waiting time delay. Hence, we do not consider solutions that schedule each workflow independently as depicted in Figure \ref{fig:indepenttask} as this approach is no different from scheduling a single workflow.

Instead, we consider scheduling algorithms designed to schedule multiple workflows simultaneously as shown in Figure \ref{fig:intertask}. There are many advantages and challenges of scheduling multiple workflows in this area. The main benefits of this scheduling model are related to the possibility of idle time slots produced by a particular workflow to be used by another workflow and the reduction of waiting time from queueing delay of workflows being scheduled. On the other hand, the challenges to achieving these are not trivial. Handling the workloads heterogeneity, managing the continuous arriving workflows, implementing general scheduling approaches that deal with different requirements, and dealing with performance variability in distributed systems are questions that must be answered.

\section{Workflow as a Service Platform for Scientific Applications}

Scientific workflows are widely used to automate scientific experiments in many areas. The latest detection of gravitational waves by the LIGO project \cite{ras2016gravitational} is an example of a scientific breakthrough assisted by workflow technologies. These workflows are composed of multiple tasks and dependencies that represent the flow of data between them. Scientific workflows are usually large-scale applications that require extensive computational resources to process. As a result, distributed systems with an abundance of storage, network, and computing capacity are widely used to deploy these resource-intensive applications.

The complexity of scientific workflows execution urges scientists to rely on workflow management systems (WMS), which manage the deployment of workflows in distributed resources. Their main functionalities include but are not limited to scheduling the workflows, provisioning the resources, managing the dependencies of the tasks, and staging the input/output data. Taverna \cite{doi:10.1093/bioinformatics/bth361}, Kepler \cite{CPE:CPE994} and Pegasus \cite{DEELMAN201517} are some examples of WMS that are widely used by the scientific community. A key responsibility of WMS and the focus of this work is the scheduling of workflow tasks. In general, this process consists of two stages, i) mapping the execution of tasks on distributed resources and ii) acquiring and allocating the appropriate compute resources to support them. Both of these processes need to be carried out while considering the Quality-of-Service (QoS) requirements specified by users and preserving the dependencies between tasks. These requirements make the workflow scheduling process a challenging problem.

The advancement of e-Science infrastructure in the form of scientific applications (i.e., scientific workflows) empowers a large number of scientist around the world to start the shifting trend of scientific experiments. They are part of the broad community that is called the Long Tail of Science. The smaller group of scientists that run the scientific experiments in far tinier scale than the LIGO project but generating a more significant number of scientific data and findings \cite{10.1007/978-3-642-22351-8_31}. However, not many scientists can afford to build their dedicated computational infrastructure for their experiments. In this case, there is a potential market for providing such services to the scientific community.

Workflow as a Service (WaaS) is an emerging paradigm that offers the execution of scientific workflows as a service. The service provider lies either in the Platform as a Service (PaaS) or Software as a Service (SaaS) layer based on the cloud stack service model. WaaS providers make use of distributed computational resources to serve the enormous need for computing power in the execution of scientific workflows. They provide a holistic service to scientists starting from the user interface in the submission portal, applications installation, and configuration, staging of input/output data, workflow scheduling, and resource provisioning. WaaS platforms are designed to process multiple workflows from different users. The workload is expected to arrive continuously, and workflows must be handled as soon as they arrive due to the quality of service (QoS) constraints defined by the users. Therefore, WaaS platforms must deal with a high level of complexity derived from their multi-tenant and dynamic nature, contrary to a traditional WMS that is commonly used for managing a single workflow execution.

\begin{figure*}[!t]
	\includegraphics[width=.55\textwidth]{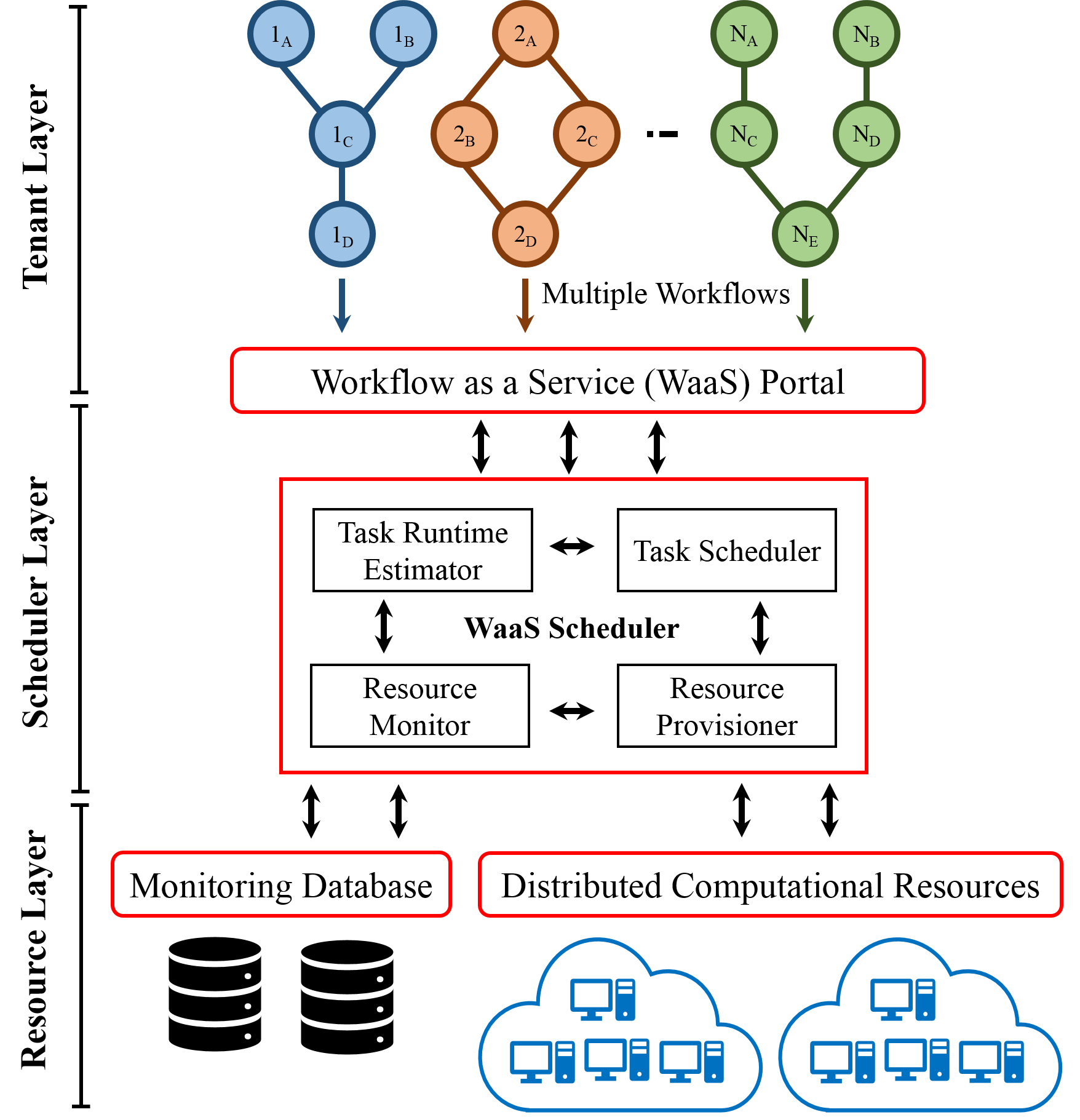}
	\caption{Workflow as a Service Architecture}    
	\label{fig:architecture}
\end{figure*}

Several variations of WaaS framework--which extend the traditional WMS architecture--are found in literature such as the work by Wang et al. \cite{WANG2014546} describing a service-oriented architecture for scientific workflows that separates the components into user management layer, scheduler, storage, and VM management. Meanwhile, a framework with a similar division that emphasizes on distributed storage is proposed by Esteves and Veiga \cite{doi:10.1093/comjnl/bxu158}. Another typical architecture for multi-tenant scientific workflows execution in clouds emplaces the proposed framework as a service layer above the Infrastructure as a Service (IaaS) layers \cite{7457258}. In general, we identified three primary layers in WaaS platforms; the tenant layer, the scheduler layer, and the resource layer. Based on three layers and the identified requirements of WaaS platforms, we propose a reference architecture for this system focusing on the scheduler component as depicted in Figure \ref{fig:architecture}.

Firstly, the tenant layer manages the workflows submission where users can configure their preferences and define the QoS of their workflows. The scheduler layer is responsible for placing the tasks on either existing or newly acquired resources and consists of four components: task runtime estimator, task scheduler, resource provisioner, and resource monitor. The task runtime estimator is used to predict the amount of time a task will take to complete in a specific computational resource (i.e., virtual machine). Another component, the task scheduler, is used to map a task into a selected virtual machine for execution. Meanwhile, resource provisioner is used to acquiring and releasing virtual machines from third-party providers. The resource monitor is used to collect the resource consumption data of a task executed in a particular virtual machine. These data are stored in a monitoring database and are used by the task runtime estimator to build a model to estimate the task's runtime. The third-party infrastructure (e.g., virtual machines, storage databases) with which the WaaS platforms interact, fall into the resource layer.

We explore the underlying problem of scheduling multiple workflows in various distributed systems that have taken place for more than a decade with explicitly focusing on the multi-tenancy aspect of the problem related to the scheduling and resource provisioning. In this study, we expect to gain some knowledge to design and develop multi-tenant WaaS platform that pretty much resembles the problem of multiple workflows scheduling in multi-tenant distributed systems.\\

\section{Taxonomy}

The scope of this study is limited to the theoretical algorithms developed for multiple workflows scheduling that represent the problem in multi-tenant WaaS platforms. In this section, we describe various challenges of scheduling multiple workflows and their relevancy for each taxonomy classification. Furthermore, the mapping and references of the algorithms to each class are presented in Section 5.

\subsection{Workload Model Taxonomy}

Multiple workflows scheduling algorithms are designed to handle workloads with a high level of heterogeneity that represent a multi-tenant characteristic in the platforms. Workload heterogeneity can be described from several aspects including the continuous arrival of workflows at different times, the various types of workflow applications that differ in computational requirements, the difference in workflow sizes, and the diversity in software libraries and dependencies. 

The different arrival time of multiple workflows in the platforms resembles the problem of streaming data processing that deals with continuous incoming tasks to be processed. In contrast with some static single workflow scheduling algorithms that make use information of the workflow structure, the runtime of tasks, and the specific computational requirements before execution time to create a near optimal schedule plan, the continuous arrival of workflows in multi-tenant platforms makes this an unsuitable approach. Furthermore, conventional techniques to achieve near-optimal schedule such as metaheuristics are computationally intensive, and the computational complexity will grow as the workflow size increases. The time for planning may take longer than the actual workflow execution. Hence, a lightweight dynamic scheduling approach is the most suitable for multi-tenant environments as the algorithms must be able to deal with the dynamicity of the workload. For instance, at peak time the concurrency of requests may be very high, whereas, at other times, the submission rate may reduce to a point where the inter-arrival time between workflows is long enough to execute each workflow in a dedicated set of resources.

The variety of application types is another issue to be addressed. A study by Juve et al. \cite{JUVE2013682} shows a variety of workflow applications with different characteristics. The Montage astronomy workflow \cite{Deelman:2008:CDS:1413370.1413421} that is used to reconstruct mosaics of the sky is considered as a data-intensive application with high I/O activities. The CyberShake workflow \cite{Maechling2007} that is used to characterize earthquake hazards using the Probabilistic Seismic Hazard Analysis (PSHA) techniques is categorized as a compute-intensive workflow with multiple reads on the same input data. The Broadband workflow that is used to integrate a collection of simulation codes and calculations for earthquake engineers has a relative balance of CPU and I/O activities in its tasks. These three samples show different types of workflow applications that may have different strategies of scheduling to be carried out. For example, a strategy of clustering tasks with a high dependency of input/output data (i.e., data-intensive) and allocating the same resource for them to minimize data transfer. 

Furthermore, the heterogeneity of workloads is also related to the size of the workflows. The size represents the number of tasks in a workflow and may differ even between instances of the same type of workflow application due to different input datasets. For example, the Montage workflow \cite{Deelman:2008:CDS:1413370.1413421} takes the parameters of width and height degree of a mosaic of the sky as input. The higher the degree, the larger the size of Montage workflow to be executed as it resembles the size and shape of the area of the sky to be covered and the sensitivity of the mosaic to produce. A large-scale workflow may raise another issue in scheduling such as high volume data transfer that may cause a bottleneck in the network which will affect other smaller scale workflows being executed in the platform.

\begin{figure*}[!t]
	\centering
	\begin{tikzpicture}[level distance=2in,sibling distance=.2in,scale=.9]
	\tikzset{edge from parent/.style= 
		{thick, draw,
			edge from parent fork right},every tree node/.style={minimum width=1.1in,text width=1.1in, align=left},grow'=right}
	\Tree 
	[.{Workload Model}
	[.{Workflow Type} 
	[.{Homogeneous} ]
	[.{Heterogeneous} ]
	]
	[.{QoS Requirement} 
	[.{Homogeneous} ]
	[.{Heterogeneous} ]
	]
	]
	\end{tikzpicture}
	\caption{Workload Model Taxonomy}
	\label{fig:workload}
\end{figure*}
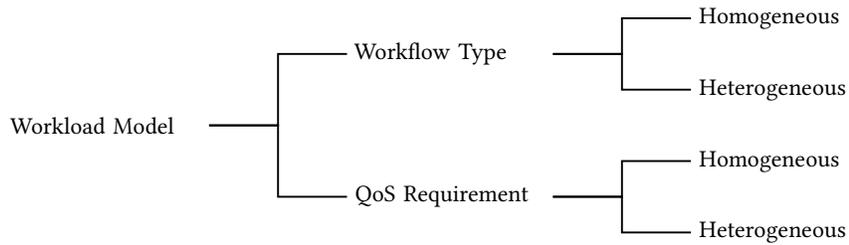

Another heterogeneity issue in the platforms is the various software libraries and dependencies required for different workflows. This problem is related to the deployment and configuration of workflows in the platforms. Deploying different software libraries and dependencies in a system requires technical efforts such as installing the software and managing conflicts between software dependencies. The most important implication related to this case is the resource sharing between workflows to utilize idle time slots produced during the scheduling. In cluster and grid environments where every user uses shared installed software systems on a physical machine, the conflicting dependencies are inevitable. This problem can be avoided by isolating applications in virtualized environments such as clouds. However, in clouds where the workflow's deployment and configuration can be isolated in a virtual machine, the possibility to share the computational power between users in a particular virtual machine is limited. This problem is due to the limitation of a virtual machine capacity (i.e., memory, storage) and possible conflicting dependencies if we want to have as much as software configured in a virtual machine instance. The trade-off between the isolation and the resource sharing in clouds can be solved using container technology as successfully implemented using Singularity and CVMFS at OSG \cite{Bryant:2018:VVC:3219104.3219125} and Shifter at Blue Waters \cite{belkin2018container}. In this case, container, a lightweight operating system level virtualization, is used to isolate the workflow application before deploying them on virtual machines. Therefore, both isolation and resource sharing objectives can be achieved. Based on the heterogeneity issue, this kind of workloads can be differentiated by their workflow type and user-defined QoS requirements as shown in Figure \ref{fig:workload}.

\subsubsection{Workflow Type}
Scheduling algorithms for multi-tenant platforms must consider the fact that the users in this system may submit differently or a single type of workflow applications. These variations can be categorized into homogeneous and heterogeneous workflow types.

A homogeneous workflow type assumes all users submit the same kind of workflow applications (e.g., multi-tenant platforms for Montage astronomical workflow). In this case, the algorithms can be tailored to handle a specific workflow application by exploiting its characteristics (e.g., topological structure, computational requirements, software dependencies, and libraries). For example, related to a topological structure, a workflow with a large number of tasks in a level may raise an issue of data transfer. This issue can potentially become a communication bottleneck when all of the tasks in a level concurrently transfer the data input needed to execute the tasks. Therefore, clustering the tasks may result in a reduction in data transfer and eliminates the bottleneck in the system.

Furthermore, the heterogeneity from the resource management perspective affects how the scheduling algorithms handle software dependencies and libraries installed in computational resources. The algorithms for a homogeneous workflow type can safely assume that all resources contain the same software for a typical workflow application. In this way, the constraints for choosing appropriate resources for particular tasks related to the software dependencies can be eliminated since all of the resources are installed and configured for the same workflow application.

On the other hand, to handle a heterogeneous workflow type, the algorithms must be able to tackle all various possibilities of workflow type submitted into the platforms. In a multi-tenant platform, where the heterogeneous workflow type is considered, tailoring the algorithms to the specific workflow application characteristics is impractical. The scheduling algorithms must be designed following a more general approach. For example, related to a topological structure, a task in a workflow is considered ready for an execution when all of its predecessors are executed, and its data input is available in a resource allocated for execution. In this way, the algorithms can exploit a simple heuristic to build a scheduling queue by throwing in all tasks with this specification to the queue.

Therefore, a variety of software dependencies and libraries required for different workflow applications increases the possible conflict of software dependencies in platforms that consider heterogeneous workflow type. Therefore, the algorithms must include some rules in the resource selection step to determine what relevant resources can be allocated for specific tasks. For example, the algorithms can define a rule that is only allowing a task to be assigned a resource based on its software dependencies and libraries availability.

\subsubsection{QoS Requirements}
Workloads in multi-tenant platforms must be able to accommodate multiple users' requirements. These requirements are represented by the Quality of Service (QoS) parameters defined when users submit their workflows to the platforms. We categorize the workloads based on the users' QoS requirements into homogeneous and heterogeneous QoS requirements.

The majority of algorithms designed for multi-tenant platforms surveyed in this study consider a homogeneous QoS requirement. They are designed to achieve the same scheduling objective (e.g., minimizing the makespan, meeting the budget) for all workflows. Meanwhile, a heterogeneous QoS requirement is addressed by the ability to be aware of various objectives and QoS parameters demanded by a particular user. The algorithms may consider several strategies within the same platforms to handle workflows with different QoS requirements. For example, to process workflows that are submitted with the deadline constraints, the algorithms may exploit the option to schedule them into the cheapest resources to minimize operational cost as long as their deadlines can be met. At the same time, the algorithms can also handle workflows with the budget constraints by using another option to lease as much as possible the fastest resources within the available budget.

\subsection{Deployment Model Taxonomy}

\begin{figure*}[!t]
	\centering
	\begin{tikzpicture}[level distance=2in,sibling distance=.2in,scale=.9]
	\tikzset{edge from parent/.style= 
		{thick, draw,
			edge from parent fork right},every tree node/.style={minimum width=1.24in,text width=1.24in, align=left},grow'=right}
	\Tree 
	[.{Platform Deployment} 
	[.{Non-virtualized} ]
	[.{Virtualized} 
	[.{VM-based} ]
	[.{Container-based} ]
	]
	]
	\end{tikzpicture}
	\caption{Deployment Model Taxonomy}
	\label{fig:deployment}
\end{figure*}
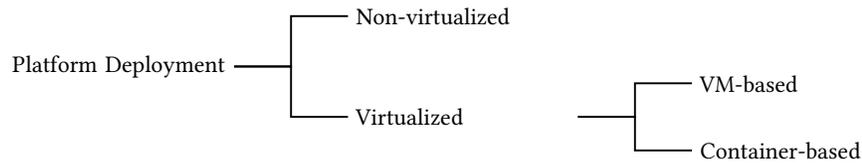

Handling the performance variability in multi-tenant distributed systems is essentials to the multiple workflows scheduling problems as the scheduling highly relies on the accurate estimation of workflow`s performance on a particular computational infrastructure. Attempts to increase the quality of scheduling by accurately estimating the time needed for completing a task, as one of the strategies for taking care of the uncertainty, has been extensively studied \cite{WITT201933}. Specific work designed for scientific workflow includes a work by Nadeem and Fahringer \cite{5071887} that utilized the template to predict the scientific workflow applications execution time. Another work by da Silva et al. \cite{doi:10.1142/S0129626415410030} introduced an online approach to estimate the resource consumption for scientific workflows. Meanwhile, Pham et al. \cite{8013738} worked on machine learning techniques to predict task runtime in workflows using a two-stage prediction approach.

When we specifically discuss cloud environments, the uncertainty becomes higher than cluster and grid environments. The virtualization that is the backbone of clouds is the primary source of the performance variability as reported by Leitner and Cito \cite{leitner2016performance} and also previously discussed by Jackson et al. \cite{jackson2010performance}. The cloud instances performance varies over time due to several aspects including the virtualization overhead, the geographical location of the data center, and especially the multi-tenancy of clouds. For example, it is not uncommon for a task to have a longer execution time during a specific time in cloud instances (i.e., peak hours) due to the number of users served by a particular cloud provider at that time. The main conclusions from their works substantiate our assumption that the performance and predictability of cloud environments is something that is not easy to address. 

Another variable of uncertainty in clouds is the provisioning and de-provisioning delays of VMs. When the user requests to launch an instance in a cloud, there is a delay between the request and when the VM is ready to use, called provisioning delay. There also exists a delay in releasing the resource, namely de-provisioning delay. Not considering the provisioning and de-provisioning delays in the scheduling phase may cause a miscalculation of when to acquire and to release the VM. This error may cause an overcharging bill of the cloud services. A study by Mao and Humphrey \cite{6253534} reported that the average provisioning delay of a VM, observed from three cloud providers--Amazon EC2, Windows Azure, and Rackspace--was 97 seconds while more recently, Jones et al. \cite{7761629} presented a study which shows that three different cloud management frameworks--OpenStack, OpenNebula, and Eucalyptus--produced VM provisioning delays between 12 to 120 seconds. However, delays are not only derived from acquiring and releasing instances. As most of the WMS treat cloud instances (i.e., virtual machines) as virtual clusters using third-party tools (e.g., HTCondor\footnote{https://research.cs.wisc.edu/htcondor/}), there exists a delay in integrating a provisioned VM from cloud providers into a virtual cluster. An upper bound delay of 60 seconds for this process was observed by Murphy et al. \cite{Murphy:2009:DPV:1577849.1577925} for an HTCondor virtual cluster. These delays are one of the sources of uncertainty in clouds, and therefore, the algorithms should consider then to get an accurate scheduling result.

Hence, the scheduling algorithms for multi-tenant platforms can be differentiated based on its deployment model. We identified two types of algorithms for multi-tenant platforms based on their deployment model as illustrated in Figure \ref{fig:deployment}. Several issues and challenges that arise from these deployment models are worthy of being considered by the scheduling algorithms.

\subsubsection{Non-virtualized}
The majority of works in our survey design are scheduling algorithms for cluster and grid environments. These two environments are the traditional way of establishing multi-tenant distributed systems where a large number of computational resources are connected through a fast network connection so that many users in a shared fashion can utilize it. However, in this way, there is no isolation between software installed related to their dependencies and libraries within the same physical machine.

Accommodating multi-tenant users in a non-virtualized environment is limited by the computational infrastructure static capacity. This staticity makes it very hard to auto-scale the resources in non-virtualized environments. Thus, the algorithms cannot efficiently serve a dynamic workload without having a queueing mechanism to schedule overloaded requests at a particular time. For example, adding a node into an established cluster infrastructure is possible but may involve technicalities that cannot be addressed in a short period. This environment also does not allow the users to shrink and expand their allocated resources easily since it needs to go through the administrator intermediaries. Therefore, the primary concern of scheduling algorithms designed for this environment is to ensure for maximum utilization of available resources, so the algorithms can reduce the queue of users waiting to execute their workflows. In this case, the techniques--such as task rearrangement and backfilling--can be used to fill the gaps produced by scheduling a particular workflow, by allocating these idle slots to other workflows.

\subsubsection{Virtualized}
The algorithms designed for virtualized environments (i.e., cloud computing environments) can gain advantages from a flexible configuration of VM as it isolates specific software requirements needed by a user in a virtualized instance. A fully configured virtual machine can be used as a template and can be shared between multiple users to run the same workflows. This isolation ensures little disturbance to the platforms and the other users whenever a failure occurs. However, in this way, the possible computational sharing of a virtual machine is limited. It is not plausible to configure a virtual machine for several workflow applications at the same time. In this case, containers can be used to increase the configuration flexibility in virtualized environments. The container is an operating-system-level virtualization method to run multiple isolated processes on a host using a single kernel. The container is initially a feature built for Linux (i.e., LXC) that is further developed and branded as a stand-alone technology (e.g., Docker\footnote{https://www.docker.com/}) that not only it can run on Unix kernel but also on Windows NT kernel (e.g., windows container\footnote{https://docs.microsoft.com/en-us/virtualization/windowscontainers/}). A full workflow configuration can be created in a container before deploying it on top of virtual machines. In this way, the computational capacity of VMs can be shared between users with different workflows.

In the context of scalability, algorithms designed for virtualized environments can comfortably accommodate multi-tenant requirements. The algorithms can acquire more resources in on-demand fashion whenever requests are overloading the system. Furthermore, this on-demand flexibility supported by pay-as-you-go pricing scheme reduces the burden for the providers to make upfront payments for reserving a large number of resources that may only be needed at a specific time (i.e., peak hours). Even if a particular cloud provider cannot meet the demand of the providers, the algorithms can provision resources from different cloud providers.

However, this environment comes with a virtualization overhead that implies a significant performance variability. The overhead not only occurs from the communication bottleneck when a large number of users deal with high volume data but also the possible degradation of CPU performance since the computational capacity is shared between several virtual machines in the form of virtual CPU. The other overheads are the delay in provisioning and de-provisioning virtual machines and the delay in initiating and deploying the container. The scheduling algorithms have to deal with these delays and consider them in the scheduling to ensure the accurate scheduling result.

\subsection{Priority Assignment Model}

Fairness and priority issues are inevitable in multiple workflows scheduling. Given two workflows that arrive at the same time, the decision to execute a particular workflow first must be determined based on some policy. The priority to be assigned for each workflow can be derived from several aspects including the QoS defined by users, the type of workflow application, the user's preference, and the size of the workflows.

Priority assignment can be determined based on the user-defined QoS. It is evident for scheduling algorithms to prioritize workflow with the earliest deadline as this can ensure the fulfillment of QoS requirements. In this way, algorithms may introduce a policy based on the deadline that delays the scheduling of a workflow with a more relaxed deadline to improve the resource sharing in the system without violating the fairness aspect. On the other hand, the priority assignment can be defined from the budget available. In real-world practice, it is common to prioritize the users with more budget available to do a particular job compared to the lower one (e.g., priority check-in for business class passenger). This policy also can be implemented in multiple workflows scheduling.

Assigning priority based on the type of application can be done by defining application or user classes. For example, workflows submitted for education or tutorial purpose may have a lower priority than the workflows executed in a scientific research project. Meanwhile, a workflow that is used to predict the typhoon occurrence may be performed first compared to a workflow for creating the mosaic of the sky. This policy can be defined out of the scheduling process based on some policy adopted by the providers.

Moreover, the priority assignment can also be determined based on the size of workflows. This approach is the most traditional way of priority scheduling that has been widely implemented such as the Shortest Job First (SJF) policy which prioritizes the smaller workflows over a larger one to avoid starvation. Another traditional scheduling algorithm like the Round-Robin (RR) also can be constructed based on the size of the workflows to ensure both of the small and large-scale workflows get fair treatment in the systems.

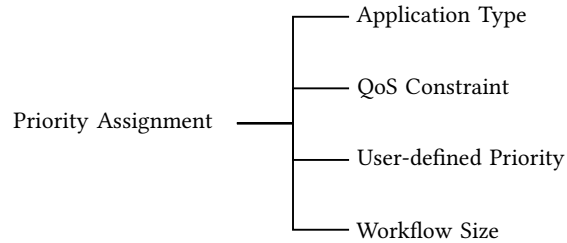
\begin{figure*}[!t]
	\centering
	\begin{tikzpicture}[level distance=2in,sibling distance=.2in,scale=.9]
	\tikzset{edge from parent/.style= 
		{thick, draw,
			edge from parent fork right},every tree node/.style={minimum width=1.25in,text width=1.25in, align=left},grow'=right}
	\Tree 
	[.{Priority Assignment} 
	[.{Application Type} ]
	[.{QoS Constraint} ]
	[.{User-defined Priority} ]
	[.{Workflow Size} ]
	]
	\end{tikzpicture}
	\caption{Priority Assignment Model Taxonomy}
	\label{fig:priority}
\end{figure*}

Fairness between users in multi-tenant platforms can be achieved through priority assignment in scheduling algorithms. This assignment is essential as the ultimate goal of the providers is to fulfill each user's QoS requirement. We identify various priority assignment model from surveyed algorithms that consider the type of workflow application, users QoS constraints, user self-defined priority, and size of workflows in their design, as shown in Figure \ref{fig:priority}.

\subsubsection{Application Type}
Different types of workflow application can be used to define the scheduling priority based on their context and critical functionality. The same workflow application can differ in priority when it is used in a different environment. Montage Astronomy workflow used for an educational purpose may have a lower priority than a solar research project using the same workflow. Meanwhile, considering the different critical functions of workflows and some events (e.g., an earthquake event occurs on some sites) some workflow applications can be prioritized from the other. For example, CyberShake workflow to predict the ground motion after an earthquake may be prioritized compared to Montage workflow that is used to create a mosaic of the sky image. This priority assignment needs to be designed in a specific policy of the providers that can regulate the fairness of the scheduling process.

\subsubsection{QoS Constraint}
Deriving priority assignment from users' QoS constraint can be done within the scheduling algorithms. This assignment is included in the logic of algorithms to achieve the scheduling objectives. For example, an algorithm that aims to minimize cost while meeting the deadline may consider to de-prioritize and delay the task of a particular workflow that has a more relaxed deadline to re-use the cheapest resources available. In this way, the algorithms must be designed to be aware of the QoS constraints of the tasks to derive these parameters into a priority assignment process during the scheduling.

Furthermore, the challenge of deriving priority assignment from QoS constraints may come from a heterogeneous QoS requirement workload. The algorithms must be able to determine a priority assignment for multiple workflows with different QoS requirements. For example, given two workflows with different QoS parameters, a workflow was submitted with a deadline, while another was included with a budget. The priority assignment can be done by combining these constraints with its application type, user-defined priority or workflow structure.

\subsubsection{User-defined Priority}
On the contrary to the application type priority model that may be arranged through a specific policy, the priority assignment must also consider the user-defined priority in scheduling algorithms. This priority can be defined by users with appropriate compensations for the providers. For example, it is not uncommon in the real-world practice to spend more money to get a prioritized treatment that affects the speed of process and quality of service (e.g., regular and express postal service). It is possible in multi-tenant platforms to accommodate such a mechanism where the users are given the option to negotiate their priority through a monetary cost compensation for the providers. This mechanism is a standard business practice adopted in multi-tenant platforms (e.g., pricing schemes of reserved, on-demand, and spot instances).

\subsubsection{Workflow Size}
Another approach on priority assignment is based on the structure of workflows (e.g., size, parallel tasks, critical path). Prioritizing workflows based on their sizes resembles a traditional way of priority scheduling, such as Shortest Job First (SJF) policy that gives priority to the shortest tasks, and Round Robin (RR) policy that attempts to balance the fairness between tasks with different sizes. This prioritization can be combined with the QoS constraint to produce better fairness between users. For example, a large-scale workflow may have a very extended deadline. Therefore, smaller workflows with tight deadlines can be scheduled between the execution of tasks from this larger workflow.

\subsection{Task Scheduling Model}

Task-resource mapping is the primary activity of scheduling. All of the workflow scheduling algorithms have the purpose of finding the most optimal configuration of task-resource mapping. However, each scheduling problems may have different requirements regarding the quality of service (QoS). In general, there are two standard QoS requirements in workflow scheduling, namely time and cost. The majority of the cases require the algorithms to minimize the overall execution time of the workflow  (i.e., makespan).

On the other hand, the cost of executing the workflows significantly affects the scheduling decisions in utility-based computational infrastructures such as utility grids and cloud environments. It is evident that every user wants to minimize the cost of executing their workflows. These two objectives have opposing goals and a trade-off between them must be considered. This trade-off then is derived into various scheduling objectives such as minimizing cost while meeting the deadline (i.e., the time limit for execution), minimizing makespan while meeting the budget (i.e., the cost limit for execution), or a more loose objective, meeting deadline, and budget.

In multi-tenant platforms, QoS diversity is prevalent due to the different needs of users to execute their workflows. The variety is not only related to the QoS values the users define but also may raise in the form of different scheduling objectives. The various user-defined QoS requirements must be handled in a way that each user's need can be fulfilled without sacrificing the other users served by the systems.

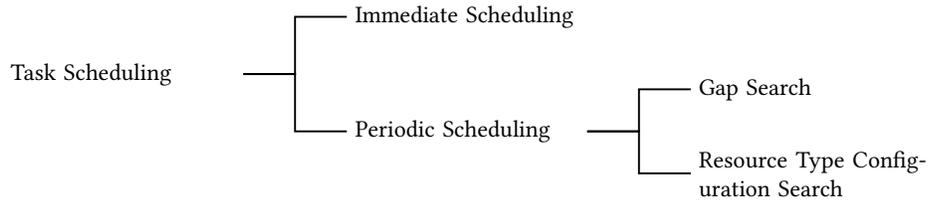
\begin{figure*}[!t]
	\centering
	\begin{tikzpicture}[level distance=2in,sibling distance=.2in,scale=.9]
	\tikzset{edge from parent/.style= 
		{thick, draw,
			edge from parent fork right},every tree node/.style={minimum width=1.3in,text width=1.3in, align=left},grow'=right}
	\Tree 
	[.{Task Scheduling} 
	[.{Immediate Scheduling} 
	]
	[.{Periodic Scheduling} 
	[.{Gap Search} ]
	[.{Resource Type Configuration Search} ]
	]
	]
	\end{tikzpicture}
	\caption{Task Scheduling Model Taxonomy}
	\label{fig:task-resource}
\end{figure*}

All of the surveyed algorithms avoid the meta-heuristics approaches that are known for its computing intensiveness to plan the schedule before runtime. This planning creates an overhead waiting delay as the continuous arriving workflows have to wait for pre-processing before the actual scheduling takes place. Therefore, they use dynamic approaches which reduce the need for intensive computing at the planning phase and aim to achieve a fast scheduling decision by considering the current status of the systems. These approaches can be divided into immediate and periodic scheduling as illustrated in Figure \ref{fig:task-resource}.

\subsubsection{Immediate Scheduling}
Immediate scheduling or just-in-time scheduling is a dynamic scheduling approach in which tasks are scheduled whenever they are ready for scheduling. In the case of multiple workflows, this scheduling approach collects all of the ready tasks from different workflows in a task pool before deciding to schedule based on some particular rules. The immediate scheduling tries to overcome the fast dynamic changes in the environments by adapting the decision based on the current status of the system. However, as the algorithm schedules the tasks based on a limited amount of information (i.e., limited view of the previous and future information), this approach cannot achieve an optimal scheduling plan, but it is an efficient way for multi-tenant platforms that deal with uncertain and dynamic environments.

The immediate scheduling resembles list-based heuristics scheduling. This scheduling approach, in general, has three scheduling phases, task prioritization, task selection, and resource selection. The algorithms repeatedly select a particular task from the scheduling queue that is constructed based on some prioritization method and then picks the appropriate resource for that specific task. For example, in deadline constraint-based heuristics algorithms that aim to minimize the cost while meeting the deadline, the scheduling queue is constructed based on the earliest first deadline (EDF) of the tasks and the cheapest resources that can meet the deadline are chosen to ensure the cost minimization. The time complexity for heuristic algorithms is low. Therefore, it is suitable for multiple workflows scheduling algorithms that deal with the speed to manage the scheduling process for multi-tenant systems.

\subsubsection{Periodic Scheduling}
Periodic scheduling approach schedules the tasks periodically to exploit the possibility to optimize the scheduling of a set of tasks within a period. While in a general batch scheduling, a particular set is constructed based on the size of workload (i.e., schedule the tasks after reaching a certain number), and periodic scheduling schedules the tasks in a set of timeframe. In this case, the periodic scheduling acts as a hybrid approach between static and dynamic scheduling methods. Static, in a way that the algorithms exploit the information of a set of tasks (i.e., structures, estimated runtime) to create an optimal plan, but it does not need to wait for a full workload of tasks to be available. The dynamic sense of algorithms adapts and changes the schedule plan periodically. Hence, periodic scheduling refers to a scheduling technique that utilizes the schedule plan of a set of tasks available in a certain period to produce a better scheduling result. This method is more adaptable to changes and has faster pre-runtime computation than static scheduling techniques since it only includes a small fraction of workload to be optimized rather than the whole workload before runtime. On the other hand, this approach can achieve a better result from having an optimized schedule plan than a typical just-in-time (i.e., immediate) scheduling but with less speed--as a trade-off--to schedule the tasks.

One of the approaches in periodic scheduling is identifying the gaps between tasks during the schedule. The identification uses an estimated runtime of tasks and their possible position in a resource during runtime. The most common techniques to fill the gap identified in the scheduling plan are task rearrangement and backfilling strategy. Task arrangement strategy re-arranges the tasks scheduling plan to ensure the minimal gap in a schedule plan while backfilling allocates the ordering list of tasks and then backfills the holes produced between the allocation using the appropriate tasks. Both strategies do not involve an optimization algorithm that requires the intensive computation since the multi-tenant platforms consider speed in the schedule to cope with the users' QoS requirements.

While gap search is related to the strategy for improving resource utilization, another approach utilizes resource type configuration to optimize the cost spent on leasing computational resources. In a heterogeneous environment where resources are leased from third-party providers with some monetary costs (i.e., utility grids, clouds), determining resource type configuration to optimize the cost of leasing resources is necessary. For example, Dyna algorithm \cite{7044594} that considers the combination use of on-demand and spot instances (i.e., VM pricing schemes) in Amazon EC2, utilizes heuristics to find the optimal resource type configuration to minimize the cost.

\subsection{Resource Provisioning Model}

\begin{figure*}[!t]
	\centering
	\begin{tikzpicture}[level distance=2in,sibling distance=.2in,scale=.9]
	\tikzset{edge from parent/.style= 
		{thick, draw,
			edge from parent fork right},every tree node/.style={minimum width=1.35in,text width=1.35in, align=left},grow'=right}
	\Tree 
	[.{Resource Provisioning} 
	[.{Static Provisioning} 
	]
	[.{Dynamic Provisioning} 
	[.{Workload-aware} ]
	[.{Performance-aware} ]
	]
	]
	\end{tikzpicture}
	\caption{Resource Provisioning Model Taxonomy}
	\label{fig:resprov}
\end{figure*}
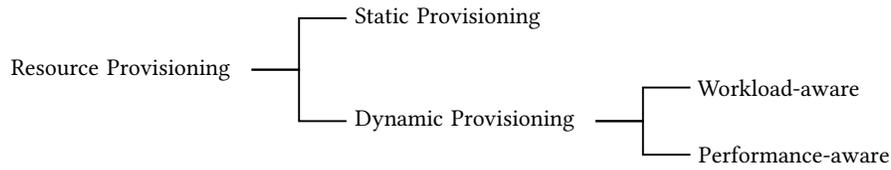

Resource provisioning forms an essential pair with task scheduling. In this stage, scheduling algorithms acquire and allocate resources to execute the scheduled tasks. We derive the categorization of resource provisioning based on the ability of scheduling algorithms to expand and shrink the number of resources within the platforms to accommodate the dynamic workloads of multi-tenant platforms as shown in Figure \ref{fig:resprov}.

\subsubsection{Static Provisioning}
The static provisioning refers to scheduling algorithms where the number of resources used is relatively static along the scheduling process. Therefore, the primary issue in a static resource provisioning is related to the ability of algorithms to optimize the available resources to accommodate multiple users submitting their workflows into the platforms. This condition can be observed from the algorithms that emphasize heavily on the prioritization technique for workflows to be scheduled due to the limited available resources contested by many users. Another aspect is the improvement of resource utilization of the systems which describes the ability of algorithms to allocate a limited number of resources efficiently.

This static provisioning is not exclusive to the non-virtualized environment (e.g., clusters, grids), where it is evident that the number of resources is hardly changing over time. This case also prevails in cloud computing environments where the providers determine the number of VMs to be leased before initiating the platforms and the number remains unchanged over time. In this scenario, the scheduling algorithms do not consider any resource provisioning strategy to scale up and down resources when facing a dynamic workload of workflows.

\subsubsection{Dynamic Provisioning}
As the clouds provide elastic provisioning of virtual machines, the scheduling algorithms of multi-tenant platforms in clouds take advantage of dynamic provisioning approach. The automated scaling of resources that can be easily implemented in clouds has been widely adopted in scheduling algorithms that deal with a dynamic workload where the need of resources can be high at a point (i.e., peak hours), while at the same time the operational cost must be kept at the minimum. To minimize the cost of leasing VMs, they have to be released when the request is low. From the existing algorithms, at least, there are two different approaches to auto-scaling the cloud instances, workload-aware and performance-aware.

Workload-aware dynamic provisioning is related to the ability of algorithms to become aware of the workload status in multi-tenant platforms, and then to act according to the situation. For example, they are acquiring more VMs to accommodate the peak condition. One of the heuristics used in this scenario is based on the deadline constraints of the workload. For example, the algorithms use a task's deadline to decide whether a task should re-use available VMs, provision a new VM that can finish before the deadline, or delay the schedule to re-use future available VMs as long as it does not violate the deadline. This decision is essential as the dynamic workload is common in multi-tenant platforms where the systems cannot predict the future status of the workload. Using this heuristic provisioning additional VMs is more accurate as the acquired new VM is based on the requirement of a particular task being scheduled.

On the other hand, the performance-aware dynamic provisioning refers to an approach of auto-scaling the VMs based on total resource utilization of current provisioned VMs. The algorithms monitor the status of the systems and acquire additional VMs when the usage is high and release several idle VMs when the utilization is low. Maintaining resource utilization at a certain threshold ensures the efficiency of multi-tenant platforms in the scheduling process. The majority of works considering this approach are the ones that consider only homogeneous VM type in their systems. In this way, the algorithms do not need to perform the complicated selection process of the VM types to be chosen.

\section{Survey}

This section discusses a number of surveyed multiple workflows scheduling algorithms from 2008 to 2019 that are relevant to our scope. Each algorithm is classified based on the taxonomy presented in the previous section. Furthermore, the description of the algorithms and their first author affiliation is shown in Table \ref{table:sol} while the classification of existing algorithms is depicted in Table \ref{table:workload} and \ref{table:attribute}.

\subsection{Planner-Guided Scheduling for Multiple Workflows}
RANK\_HYBD algorithm \cite{4626773} was introduced to overcome the impracticality of the ensemble approach (i.e., merging multiple workflows) to handle different submission time of workflows to the system by scheduling individual tasks dynamically. RANK\_HYBD algorithm put together all ready tasks from different workflows into a pool. Then, the algorithm used a modified upward ranking \cite{993206} which calculates the weight of a task based on its relative position from the exit tasks and estimated computational length to assign individual tasks priorities. This task ranking time complexity was $\mathcal{O}\,$($\,T_w\,.\,P_s\,$) for all tasks in a workflow $T_w$ given a set of static processors $P_s$. In contrast with the original upward ranking implementation in HEFT algorithm that chose tasks with higher rank value, RANK\_HYBD preferred tasks with the lowest rank in the pool which created a time complexity of $\mathcal{O}\,$($\,T_r\,.\,P_s\,$) for re-prioritizing all ready tasks $T_r$. In this case, HEFT preferred the tasks from later arriving workflows and the tasks with the most extended estimated runtime, which created an unfair pre-emptive policy for the running workflows. By using the opposite approach, RANK\_HYBD algorithm avoided the pre-emptive scheduling delay of a nearly finished workflow if a new workflow is submitted in the middle of the execution. Finally, RANK\_HYBD schedules each task to the processor that can give the earliest processing time with $\mathcal{O}\,$($\,T_r\,$($\,T_r\,.\,P_s\,$)$\,$) time complexity.

In general, the time complexity of RANK\_HYBD was quadratic to the number of tasks processed. The report showed that RANK\_HYBD outperformed the makespan of RANDOM and FIFO algorithms by an average of 43.6\% on workloads up to 25 multiple workflows. This algorithm was the first solution for multiple workflows scheduling. The approach to tackling dynamic workload of workflows using dynamic prioritization for tasks within a workflow and between workflows has been adopted by many algorithms later. Even though many aspects such as QoS constraints, performance variability, and real workflow applications have not been included in the experiment, this pioneering work became an important benchmark for the following algorithms.

\begin{table*}[!t]
	\begin{center}
		\caption{Description of Multiple Workflows Scheduling Algorithms}
		\label{table:sol}
		\resizebox{\textwidth}{!}{\begin{tabular}{ r l c c }
				\hline \noalign{\vskip 1mm}
				\multicolumn{1}{c}{\textbf{Algorithms}} &\multicolumn{1}{c}{\textbf{References}} &
				\multicolumn{1}{c}{\textbf{First Author Affiliation}}&\multicolumn{1}{c}{\textbf{Keyword}}  \\
				
				\hline \noalign{\vskip 1mm}
				\multirow{1}{*}{RANK\_HYBD} &\multirow{1}{*}{\cite{4626773}}&\specialcell[t]{Wayne State University, United States} & \multirow{1}{*}{Dynamic-guided scheduling} \\ \hline \noalign{\vskip 1mm}
				\multirow{1}{*}{MQMW}&\multirow{1}{*}{\cite{5207867}} & \multirow{2}{*}{Shandong University, China} &\multirow{2}{*}{Multi-QoS scheduling} \\
				\multirow{1}{*}{MQSS}&\multirow{1}{*}{\cite{5420119}} & &\multirow{1}{*}{} \\ \hline \noalign{\vskip 1mm}
				\multirow{1}{*}{P-HEFT} &\multirow{1}{*}{\cite{BARBOSA2011428}}&\multirow{5}{*}{\specialcell[t]{Universidade do Porto, Portugal}} & \multirow{1}{*}{Dynamic-parallel scheduling}\\
				\multirow{1}{*}{FDWS}&\multirow{1}{*}{\cite{arabnejad:hal-00926460}} & &\multirow{1}{*}{Fairness \& priority} \\
				\multirow{1}{*}{MW-DBS}&\multirow{1}{*}{\cite{ARABNEJAD2017120}}  & &\multirow{1}{*}{\specialcell[t]{Deadline-budget constraints}} \\
				\multirow{1}{*}{MQ-PAS}&\multirow{1}{*}{\cite{ARABNEJAD2017211}} & & \multirow{1}{*}{Profit-aware}\\
				\hline \noalign{\vskip 1mm}
				\multirow{1}{*}{OWM}&\multirow{1}{*}{\cite{HSU2011860}} & \multirow{2}{*}{\specialcell[t]{National Chiao-Tung University, Taiwan}} & \multirow{2}{*}{Scheduling framework}\\ 
				\multirow{1}{*}{MOWS}&\multirow{1}{*}{\cite{WANG201635}} & &\\
				\hline \noalign{\vskip 1mm}
				\multirow{1}{*}{EDF\_BF}&\multirow{1}{*}{\cite{STAVRINIDES2011540}} &\multirow{3}{*}{\specialcell[t]{Aristotle University of Thessaloniki, Greece}} &\multirow{2}{*}{Exploit schedule gaps} \\
				\multirow{1}{*}{EDF\_BF\_IC}&\multirow{1}{*}{\cite{7300823}} & & \\
				\multirow{1}{*}{EDF\_BF \textit{In-Mem}}&\multirow{1}{*}{\cite{STAVRINIDES2017120}} & & \multirow{1}{*}{Data-locality perspective}\\ \hline \noalign{\vskip 1mm}
				\multirow{1}{*}{OPHC-TR} &\multirow{1}{*}{\cite{7037702}}&\multirow{2}{*}{\specialcell[t]{The University of Sydney, Australia}} &\multirow{1}{*}{Privacy constraint} \\
				\multirow{1}{*}{DGR}&\multirow{1}{*}{\cite{Chen2015}} & & \multirow{1}{*}{Task rearrangement}\\ \hline \noalign{\vskip 1mm}
				\multirow{1}{*}{Adapt. dual-criteria}&\multirow{1}{*}{\cite{Tsai2015}} &\multirow{1}{*}{\specialcell[t]{National Taichung University of Education, Taiwan}} & \multirow{1}{*}{Partition-based scheduling}\\ \hline \noalign{\vskip 1mm}
				\multirow{1}{*}{MLF\_ID}&\multirow{1}{*}{\cite{CPE:CPE3582}} & \multirow{1}{*}{Fuzhou University, China} & \multirow{1}{*}{Partition-based scheduling} \\ \hline \noalign{\vskip 1mm}
				\multirow{1}{*}{FASTER}&\multirow{1}{*}{\cite{7435325}} & \multirow{7}{*}{\specialcell[t]{National University of Defense Technology, China}} & \multirow{1}{*}{Fault-tolerant} \\
				\multirow{1}{*}{PRS}&\multirow{1}{*}{\cite{7820319}} & &\multirow{1}{*}{Uncertainty-aware}  \\
				\multirow{1}{*}{EONS}&\multirow{1}{*}{\cite{7576489}} & & \multirow{1}{*}{Energy-efficient} \\
				\multirow{1}{*}{EDPRS}&\multirow{1}{*}{\cite{Chen2017}} & & \multirow{1}{*}{Uncertainty-aware} \\
				\multirow{1}{*}{ROSA}&\multirow{1}{*}{\cite{8443134}} & & \multirow{1}{*}{Uncertainty-aware} \\
				\multirow{1}{*}{CERSA}&\multirow{1}{*}{\cite{Chen2018}} & & \multirow{1}{*}{Real-time scheduling} \\ \hline \noalign{\vskip 1mm}
				\multirow{1}{*}{EnReal}&\multirow{1}{*}{\cite{7276993}} &\multirow{1}{*}{\specialcell[t]{Nanjing University, China}} & \multirow{1}{*}{Energy-efficient} \\ \hline \noalign{\vskip 1mm}
				\multirow{1}{*}{Dyna}&\multirow{1}{*}{\cite{7044594}} & \specialcell[t]{Nanyang Technological University, Singapore} & \multirow{1}{*}{Cloud spot instances} \\ \hline \noalign{\vskip 1mm}
				\multirow{1}{*}{FSDP}&\multirow{1}{*}{\cite{7951947}} & \multirow{1}{*}{\specialcell[t]{Dalian University of Technology, China}} & \multirow{1}{*}{Fairness \& priority} \\ \hline \noalign{\vskip 1mm}
				\multirow{1}{*}{F\_DMHSV} &\multirow{1}{*}{\cite{CPE:CPE3782}}&\multirow{2}{*}{\specialcell[t]{Hunan University, China}} & \multirow{1}{*}{Fairness \& priority} \\
				\multirow{1}{*}{DPMMW\&GESMW}&\multirow{1}{*}{\cite{XIE2017}} & & \multirow{1}{*}{Energy-efficient} \\
				\hline \noalign{\vskip 1mm}
				\multirow{1}{*}{CWSA}&\multirow{1}{*}{\cite{7457258}} & \specialcell[t]{Institut National de la Recherche Scientifique, Canada} & \multirow{1}{*}{\specialcell[t]{Exploiting schedule gaps}} \\ \hline \noalign{\vskip 1mm}
				\multirow{1}{*}{EPSM} &\multirow{1}{*}{\cite{RODRIGUEZ2018739}}& \specialcell[t]{The University of Melbourne, Australia} & \multirow{1}{*}{\specialcell[t]{Container service}} \\ \hline \noalign{\vskip 1mm}
				\multirow{1}{*}{MW-HBDCS} &\multirow{1}{*}{\cite{Zhou2018}}& \specialcell[t]{Guangzhou University, China} & \multirow{1}{*}{\specialcell[t]{Deadline-budget constraints}} \\ \hline \noalign{\vskip 1mm}
				\multirow{1}{*}{NOSFA} &\multirow{1}{*}{\cite{8669862}}& \specialcell[t]{Central South University, China} & \multirow{1}{*}{\specialcell[t]{Uncertainty-aware}} \\ \hline \noalign{\vskip 1mm}
		\end{tabular}}
	\end{center}
\end{table*}

\subsection{Multiple QoS Constrained for Multiple Workflows Scheduling}
Multiple QoS Constrained Scheduling Strategy of Multi-Workflows (MQMW) algorithm \cite{5207867} incorporated a similar strategy of RANK\_HYBD to schedule multiple workflows. MQMW prioritized tasks dynamically based on several parameters including resource requirement of a task, time and cost variables, and covariance value between time and cost constraint. This task ranking time complexity was $\mathcal{O}\,$($\,T_w\,.\,C_s\,$) for all tasks in a workflow $T_w$ given a set of static cloud instances $C_s$. The algorithm preferred the tasks with a minimum requirement of resource to execute, minimum time and cost limit, and a task with minimum covariance between its time and cost limit (i.e., when time limit decreases, the cost will highly increase). Each time the scheduling takes place, MQMW re-compute all ready tasks $T_r$ by $\mathcal{O}\,$($\,T_r\,.\,C_s\,$) time complexity. Finally, MQMW schedules each task to the best fit idle cloud instances with $\mathcal{O}\,$($\,T_r\,$($\,T_r\,.\,C_s\,$)$\,$) time complexity.

In general, the time complexity of MQMW was quadratic to the number of tasks processed. It is tested against RANK\_HYBD, even though the RANK\_HYBD was not considered the cost in the scheduling constraint. The evaluation results showed that MQMW outperformed the success rate of RANK\_HYBD \cite{4626773} algorithm by 22.7\%. MQMW was the first attempt to provide the solution of multiple workflows scheduling on the cloud computing environment. However, its cloud model did not resemble the real characteristics that are inherent in clouds such as elastic scalability of instances, on-demand resources, pay-as-you-go pricing schemes, and performance variability of cloud environments.

The MQSS \cite{5420119} algorithm was proposed to overcome shortcomings from the previous MQMW algorithm that include the number of QoS considered in the scheduling and the adoption of a more optimal scheduling strategy. With the relatively same approaches, the MQSS includes other QoS parameters into the scheduling's attributes (e.g., time, cost, availability, reputation, and data quality). In general, MQSS has the same time complexity with MQMW with 12.47\% success rate improvement for the same workloads.

\subsection{Fairness in Multiple Workflows Scheduling}
Parallel Task HEFT (P-HEFT) algorithm \cite{BARBOSA2011428} was the first work of a group from the Universidade do Porto. P-HEFT modeled the non-monotonistic tasks (i.e., the execution time of a task might differ on the different number of resource usage) in their work. The algorithm used a relative length position of a task from the entry task (i.e., t-level/top-level) and exit task (i.e., b-level/bottom-level) to assign the priorities between tasks. This ranking time complexity was $\mathcal{O}\,$($\,T_w\,.\,P_s\,$) for all tasks in a workflow $T_w$ given a set of static processors $P_s$. In their case, the task model was different from other works as it allowed parallel execution of a task in several processors. Furthermore, the processor selection and task scheduling for all ready tasks $T_r$ was $\mathcal{O}\,$($\,T_r\,(\,T_r\,.\,P_s\,)\,$). In general, the complexity of P-HEFT was quadratic to the number of tasks processed. The evaluation results showed that P-HEFT outperformed the schedule length (i.e., makespan) of static algorithm HPTS \cite{4154152} by 34.3\% on workloads of 12 multiple workflows.

The next work from this group was the Fairness Dynamic Workflow Scheduling (FDWS) algorithm \cite{arabnejad:hal-00926460}. FDWS chose a single ready task from each workflow into the pool instead of putting all ready tasks together. Local prioritization within a workflow utilized upward-rank mechanism while the task selection from different workflows to schedule used a percentage of the remaining task number of workflow the task belongs to (PRT) and a task position in its workflow's critical path (CPL). This prioritization time complexity was $\mathcal{O}\,$($\,T_w\,.\,P_s\,$) for all tasks in a workflow $T_w$ given a set of static processors $P_s$. Meanwhile, the resource selection and task scheduling for all ready tasks $T_r$ in FDWS was $\mathcal{O}\,$($\,T_r\,(\,T_r\,.\,P_s\,)\,$). In general, the complexity of FDWS was quadratic to the number of tasks processed. The evaluation results showed that FDWS outperformed the turnaround time  (i.e., makespan plus waiting time) of RANK\_HYBD \cite{4626773} and OWM \cite{HSU2011860} by 5.9\% and 13\% respectively on workloads of 50 multiple workflows.

\begin{table*}[!t]
	\begin{center}
		\caption{Taxonomy of Workload and Deployment Model}
		\label{table:workload}
		\resizebox{\textwidth}{!}{\begin{tabular}{@{\extracolsep{4pt}} r l c c c c c c c@{}}
				\hline \noalign{\vskip 1mm}
				\multicolumn{1}{c}{\multirow{3}{*}{\textbf{Algorithms}}}& \multicolumn{1}{c}{\multirow{3}{*}{\textbf{References}}} &\multicolumn{4}{c}{\textbf{Workload Model}} & \multicolumn{3}{c}{\textbf{Deployment Model}} \\ \cline{3-6} \cline{7-9} \noalign{\vskip 1mm}
				
				& &\multicolumn{2}{c}{\textbf{Workflow Type}} & \multicolumn{2}{c}{\textbf{QoS Requirements}}&\multicolumn{1}{c}{\multirow{2}{*}{\textbf{Non-virtualized}}}&\multicolumn{2}{c}{\textbf{Virtualized}} \\ \cline{3-4} \cline{5-6} \cline{8-9} \noalign{\vskip 1mm}
				
				& &\textbf{Homogen} & \textbf{Heterogen} &
				\textbf{Homogen} & \textbf{Heterogen}& & \textbf{VM-based} & \textbf{Container-based}  \\
				
				\hline \noalign{\vskip 1mm}
				\multirow{1}{*}{RANK\_HYBD}&\cite{4626773} &-&\checkmark &\checkmark&- &\checkmark&-&-  \\ \hline \noalign{\vskip 1mm}
				\multirow{1}{*}{MQMW}&\cite{5207867} &-&\checkmark &\checkmark &-&-&\checkmark&- \\
				\multirow{1}{*}{MQSS}&\cite{5420119} &-&\checkmark &\checkmark &-&\checkmark&-&- \\ \hline \noalign{\vskip 1mm}
				\multirow{1}{*}{P-HEFT}&\cite{BARBOSA2011428} &-&\checkmark &\checkmark &-&\checkmark&-&- \\
				\multirow{1}{*}{FDWS}&\cite{arabnejad:hal-00926460} &- &\multirow{1}{*}{\checkmark} &\multirow{1}{*}{\checkmark}&-&\checkmark&-&- \\
				\multirow{1}{*}{MW-DBS}&\cite{ARABNEJAD2017120} &- &\multirow{1}{*}{\checkmark} &\multirow{1}{*}{\checkmark}&-&\checkmark&-&- \\
				\multirow{1}{*}{MQ-PAS}&\cite{ARABNEJAD2017211} &- &\checkmark &\checkmark &-&-&\checkmark&- \\
				\hline \noalign{\vskip 1mm}
				\multirow{1}{*}{OWM}&\cite{HSU2011860} &- &\checkmark&- &\checkmark &\checkmark&-&- \\ 
				\multirow{1}{*}{MOWS}&\cite{WANG201635} &- &\checkmark&- &\checkmark &\checkmark&-&- \\
				\hline \noalign{\vskip 1mm}
				\multirow{1}{*}{EDF\_BF}&\cite{STAVRINIDES2011540} &- &\checkmark &\checkmark &-&\checkmark&-&- \\
				\multirow{1}{*}{EDF\_BF\_IC}&\cite{7300823} &- &\checkmark &\checkmark &-&\checkmark&-&- \\
				\multirow{1}{*}{EDF\_BF \textit{In-Mem}}&\cite{STAVRINIDES2017120} &- &\checkmark &\checkmark &-&\checkmark&-&- \\ \hline \noalign{\vskip 1mm}
				\multirow{1}{*}{OPHC-TR}&\cite{7037702} &\checkmark &- &\checkmark &-&-&\checkmark&- \\
				\multirow{1}{*}{DGR}&\cite{Chen2015} &- &\checkmark &\checkmark &-&\checkmark&-&- \\ \hline \noalign{\vskip 1mm}
				\multirow{1}{*}{Adapt. dual-criteria}&\cite{Tsai2015} &- &\checkmark &\checkmark&- &\checkmark&-&- \\ \hline \noalign{\vskip 1mm}
				\multirow{1}{*}{MLF\_ID}&\cite{CPE:CPE3582} &- &\checkmark &\checkmark &-&\checkmark&-&- \\ \hline \noalign{\vskip 1mm}
				\multirow{1}{*}{FASTER}&\cite{7435325} &- &\checkmark &\checkmark &-&-&\checkmark&- \\
				\multirow{1}{*}{PRS}&\cite{7820319} &- &\checkmark &\checkmark &-&-&\checkmark&- \\
				\multirow{1}{*}{EONS}&\cite{7576489} &- &\checkmark &\checkmark &-&-&\checkmark&- \\
				\multirow{1}{*}{EDPRS}&\cite{Chen2017} &- &\checkmark &\checkmark &-&-&\checkmark&- \\
				\multirow{1}{*}{ROSA}&\cite{8443134} &- &\checkmark &\checkmark &-&-&\checkmark&- \\
				\multirow{1}{*}{CERSA}&\cite{Chen2018} &- &\checkmark &\checkmark &-&-&\checkmark&- \\ \hline \noalign{\vskip 1mm}
				\multirow{1}{*}{EnReal}&\cite{7276993} &- &\checkmark &\checkmark &-&-&\checkmark&- \\ \hline \noalign{\vskip 1mm}
				\multirow{1}{*}{Dyna}&\cite{7044594} &- &\checkmark &\checkmark &-&-&\checkmark&- \\ \hline \noalign{\vskip 1mm}
				\multirow{1}{*}{FSDP}&\cite{7951947} &- &\checkmark &\checkmark &-&\checkmark&-&- \\ \hline \noalign{\vskip 1mm}
				\multirow{1}{*}{F\_DMHSV}&\cite{CPE:CPE3782} &- &\checkmark &\checkmark &-&\checkmark&-&- \\
				\multirow{1}{*}{DPMMW\&GESMW}&\cite{XIE2017} &- &\checkmark &\checkmark &-&-&\checkmark&- \\
				\hline \noalign{\vskip 1mm}
				\multirow{1}{*}{CWSA}&\cite{7457258} &- &\checkmark &\checkmark &-&-&\checkmark&- \\ \hline \noalign{\vskip 1mm}
				\multirow{1}{*}{EPSM}&\cite{RODRIGUEZ2018739} &- &\checkmark &\checkmark &-&-&\checkmark&- \\ \hline \noalign{\vskip 1mm}
				\multirow{1}{*}{MW-HBDCS}&\cite{Zhou2018} &- &\checkmark &\checkmark &-&\checkmark&-&- \\ \hline \noalign{\vskip 1mm}
				\multirow{1}{*}{NOSFA}&\cite{8669862} &- &\checkmark &\checkmark &-&-&\checkmark&- \\ \hline \noalign{\vskip 1mm}
		\end{tabular}}
	\end{center}
\end{table*}

Multi-Workflow Deadline-Budget Scheduling Algorithm (MW-DBS) \cite{ARABNEJAD2017120} was their work that addressed the utility aspect of a heterogeneous multi-tenant distributed system. This algorithm included deadline and budget as constraints. Furthermore, local priority was assigned using the same method from FDWS which creates $\mathcal{O}\,$($\,T_w\,.\,P_s\,$) time complexity. However, instead of using PRT and CPL, MW-DBS used task's deadline and workflow's scheduled tasks ratio for assigning global priority. Finally, MW-DBS modified the processor selection phase, in which it included a budget limit for task processing as a quality measure with $\mathcal{O}\,$($\,T_r\,.\,P_s\,$) time complexity. Furthermore, the complexity of resource selection and task scheduling for all ready tasks $T_r$ was $\mathcal{O}\,$($\,T_r\,(\,T_r\,.\,P_s\,)\,$). In general, the complexity of MW-DBS was quadratic to the number of tasks processed. The evaluation results showed that MW-DBS outperformed the success rate of FDWS \cite{arabnejad:hal-00926460} and its variants by 43\% on workloads of 50 multiple real-world application workflows.

The latest work on multiple workflow scheduling was the Multi-QoS Profit-Aware Scheduling algorithm (MQ-PAS) \cite{ARABNEJAD2017211}. MQ-PAS was designed not only for the cloud computing environment but also for the general utility-based distributed system. Task ranking and selection complexity in MQ-PAS was $\mathcal{O}\,$($\,T_w\,.\,C_s\,$) for all tasks in a workflow $T_w$ given a set of static cloud instances $C_s$. Meanwhile, the quality measure in cloud instances selection was $\mathcal{O}\,$($\,T_r\,.\,C_s\,$). Furthermore, the complexity of resource selection and task scheduling for all ready tasks $T_r$ was $\mathcal{O}\,$($\,T_r\,(\,T_r\,.\,C_s\,)\,$). In general, the complexity of MQ-PAS was quadratic to the number of tasks processed. The evaluation results showed that MQ-PAS outperformed the success rate of FDWS \cite{arabnejad:hal-00926460} by only 1\% but significantly 20\% improvement of profit on workloads of 50 multiple real-world application workflows.

Several variations of multiple workflows scheduling scenarios were covered in their works. One of the specific signatures from this group is the strategy of choosing a single ready task from workflow to compete in the scheduling cycle with the other workflows. This strategy represents the term "Fairness" that becomes the primary concern in most of their works. However, with their broad scenarios of algorithms that were intended to cover as general as possible the process in a multi-tenant distributed systems, specific requirements in clouds that were different from utility grids (e.g., billing period schemes, dynamic and uncertain environment) were not considered in their works.

\subsection{Online Multiple Workflows Scheduling Framework}
A group from The National Chiao-Tung University focused on developing a scheduling framework for multiple workflows scheduling. Their first algorithm, called the Online Workflow Management (OWM) \cite{HSU2011860} consisted of four phases, critical path workflow Scheduling (CPWS), task scheduling, multi-processor task re-arrangement, and adaptive allocation (AA). The CPWS phase ranked all tasks $T_w$ based on their relative position in their workflows before they were submitted to the scheduling queue to create a schedule plan with $\mathcal{O}\,$($\,T_w\,.\,P_s\,$) time complexity. If there were some gaps in a schedule plan, task re-arrangement took place to fill the gaps to improve resource utilization that creates $\mathcal{O}\,$($\,{T_r}^2\,$) time complexity. Furthermore, AA scheduled the highest priority task from the queue that was constructed based on the near-optimal schedule plan which time complexity was $\mathcal{O}\,$($\,T_r\,(\,T_r\,.\,P_s\,+\,P_s\,)\,$. In general, the complexity of OWM was quadratic to the number of tasks processed. The evaluation results showed that OWM outperformed the makespan of RANK\_HYBD \cite{4626773} by an average of 13\% on workloads of 100 multiple workflows.

In their following work, they extended OWM into Mixed-Parallel Online Workflow Scheduling (MOWS) \cite{WANG201635}. They modified the CPWS phase using the Shortest-Workflow-First (SWF) policy combined with the critical path prioritization. Then, MOWS used the priority-based backfilling to fill the hole of a schedule in the task re-arrangement stage. The pre-emptive task scheduling policy was introduced in the AA phase, so the algorithm allowed the system to schedule the shortest workflow with pre-emptive strategy and stopped it when the higher priority workflow was ready to run. The difference between MOWS and OWM is the task re-arrangement phase which used the priority-based backfilling that takes a lower time complexity of $\mathcal{O}\,$($\,{T_r}\,$). In general, the complexity of MOWS was quadratic to the number of tasks processed. The evaluation results showed that MOWS outperformed the turnaround time (i.e., makespan plus waiting time) of OWM \cite{HSU2011860} by an average of 20\% on workloads of 100 multiple workflows.

Both OWM and MOWS utilized periodic scheduling, in which it periodically created a schedule plan for a set of ready tasks before submitting it to the scheduling queue. In this way, the algorithm can produce better scheduling results without having a compute-intensive optimization beforehand. However, this approach may still create a bottleneck of pre-processing computation if the number of ready tasks in the pool increases. Implementing a strategy to create a fairness scenario when selecting ready tasks to reduce the complexity of calculating a schedule plan may work to enhance this scheduling framework.

\subsection{Real-time Multiple Workflows Scheduling}
One of the active groups that focused on real-time and uncertainty aspects of multiple workflows scheduling was the Stavrinides Group from The Aristotle University of Thessaloniki, Greece. They did impressive works on multiple workflows scheduling that explicitly addressed the uncertainty in cloud computing environments.

Their first work was the Earliest Deadline First with Best Fit (EDF\_BF) algorithm \cite{STAVRINIDES2011540}. The EDF policy was used for the task selection phase, and the BF component was the strategy for exploiting the schedule gap. EDF\_BF incorporated schedule gap exploitation that can be identified through the estimated position of a task's execution in a specified resource. From all of the possible positions, the algorithm exploited the holes using a bin packing technique to find the best fit for a task's potential position in a particular resource. Later, the result can also be used to determine which resource should be selected for that specific task. Given a set of ready tasks $T_r$ processed each time and static processor $P_s$ available, task selection complexity in EDF\_BF was

\begin{landscape}
	\begin{table*}[p!]\centering
		\caption{Taxonomy of Priority Assignment, Task Scheduling, and Resource Provisioning Model}
		\label{table:attribute}
		\resizebox{1.22\textwidth}{!}{\begin{tabular}{@{\extracolsep{4pt}} r l c c c c c c c c c c c@{}}
				\hline \noalign{\vskip 1mm}
				\multicolumn{1}{c}{\multirow{2}{*}{\textbf{Algorithms}}}&\multicolumn{1}{c}{\multirow{2}{*}{\textbf{References}}} &\multicolumn{4}{c}{\textbf{Priority Assignment Model}}&\multicolumn{3}{c}{\textbf{Task Scheduling Model}}&\multicolumn{3}{c}{\textbf{Resource Provisioning Model}} \\ \cline{3-6} \cline{7-9} \cline{10-12} \noalign{\vskip 1mm}
				
				& &\multirow{2}{*}{\textbf{App Type}} & \multirow{2}{*}{\textbf{QoS Constraint}}&\multirow{2}{*}{\textbf{User-defined}}&\multirow{2}{*}{\textbf{Wf Structure}}&\multirow{2}{*}{\textbf{Immediate}}&\multicolumn{2}{c}{\textbf{Periodic}}&\multirow{2}{*}{\textbf{Static}}&\multicolumn{2}{c}{\textbf{Dynamic}} \\ \cline{8-9} \cline{11-12} \noalign{\vskip 1mm}
				
				& & & & & & &\textbf{Gap Search} &\textbf{Conf. Search} & &\textbf{Workload}&\textbf{Performance} \\
				
				\hline \noalign{\vskip 1mm}
				\multirow{1}{*}{RANK\_HYBD}&\cite{4626773}&- &- &- &\checkmark&\checkmark&-&-&\checkmark&-&-\\ \hline \noalign{\vskip 1mm}
				\multirow{1}{*}{MQMW}&\cite{5207867}&- &\checkmark &- &-&\checkmark&-&-&\checkmark&-&- \\
				\multirow{1}{*}{MQSS}&\cite{5420119}&- &\checkmark &- &-&\checkmark&-&-&\checkmark&-&- \\ \hline \noalign{\vskip 1mm}
				\multirow{1}{*}{P-HEFT}&\cite{BARBOSA2011428}&- &- &- &\checkmark&\checkmark&-&-&\checkmark&-&-\\
				\multirow{1}{*}{FDWS}&\cite{arabnejad:hal-00926460}&- &- &- &\multirow{1}{*}{\checkmark}&\checkmark&-&-&\checkmark&-&- \\
				\multirow{1}{*}{MW-DBS}&\cite{ARABNEJAD2017120}&- &\multirow{1}{*}{\checkmark} &- &\checkmark&\checkmark&-&-&\checkmark&-&- \\
				\multirow{1}{*}{MQ-PAS}&\cite{ARABNEJAD2017211}&- & \checkmark&- &\checkmark&\checkmark&-&-&\checkmark&-&- \\
				\hline \noalign{\vskip 1mm}
				\multirow{1}{*}{OWM}&\cite{HSU2011860}&- &- &- & \checkmark&-&\checkmark&-&\checkmark&-&-\\ 
				\multirow{1}{*}{MOWS}&\cite{WANG201635}&- &- &- & \checkmark&-&\checkmark&-&\checkmark&-&-\\
				\hline \noalign{\vskip 1mm}
				\multirow{1}{*}{EDF\_BF}&\cite{STAVRINIDES2011540}&- & \checkmark&- &-&-&\checkmark&-&\checkmark&-&- \\
				\multirow{1}{*}{EDF\_BF\_IC}&\cite{7300823}&- &\checkmark &- &-&-&\checkmark&-&\checkmark&-&- \\
				\multirow{1}{*}{EDF\_BF \textit{In-Mem}}&\cite{STAVRINIDES2017120}&- &\checkmark &- &-&-&\checkmark&-&\checkmark&-&- \\ \hline \noalign{\vskip 1mm}
				\multirow{1}{*}{OPHC-TR}&\cite{7037702}& \checkmark&\checkmark &- &-&\checkmark&-&-&-&\checkmark&- \\
				\multirow{1}{*}{DGR}&\cite{Chen2015}&- &\checkmark &- &-&-&\checkmark&-&\checkmark&-&- \\ \hline \noalign{\vskip 1mm}
				\multirow{1}{*}{Adapt. dual-criteria}&\cite{Tsai2015}&- &\checkmark &- & \checkmark&-&\checkmark&-&\checkmark&-&-\\ \hline \noalign{\vskip 1mm}
				\multirow{1}{*}{MLF\_ID}&\cite{CPE:CPE3582}&- &\checkmark &- &-&\checkmark&-&-&-&\checkmark&- \\ \hline \noalign{\vskip 1mm}
				\multirow{1}{*}{FASTER}&\cite{7435325}&- &\checkmark &- &-&-&\checkmark&-&-&\checkmark&- \\
				\multirow{1}{*}{PRS}&\cite{7820319}&- &\checkmark &- &-&\checkmark&-&-&-&\checkmark&- \\
				\multirow{1}{*}{EONS}&\cite{7576489}&- &- &- &-&\checkmark&-&-&-&\checkmark&- \\
				\multirow{1}{*}{EDPRS}&\cite{Chen2017}&- &\checkmark &- &-&\checkmark&-&-&-&\checkmark&- \\
				\multirow{1}{*}{ROSA}&\cite{8443134}&- &\checkmark &- &-&\checkmark&-&-&-&\checkmark&- \\
				\multirow{1}{*}{CERSA}&\cite{Chen2018}&- &\checkmark &\checkmark &-&\checkmark&-&-&-&\checkmark&- \\ \hline \noalign{\vskip 1mm}
				\multirow{1}{*}{EnReal}&\cite{7276993}&- &\checkmark &- &-&-&-&\checkmark&-&\checkmark&- \\ \hline \noalign{\vskip 1mm}
				\multirow{1}{*}{Dyna}&\cite{7044594}&- &\checkmark &- &-&-&-&\checkmark&-&\checkmark&- \\ \hline \noalign{\vskip 1mm}
				\multirow{1}{*}{FSDP}&\cite{7951947}&- &\checkmark &- &-&\checkmark&-&-&\checkmark&-&- \\ \hline \noalign{\vskip 1mm}
				\multirow{1}{*}{F\_DMHSV}&\cite{CPE:CPE3782}&- &- &- & \checkmark&-&\checkmark&-&\checkmark&-&-\\
				\multirow{1}{*}{DPMMW\&GESMW}&\cite{XIE2017}&- & \checkmark&- &-&-&\checkmark&-&-&\checkmark&- \\
				\hline \noalign{\vskip 1mm}
				\multirow{1}{*}{CWSA}&\cite{7457258}&- & \checkmark&- &-&-&\checkmark&-&-&-&\checkmark \\ \hline \noalign{\vskip 1mm}
				\multirow{1}{*}{EPSM}&\cite{RODRIGUEZ2018739}&- & \checkmark&- &-&\checkmark&-&-&-&\checkmark&- \\ \hline \noalign{\vskip 1mm}
				\multirow{1}{*}{MW-HBDCS}&\cite{Zhou2018}&- & \checkmark&- &\checkmark&\checkmark&-&-&\checkmark&-&- \\ \hline \noalign{\vskip 1mm}
				\multirow{1}{*}{NOSFA}&\cite{8669862}&- & \checkmark&- &\checkmark&\checkmark&-&-&-&\checkmark&- \\ \hline \noalign{\vskip 1mm}
		\end{tabular}}
	\end{table*}
\end{landscape}

\noindent
$\mathcal{O}\,$($\,T_r\,.\,P_s\,$). Meanwhile, the processor selection and schedule gap exploitation was $\mathcal{O}\,$($\,T_r\,(\,T_r\,+\,T_r\,.\,P_s\,$). In general, the complexity of EDF\_BF was quadratic to the number of tasks processed. The evaluation results showed that EDF\_BF outperformed the job guarantee ratio (i.e., success rate) of its variants with HLF (Highest-Level First) and LSFT (Least-Space-Time First) policy on task selection by an average of 10\%.

Another work was the Earliest Deadline First with Best Fit and Imprecise Computation (EDF\_BF\_IC) algorithm \cite{7300823}, which extended the previous algorithm with imprecise computation. The imprecise computation was firstly introduced in \cite{STAVRINIDES20101004} to tackle the problem in a real-time environment that was often needed to produce an early proximate result within a specified time limit. The imprecise computation model was implemented by dividing the task's components into a mandatory and optional component. A task was considered meeting the deadline if its mandatory part was completed, while the optional component may be fully executed, partially executed, or skipped. The evaluation results showed that EDF\_BF\_IC outperformed the success rate of its baseline EDF by an average of 16\% and cost-saving improvement by 12\%.

Furthermore, this group explored data-locality and in-memory processing for multiple workflow scheduling \cite{STAVRINIDES2017120}. In this case, they combined EDF\_BF algorithm with a distributed in-memory storage solution called Hercules \cite{Duro:2013:HPS:2488551.2488598} to evaluate a different way of communication of workflow tasks. They considered two different communication scenarios, communication through a network and via temporary files utilizing the Hercules in-memory storage solution. The results showed that scheduling performance increased when the I/O to computation ratio reduced by using in-memory storage which enforced the locality of data. The evaluation results showed that the application completion ratio (i.e., success rate) improves as the tardiness bound (i.e., soft deadline ratio) increases while the average makespan deteriorate. Besides, the average makespan of the completed workflows improves as the I/O activities decrease.

Despite the variation, their algorithm's main idea was to schedule all of the ready tasks using EDF policy for resources that can allow the tasks to finish in their earliest time. The algorithm maintained a local queue for each resource, and then, optimized the local allocated queue using gaps filling techniques and, in one of the works, manipulated a small portion of the tasks that may have a little significance (i.e., imprecise computation). Their algorithms were designed for a multi-tenant system with a static number of resources. Therefore, the design may not be suitable for cloud computing environments--in which were suffered most by the uncertainty problems--where the auto-scaling of resources is possible.

\subsection{Adaptive and Privacy-aware Multiple Workflows Scheduling}
A group from The University of Sydney introduced an excellent work of multiple workflows scheduling that concerned on the privacy of users \cite{7037702}. They developed two algorithms; Online Multiterminal Cut for Privacy in Hybrid Clouds using PCP Ranking (OMPHC-PCPR) and Online Scheduling for Privacy in Hybrid Clouds using Task ranking (OPHC-TR). OMPHC-PCPR was merging multiple workflows into one single workflow before scheduling. Hence, this solution is out of our scope but the other one, OPHC-TR, used an approach that is inclusive of our study. Both algorithms calculated the privacy level of each workflow before they decided to schedule them in private or public clouds. The private clouds were used mainly for the workflow that comprised a high level of privacy parameters. The main differences between the two algorithms were their input. While OMPHC-PCPR considered a merged single workflow from several workflows, OPHC-TR processed each task using a rank mechanism to decide which tasks were submitted into the scheduling queue. Given a set of tasks in a workflow $T_w$ and static processors $P_s$ available, task ranking and selection complexity in OPHC-TR was $\mathcal{O}\,$($\,T_w\,.\,P_s\,$). Meanwhile, the resource selection and task scheduling for all ready tasks $T_r$ was $\mathcal{O}\,$($\,T_r\,(\,T_r\,.\,P_s\,)\,$). In general, the complexity of OPHC-TR was quadratic to the number of tasks processed. The evaluation results showed that OMPHC-PCPR outperformed the cost-saving of OPHC-TR algorithm by 50\%. However, in this work, the overhead of merging several workflows in OMPHC-PCPR did not being evaluated thoroughly. Such an approach may result in a very high bottleneck when the number of workflows arriving reached a certain high particular number.

Another work from this group was the DGR algorithm \cite{Chen2015}. This algorithm used heuristics, which started the solution with the initial reservation of resources for scheduling particular tasks which time complexity was $\mathcal{O}\,$($\,T_w\,.\,P_s\,$) for all tasks in a workflow $T_w$ given a set of static processors $P_s$. During the execution, uncertainty (i.e., performance and execution time variation) may profoundly affect the initial reservation and break the schedule plan. In this case, the algorithm rescheduled the tasks to handle the broken reservation. DGR utilized task rearrangement techniques and exploited a dynamic search tree to fix this reservation with $\mathcal{O}\,$($\,T_r\,(\,P_s\, + \,T_r\,.\,P_s\,)\,$) time complexity. In general, the complexity of DGR was quadratic to the number of tasks processed. The evaluation results showed that DGR outperformed the makespan of a traditional HEFT algorithm by 30\% on workloads of 300 multiple workflows.

\subsection{Adaptive Dual-criteria Multiple Workflows Scheduling}
Another adaptive approach in scheduling multiple workflows was an adaptive dual-criteria algorithm \cite{Tsai2015}. This algorithm used heuristics that utilized scheduling adjustment via task re-arrangement. An essential strategy to this algorithm was the clustering of tasks and treated them as an integrated set in scheduling to minimize the critical data movement within tasks. Hence, any re-arrangement or adjustment to fill the schedule holes involved the set of tasks to be moved. Given a set of tasks in a workflow $T_w$ processed each time, task group after clustering $T_g$ where $T_g$ $\leq$ $T_w$, and static processors $P_s$ available, the task initial clustering process complexity was $\mathcal{O}\,$($\,{T_w}^2\,$). Meanwhile, the adjustment of idle time gap selection was $\mathcal{O}\,$($\,T_g\,(\,T_g\,+\,P_s\,)\,$), it got a higher complexity compared to a simple Best-Fit and EFT calculation of single task due to the $T_g$ constraint . Finally, the complexity of adaptive task group re-arrangement was $\mathcal{O}\,$($\,T_g\,P_s\,$). In general, the complexity of this algorithm was quadratic to the number of tasks and task groups processed. The evaluation results showed that this algorithm outperformed the makespan of the similar process using traditional Best-Fit and EFT approaches by up to 29\% in various scenarios.

Since the approach used was the periodic scheduling, the frequency of scheduling cycle becomes critical. Infrequent scheduling cycle implies to the broader set of tasks to be processed which may result in a more optimized scheduling plan but potentially required a more intensive computation for creating the plan. Meanwhile, a perpetual cycle may fasten the scheduling plan computation due to its size of tasks but may reduce the quality of a schedule. This variation was not being addressed and explored in-depth by the authors. Besides, the treatment of a cluster of tasks increases the coarse-granularity of scheduling that may widen the gaps produced. In this way, the task re-arrangement may hardly find the holes that can be fit by a coarse-grained set of clustered tasks.

\subsection{Multiple Workflows Scheduling on Hybrid Clouds}
Another work designed for hybrid cloud environments was the Minimum-Load-Longest-App-First with the Indirect Transfer Choice (MLF\_ID) \cite{CPE:CPE3582}. The term Load-Longest-App had a similar concept to the critical path. So, MLF\_ID was a heuristic algorithm that incorporated the workflows prioritization based on their critical path and exploited the use of private clouds before leasing the resources in public clouds. MLF\_ID partitioned the workflow based on a hierarchical iterative application partition (HIAP) to eliminate data dependencies between a set of tasks by clustering tasks with dependencies into the same set before scheduling them into either private or public clouds.

Given a set of tasks in a workflow $T_w$ processed each time, static private cloud resources $C_s$, and dynamic public cloud resources $C_d$, the application partition complexity was $\mathcal{O}\,$($\,T_w\,(\,C_s\,+\,C_d\,)\,$). Meanwhile, the ready tasks $T_r$ scheduling which included the decision to schedule on public cloud was $\mathcal{O}\,$($\,T_r\,.\,C_d\,$) or private cloud was $\mathcal{O}\,$($\,T_r\,(\,T_r\,+\,C_s\,$). In general, the complexity of this algorithm was quadratic to the number of tasks processed. The evaluation results showed that the combined resources of hybrid clouds could minimize the total execution cost when the number of workflows can be allocated as much as possible to the private resources. However, the private cloud capacity was restrained as the scaling process is not as simple of such an approach in public clouds.

The use of hybrid clouds in this work was emphasized to extend the computational capacity when the available on-premises (i.e., private clouds) was not enough to serve the workloads. Firstly, the tasks were scheduled to the private clouds, and whenever the capacity was not possible to process, they were being transferred to public clouds. Even though the tasks had been partitioned to make sure that the data transfer between them was minimum, the decision to move to public clouds evoked a possible transfer overhead problem. Therefore, some improvements can be made by implementing a policy to decide whether a set of tasks was considered impractical to process in private clouds that include some intelligence, which can be designed to predict the possible overhead in the future of the system. In this way, instead of directly transferring the execution to the public clouds that incite not only the additional cost but also the transfer overhead, the algorithm can decide whether it should move the execution or delay the process waiting the next available resources.

\subsection{Proactive and Reactive Scheduling for Multiple Workflows}
Another group that focused on real-time and uncertainty problems in scheduling was a group from The National University of Defense Technology, China. They proposed the algorithms that dynamically exploited proactive and reactive methods in multiple workflows scheduling.

Their first work was Proactive Reactive Scheduling (PRS) algorithm \cite{7820319}. The proactive phase calculated the estimated earliest start and the complete time of tasks and then scheduled them dynamically based on a simple list-based heuristic. This method had been incorporated into many algorithms for multiple workflows scheduling. However, using only the proactive method was unable to tackle the uncertainties (e.g., performance variation, overhead delays) that led to sudden changes in the system. Then, the PRS algorithm introduced a reactive phase whenever two disruptive events occurred (i.e., the arrival of new workflow and finishing time of a task). The reactive phase was triggered by two disruption events to re-do (i.e., update) the scheduling process based on the latest system status. Given a set of tasks in a workflow $T_w$ processed each time and dynamic cloud resources $C_d$ available, time complexity of task ranking in PRS was $\mathcal{O}\,$($\,T_w\,$). Meanwhile, the VM selection and task scheduling for all ready tasks $T_r$ was $\mathcal{O}\,$($\,T_r\,(\,T_r\,.\,C_d\,$). In general, the complexity of PRS was quadratic to the number of tasks processed. The evaluation results showed that PRS outperformed the cost-savings of modified SHEFT \cite{TANG20111083} and RTC \cite{6838754} algorithms for multiple workflows by 50.94\% and 67.23\% respectively on workloads of 1000 multiple workflows.

They then extended the PRS into Event-driven and Periodic Rolling Strategies (EDPRS) algorithm \cite{Chen2017}. EDPRS tackled a flaw in the PRS algorithm, that, if none of the two disruption events happened, the scheduling process could not be pushed forward. They introduced a periodic rolling strategy (i.e., scheduling cycle) that drove the re-iteration of the schedule. In this way, even though no disruption events occurred, the algorithm repeated their scheduling activities after a specific periodic rolling time. In general, the complexity of EDPRS was quadratic to the number of tasks processed, similar to PRS algorithm time complexity. The evaluation results showed that EDPRS outperformed the cost-savings of modified SHEFT \cite{TANG20111083} and RTC \cite{6838754} algorithms for multiple workflows by 12.98\% and 21.57\% respectively. Both PRS and EDPRS worked well in handling the uncertainty in cloud computing environments.

This group also worked on energy-efficient multiple workflow scheduling algorithms. Their work was the Energy-Efficient Online Scheduling Algorithm (EONS) \cite{7576489}. EONS was different from the other energy-efficient scheduling algorithms due to its focus on fast and real-time oriented scheduling. EONS utilized simple auto-scaling techniques to lower energy consumption instead of optimizing energy usage using techniques such as VM live migration and VM consolidation. The scaling method used simple heuristics that considered the load of the physical host and the hardware efficiency. Given a set of tasks in a workflow $T_w$ processed each time and dynamic cloud resources $C_d$ available, task ranking complexity in EONS was $\mathcal{O}\,$($\,T_w\,$). Meanwhile, the VM selection and task scheduling for all ready tasks $T_r$ was $\mathcal{O}\,$($\,T_r\,(\,T_r\,.\,C_d\,$). In general, the complexity of EONS was quadratic to the number of tasks processed. The evaluation results showed that EONS outperformed the energy-savings of modified EASA \cite{Ebrahimirad2015} and ESFS \cite{7103444} algorithms for multiple workflows by 45.64\% and 35.98\% respectively.

Another work from this group addressed the failure in multiple workflows scheduling. The algorithm, called FASTER, \cite{7435325} utilized primary backup technique to handle the failure. To the best of our knowledge, this was the only fault-tolerant algorithm for multiple workflows scheduling. As part of the pre-processing phase, they scheduled two copies of a task (i.e., primary and backup copy) based on the FCFS policy. The workflows were accepted for execution when their both primary and backup copies were successfully met their deadlines in the estimation phase. Whenever a task was not able to meet its deadline, the algorithm re-calculates its earliest start time. This estimation takes $\mathcal{O}\,$($\,{T_w}^2\,$) given a set of tasks in a workflow $T_w$ processed each time. FASTER then ensured that the primary copy was distributed among all available host as part of its fault-tolerant strategy. This heuristic requires periodic scanning of all VMs $C_d$ within the available physical host $H_d$ in the system. The complexity of host monitoring phase was $\mathcal{O}\,$($\,H_d\,(\,C_d\,.\,T_w\,)\,$) for each primary and backup type of tasks. In general, the complexity of FASTER was quadratic to the number of tasks processed. The evaluation results showed that FASTER outperformed the modified eFRD \cite{QIN2006331} algorithm for multiple workflows by 239.66\% in terms of guarantee ratio and 63.79\% in terms of resource utilization.

Their next algorithms were called ROSA \cite{8443134} and CERSA \cite{Chen2018} algorithms. These algorithms were the improvement of the PRS and EDPRS algorithms that specifically tackle the uncertainties in executing multiple real-time workflows. While their previous algorithm EDPRS relies on a periodic trigger to clear a tasks pool beside the arrival of new workflows, ROSA and CERSA initiate the scheduling based on specific disturbance events. ROSA defined the triggering events like the arrival of new workflows and the completion of a task in a particular cloud instance. On the other hand, CERSA added the arrival of the urgent task as one of the triggering events. In general, the time complexity of CERSA and ROSA are quadratic, similar to their previous algorithms (i.e., PRS and EDPRS). The evaluation results showed that in terms of monetary cost, ROSA outperformed EPSM \cite{RODRIGUEZ2018739} and CWSA \cite{7457258} by 10.07\% and 23.18\% while CERSA outperformed CWSA \cite{7457258} and OPHC-TR \cite{7037702} by 8.31\% and 17.22\% respectively.

The algorithms emphasize a specific strategy to handle real-time scenario by using an immediate scheduling approach which includes the update strategy to adapt to changes dynamically. However, this dynamic approach, especially on the energy-efficient and fault-tolerant problem, can be improved by optimizing the VM placement on the physical machine since the algorithms may have access to the information of the physical infrastructure.

\subsection{Energy Aware Scheduling for Multiple Workflows}
A group from Nanjing University, China proposed an algorithm for multiple workflows scheduling that was called EnReal--an energy-aware resource allocation method for workflow in the cloud environment \cite{7276993}. While the previous energy-aware algorithm--EONS--utilized auto-scaling techniques to lower the energy consumption, EnReal exploited the VM live migration-based policy. The algorithm partitioned all of the ready tasks in the scheduling queue based on their requested start time and allocated them to the resources on the same physical machine. The adjustment was made whenever a load of the physical machine was exceeding the threshold, and then, VM live migration policy took place.

EnReal also adjusted the VM allocation dynamically whenever a task was finished. Combined with the physical machine resource monitoring, the global resource allocation method emphasized the energy saving of the platform. However, the partitioning method in EnReal did not consider the data dependencies between the tasks that imply a data transfer overhead between tasks when they were allocated to different physical machines. The energy-aware resource allocation policy in EnReal should have complemented by an ability to aware of data-locality. This policy will not only minimize energy consumption but also improves the scheduling results in term of total execution cost and makespan. In general, the most intensive phase in EnReal was the resource monitoring that takes quadratic time complexity. Furthermore, the performance evaluation results showed that EnReal outperformed the modified energy-aware Greedy-D \cite{6681002} algorithm in terms of energy efficiency by 18\% in average.

\subsection{Monetary Cost Optimization for Multiple Workflows on Commercial Clouds}
A group from the National University of Singapore, proposed Dyna \cite{7044594}, an algorithm that concerned on the clouds dynamicity nature. They introduced a probabilistic guarantee of any defined SLAs of workflow users as it was the closest assumption to uncertainty environment in clouds. This approach was a novel contribution since the majority of the works assumed deterministic SLAs in their algorithms. Dyna aimed to minimize multiple workflows scheduling execution cost by utilizing VMs with spot instances pricing scheme in Amazon EC2 along with its on-demand instances. Dyna started with the initial configuration of different cloud instance types and refined the configuration iteratively to get the better scenario that minimizes the cost while meeting the deadline. In general, the complexity of Dyna was quadratic to the number of tasks processed. The evaluation results showed that Dyna outperformed the monetary cost of modified MOHEFT \cite{6427573} algorithm by 74\% in average.

Dyna presented an exploration of possible cost reduction in executing multiple workflows by utilizing the spot instances in Amazon EC2. Since multi-tenant platforms that were assumed in their work acted as a service provider for many users, the use of reserved instances in Amazon EC2 may further reduce the cost of running the platform. Comparison between on-demand, spot, and reserved instances in Amazon EC2 needs to be done further to deepen the plausible scenario on minimizing the execution cost of multiple workflows in clouds.

\subsection{Fairness Scheduling for Multiple Workflows}
Fairness Scheduling with Dynamic Priority for Multi Workflow (FSDP) \cite{7951947} was an algorithm proposed by a group from Dalian University of Technology, China. FSDP emphasized the fairness aspect as it incorporated slowdown metrics to their algorithm's policy. Slowdown value was the ratio of the makespan of a workflow when it was being scheduled in dedicated service to the makespan of it being scheduled in a shared environment with the other workflows. The closest slowdown to 1, the fairest the algorithm scheduled the workflows in the system. FSDP also included urgency metric, a value that represented the priority of each workflow based on its deadline. The slowdown and urgency were updated periodically when a workflow finished ensuring the refinement in the scheduling process.

However, the fairness scenario was not explored in-depth by the authors. FSDP algorithm is only evaluated using two different workflows on a various number of resources (i.e., processor). The issue of fairness would arise when the number of submitted workflows was high enough to represent the condition of peak hour in multi-tenant distributed systems. In general, the complexity of FSDP was quadratic to the number of tasks processed. The evaluation results showed that FSDP slightly outperformed the overall makespan of MMHS \cite{tian2012hybrid} algorithm.

\subsection{Scheduling Trade-off of Dynamic Multiple Workflows}
A group from Hunan University presented two algorithms for multiple workflows scheduling. The first one was the Fairness-based Dynamic Multiple Heterogeneous Selection Value (F\_DMHSV) \cite{CPE:CPE3782}. The algorithm consisted of six steps which were task prioritization, task selection, task allocation, task scheduling, the arrival of new workflow handling, and task monitoring. The task prioritization used a descending order of heterogeneous priority rank value (HPRV) \cite{6838775}, which included the out-degree (i.e., number of successors) of the task. This task prioritization time complexity was $\mathcal{O}\,$($\,T_w\,.\,P_s\,$) for all tasks in a workflow $T_w$ given a set of static processors $P_s$. The task was selected from the ready tasks pool based on the maximum HPRV. Furthermore, the task was allocated to the processor with minimum heterogeneous selection value (HSV) \cite{6838775} that optimized the task allocation criteria using the combination of upward and downward rank which created a time complexity of $\mathcal{O}\,$($\,T_r\,.\,P_s\,$) for all ready tasks $T_r$. The task, then, was scheduled to the earliest available processor with minimum HSV. The performance evaluation results showed that F\_DMHSV outperformed the overall makespan of RANK\_HYBD \cite{4626773}, OWM \cite{HSU2011860}, and FDWS \cite{arabnejad:hal-00926460} algorithms by 27\%, 10\%, and 3\% respectively.

In the same year, this group published energy-efficient algorithms for multiple workflows scheduling, which combined the Deadline-driven Processor Merging for Multiple Workflow (DPMMW) that aimed to meet the deadline, and the Global Energy Saving for Multiple Workflows (GESMW) sought to lower the energy consumption \cite{XIE2017}. DPMMW was a clustering algorithm which allocated the clustered tasks in a minimum number of processors so that the algorithm can put idle processors into sleep mode. Meanwhile, GESMW reassigned and adjusted the tasks to any processor with minimum energy consumption in the global scope. The combination of DPMMW\&GESMW was exploited to get lower energy consumption. This approach was different from the previous two energy-efficient algorithms that focused on virtual machine level manipulation. In general, the most intensive phase in this algorithm was the invoking of the HEFT algorithm to create a baseline scheduling plan and traversing all processors which take quadratic time complexity. Furthermore, the performance evaluation results showed that DPMMW\&GESMW outperformed the energy saving of the reusable DEWTS, a modified version of DEWTS \cite{Tang2016} algorithm by 8.07\% in average

This group presented two opposite approaches to scheduling with different objectives. However, in both methods, the works emphasize a similar strategy of resource selection. In their first work, the algorithm focuses on selecting various resources to minimize the makespan, while in the second one, it is choosing the different machine with various energy efficiency to reduce energy consumption. These resource selection strategies can improve the scheduling result by combining them with efficient task scheduling approaches.

\subsection{Workflow Scheduling in Multi-Tenant Clouds}
Another algorithm for multiple workflows scheduling was Cloud-based Workflow Scheduling (CWSA) \cite{7457258}. This work used the term "multi-tenant clouds" in its paper for describing the multi-tenancy aspect that was generally considered in cloud computing environments. Whereas the definition itself was similar to which the multiple workflows we used in this survey. The algorithm was intended for compute-intensive workflows applications. Hence, CWSA ignored data-related overhead and focused on compute resource management. The algorithm aimed to minimize the total makespan of the workflows which in the result, decreasing the cost of execution. CWSA was an extension to multi-tenant workflow management system \cite{5541997} which exploited the schedule gap. In general, the time complexity of CWSA is $\mathcal{O}\,$($\,T_w\,.\,C_d\,$), given a set of workflow tasks $T_w$ and a number of dynamic cloud instances $C_d$. The performance evaluation results showed that the CWSA outperformed both the makespan and cost of standard FCFS, EASY Backfilling, and Minimum Completion Time (MCT) policy.

However, CWSA did not further optimize their cost minimization strategy using a specific cost-aware resource provisioning technique. CWSA auto-scaled the resources using a resource utilization threshold, in which it acquired and released the resources if their utilization exceeded or was below a specific number. For example, they implemented the rule such as if the usage was $\geq$ 70\% for 10 minutes, then it was scaled-up by adding 1 VM of small size. In this case, the algorithms with cost-aware auto-scaling strategy--that specifically acquires and releases particular VMs based on the workload--may outperform CWSA that only considers overall system utilization based auto-scaling. This type of auto-scaling strategy is not provisioning resources that are specifically tailored to the need of the workloads.

\subsection{Multi-tenant Workflow as a Service Platform}
The latest solution for multiple workflow scheduling was Elastic Resource Provisioning and Scheduling Algorithm for Multiple Workflows designed for Workflow as a Service Platforms (EPSM) \cite{RODRIGUEZ2018739}. This work used a specific term of "multi-tenant workflow as a service" in its paper to describe the platform for executing multiple workflows in the clouds. However, the "multi-tenant workflow as a service" term and "multiple workflows" we defined, can be used interchangeably in this case. The EPSM algorithm introduced scheduling algorithm for multi-tenant platforms that utilized a container to bundle workflow's application before deploying it into VMs. In this way, the users can share the same VMs without having any problem related to software dependencies and libraries.

The algorithm consisted of two-phase, resource provisioning which included a flexible approach of scaling up and down the resources to cope up with the dynamic workload of workflows, and scheduling which exploited a delay policy based on the task's deadline to re-use the cheapest resources as much as possible to minimize the cost. In the resource provisioning phase, EPSM incorporated an overhead detection in the form of provisioning delay (i.e., acquisition delay) and de-provisioning delay (i.e., release delay) of the VMs. This strategy was proven to be able to reduce unnecessary cost due to violating a coarse-grain billing period of clouds. The algorithm made an update of unscheduled tasks' deadline whenever a task finished the execution. In this way, the algorithm dynamically adapted the gap between the estimated and actual execution plan to ensure the scheduling objectives. In the scheduling phase, EPSM considered re-using available VMs before provisioning the new one to minimize the delay overhead of acquiring new VMs and possible cost minimization by re-using the cheapest VMs available. In general, the time complexity of the EPSM algorithm is quadratic to the number of tasks processed. Furthermore, the performance evaluation results showed that this algorithm outperformed the cost of Dyna \cite{7044594} algorithm by 19\% on average for various scenarios.

\subsection{Concurrent Multiple Workflows Scheduling}

The latest work on deadline- and budget-constrained multiple workflows scheduling was Multi-workflow Heterogeneous budget-deadline-constrained Scheduling (MW-HBDCS) algorithm \cite{Zhou2018} that was introduced by a group from Guangzhou University, China. This work used the term "concurrent multiple workflows" as it emphasized on the concurrent condition of multiple workflows, means tackling several workflows that arrived at the same time or overlapped on a dense condition. MW-HBDCS algorithm was designed to improve the flaw on the previous similar algorithm, MW-DBS \cite{ARABNEJAD2017120}. Significant enhancement was the inclusion of a budget in the ranking process to prioritize the tasks for scheduling. MW-HBDCS was also designed to tackle uncertainties in the environments. In this work, the authors used the terms "consistent" and "inconsistent" environments to describe various dynamicity in multi-tenant distributed systems. In general, the complexity of MW-HBDCS was quadratic to the number of tasks processed, similar to the time complexity of MW-DBS. The evaluation results showed that MW-HBDCS outperformed the success rate of MW-DBS by 46\% and 52\% on synthetic and real-world workflow applications respectively.

MW-HBDCS tackled many flaws that were not considered in the previous deadline- and budget-constrained scheduling algorithms. These enhancements were the model of multi-tenant distributed systems that incorporated high uncertainties and dynamicity, the improvement of task's ranking mechanism that enclosed the budget as one of the primary constraints besides the deadline while previously only acted as a complementary constraint. Since the authors were highly considered the budget as crucial as the deadline, it is essential to include the trade-off analysis between the values of budget and deadline related to the success rate of workflows execution. One of the techniques to such an approach is the Pareto analysis that is used for multi-objective workflow scheduling (e.g., MOHEFT \cite{6427573}). Furthermore, this algorithm considers static resource provisioning. Therefore, it may not achieve optimal performance in cloud computing environments where the auto-scaling of resources is possible.

\subsection{Scheduling Multiple Workflows under Uncertain Task Execution Time}

The latest deadline-aware multiple workflows scheduling algorithm is NOSFA \cite{8669862}. This algorithm adopted a similar strategy to several previous algorithms (e.g., EDPRS, EPSM, ROSA) designed to tackle the uncertainties in cloud computing environments. The NOSFA aims to minimize the cost of leasing cloud instances by optimizing the resource utilization using the sharing strategy of VM billing period. To further distribute a fair share of sub-deadlines between task, NOSFA made use of PCP to create end-to-end scheduling of several tasks during the deadline distribution process. Therefore, traversing workflows for detecting the PCP was the most intensive phase that takes quadratic time complexity.

NOSFA relies on to the strategy to maintain the minimum growth of the leased cloud instances instead of auto-scaling the resources dynamically by eliminating future idle VMs. This strategy may cause a problem of waiting overhead in a very dense workflows' arrival (i.e., high concurrent workflows). Combining dynamic auto-scaling and maintaining a low growth of VM leased may become an essential strategy for multi-tenant platforms with a quite high uncertainties situation. Furthermore, the performance evaluation results showed that NOSFA outperformed ROSA \cite{8443134} in terms of reducing the cost and deadline violation probability by an average of 50.5\% and 55.7\% respectively while improving resource utilization by 32.6\% in average.

\section{Future Directions}
Many challenges and problems in the current solutions should be considered for the future of scientific workflows as nicely discussed in a study by Deelman et al. \cite{doi:10.1177/1094342017704893}. However, this paper describes explicitly the particular aspect of multiple workflows scheduling in multi-tenant distributed systems. We capture the future direction of multi-tenancy in multiple workflows scheduling from existing solutions and rising trend technologies that have a high potential to support the enhancement of multi-tenant platforms. The range is broad from the pre-processing phase which involves the strategy to accurately estimate task execution time that is a prerequisite for scheduling process; scheduling techniques that are aware of constraints such as failure, deadline, budget, energy usage, privacy, and security; to the use of heterogeneous distributed systems that differ not only in capacity but also pricing scheme and provisional procedures. We also observe a potential use of several technologies to enhance the multi-tenancy that comes from the rising trend technologies such as containers, serverless computing, Unikernels and the broad adoption of the Internet of Things (IoT) workflows.

\subsection{Advanced Multi-tenancy Using Microservices}
Microservice is a variant of service-oriented architecture (SOA) that has a unique lightweight or even simple protocol and treated the application as a collection of loosely coupled service \cite{7742215}. In this sense, we can consider container technology, serverless computing (i.e., function as a service), and Unikernels to fall into the category.

Kozhirbayev and Sinnott \cite{KOZHIRBAYEV2017175} report that the performance of a container on a bare metal machine is comparable to a native environment since no significant overhead recording is produced in the runtime. It is a promising technology to enhance multi-tenancy features for multiple workflows scheduling as it can be used as an isolated environment for workflow application before deploying it into virtual machines in clouds. This advantage is because using a virtual machine as an isolated environment eliminates the possibility of sharing its computational capacity between different workflow applications that may have different software dependencies. We argue that, in the future, this technology will be widely used for solving multi-tenancy problem as it has been explored for executing a single scientific workflow as reported in several studies \cite{7092947}, \cite{7830693}, \cite{7600178}, \cite{Hung099010}, \cite{8171362}.

However, the main trade-off of general purpose container's performance nativity (i.e., Docker) for scientific applications is IT security \cite{7742298}. The multi-tenancy requirements inevitably invite multiple users sharing the same computational infrastructure at one time. In the case of Docker, every container process has access to Docker daemon which is spawned as a child of the root. At any rate, this activity compromised the whole IT infrastructure. To tackle this security problem, Singularity Container that targets explicitly the scientific applications have been developed \cite{10.1371/journal.pone.0177459}. It has been tested in the Comet Supercomputer at the San Diego Supercomputer Center \cite{Le:2017:PAA:3093338.3106737} and shown a promising result for handling multi-tenancy in the future workflow as a service environment.

Another promising technology is serverless computing (i.e., Function as a Service). The FaaS is a new terminology that stands on the top of cloud computing as a simplified version of the virtualization service. In this way, cloud providers directly manage resource allocation, and the users only needed to pay for the time of resource usage based on the application codes. This technology facilitates the users who need to run specific tasks from a piece of code without having a headache in managing the cloud instances. We consider to include this into the future directions since the potential of its multi-tenancy service is high to accommodate the multi-tenant workflows scheduling. Furthermore, this technology has been tested for a single scientific workflows execution as reported by Malawski \cite{malawski2016towards}, Jiang et al. \cite{Jiang2017b} and Malawski et al. \cite{MALAWSKI2017}. Notably, this function as a service can serve the workloads that consist of platform-independent workflows, which can be efficiently executed on top of this facility without having to provision a new virtual machine.

Finally, Unikernels is another rise of virtualization technology that is designed to maintain the perfect isolation of virtual machines and keep up with the lightweight of the container \cite{Madhavapeddy:2013:ULO:2451116.2451167}. The Unikernels enhance the virtualization in terms of the weight by removing a general purpose OS from the VM. In this way, Unikernels directly run the application on the virtual hardware. To make it even more interesting, recent finding shows that Unikernels do not require a virtual hardware abstraction. It can directly run as a process by exploiting an existing kernel system that is called whitelisting mechanism \cite{Williams:2018:UP:3267809.3267845}. Looking into the combination features of virtual machine and container in one single virtualization technology, we can hope for the better multi-tenancy for workflow as a service platform using this technology.

\subsection{Reserved vs. On-demand vs. Spot Instances}
The further reduction of operational cost has been a long existing issue in utility-based multi-tenant distributed systems. Notably, in cloud computing environments where the resources are leased from third-party providers based on various pricing schemes, cost-aware scheduling strategy is highly considered. While most of the algorithms for cloud computing environments use on-demand instances which ensure the reliability in a pay-as-you-go pricing model, a work by Zhou et al. \cite{7044594} explored the use of spot instances that is relatively cheaper than on-demand, but less reliable as they were only available for a limited time and could be terminated at unpredictable time by the providers. This type of resource raises a fault-tolerant issue to be considered in scheduling.

On the other hand, as multiple workflows scheduling involves a high number of workflow execution, the use of reserved instances in clouds should be explored to minimize further the total operational cost as the pricing of this bulk reservation is lower than on-demand even spot instances. The issue of using reserved instances is how accurate the algorithms can predict the workload of workflows to lease a relatively constant number of reserved instances at some point in multi-tenant platforms. This combination of reserved, on-demand and spot instances must be explored to create an efficient resource provisioning strategy in multi-tenant distributed systems.

\subsection{Multi-clouds vs. Hybrid Clouds vs. Bare-Metal Clouds}
The use of multi-cloud providers for executing scientific workflows was explored by Jrad et al. \cite{Jrad:2013:BFM:2462326.2462339} and Montes et al. \cite{6973738} by introducing algorithms that were aware of different services available from several providers. However, the only relevant works found in our study are the use of hybrid clouds for separating tasks execution based on some properties instead of multi-clouds. Furthermore, a work used hybrid clouds to treat tasks with different privacy level in healthcare services, while another research utilized public clouds to cover the computational need that could not be fulfilled using private clouds and on-premises infrastructure. In our opinion, further utilization of multi-clouds can benefit a single cloud provider may not serve the providers since the high requirements of resources in multi-tenant platforms. The other advantage of multi-clouds is the reduction of operational cost as various cloud providers employed different price for the datacenter in different geographical location. In this way, discovering relevant services can be further exploited to minimize data movements by choosing a handy datacenter location and also comparing the best ratio of price and performance from various cloud instances from multiple cloud providers.

Related to the cloud heterogeneity in multi-cloud and hybrid clouds, we have to mention a valuable service that provides a more heterogeneous infrastructure that is called bare-metal clouds. Bare-metal cloud is an emerging service in the Infrastructure as a Service business model that leases a physical machine instead of a virtual machine to the users. This service targets user that need specific hardware requirements in an intensive computation (i.e., GPU, FPGA). While one may ask the elasticity of this service against any standard cloud services, recent work shown that such agility in provisioning bare-metal clouds can be compared to general virtualization using virtual machine \cite{Omote:2015:IAE:2694344.2694349}. In this case, the specific hardware requirements of particular scientific applications can be fulfilled. However, on the other hand, the challenges to managing such an environment is an exciting list to do.

\subsection{Fast and Reliable Task Runtime Estimation in near Real-time Processing}
Predicting task runtime in clouds is non-trivial, mainly due to the problem in which clouds resources are subject to performance variability \cite{jackson2010performance}. This variability occurs due to several factors--including virtualization overhead, multi-tenancy, geographical distribution, and temporal aspects  \cite{leitner2016performance}--that affect not only computational performance but also the communication network used to transfer input/output data \cite{6848061}. The majority of algorithms rely on the estimation of tasks execution time in appropriate resources to produce an accurate schedule plan. Meanwhile, the works on task runtime estimation in scientific workflows are limited including the latest works by Pham et al. \cite{8013738} and Nadeem et al. \cite{Nadeem2017} that used machine learning techniques while previously, a work on scientific workflows profiling and characterization by Juve et al. \cite{JUVE2013682} that produced a synthetic workflows generator is being used by the majority of works on workflow scheduling.

The future task runtime prediction techniques must be able to address dynamic workloads in multi-tenant platforms that are continuously arriving in resemblance to stream data processing. The adoption of online and incremental machine learning approach may become another solution. In this approach, the algorithm does not need to learn from a model constructed from a large number of collected datasets which is generally time-consuming and compute intensive. The algorithm only sees the data once and then integrate additional information as the model incrementally built from new data. The latest work by Sahoo et al. \cite{Sahoo:2019:LSO:3301280.3299875} attempts to make OMKR, an online and incremental machine learning approach, scalable to large time-series datasets in a near real-time process. While this approach is still intensively being studied for scientific workflow area, the preliminary work on such an approach has been presented by Hilman et al. \cite{8603156} for future workflow as a service platform.

\subsection{Integrated Anomaly Detection and Fault-tolerant Aware Platforms}
Detecting anomaly in scientific workflows is one of the methods to ensure the fault-tolerance of multiple workflows scheduling in multi-tenant distributed systems. Several notable works in workflows anomaly detection are presented by Samak et al. \cite{6063014} that detailed integrated workflows and resource monitoring for STAMPEDE project. Furthermore, Gaikwad et al. \cite{7568396} used Autoregression techniques to detect the anomalies by monitoring the systems and a similar work by Rodriguez et al. \cite{RODRIGUEZ2018} that adopted Neural Network methods. On the other hand, the fault-tolerant algorithms found in our survey used replication technique \cite{7435325} and checkpointing \cite{7044594} to handle failure in workflows execution.

Future works on this area include the integration of detecting anomalies and failure-aware scheduling in multi-tenant platforms and the use of various fault-tolerant methods in failure-aware algorithms, such as resubmission and live migration. How the anomalies detection model can be combined with the task runtime prediction model to better schedule multiple tasks on heterogeneous environments. In this case, the algorithms designed must incorporate the ability to fully aware of the underlying hardware performance and their monitoring features.

\subsection{Multi-objective Constraints Scheduling}
The flexibility and ability to easily scale the number of resources (i.e., VMs) in the cloud computing environment, leads to a trade-off between two conflicting Quality of Service (QoS) requirements: time and cost. In this case, the more powerful VMs capable of processing a task faster will be more expensive than the slower less powerful ones. There has been an extensive research \cite{CPE:CPE4041} on this scheduling topic that specifically designed for cloud computing environments, with most works proposed algorithms that were aimed to minimize the total execution cost while finishing the workflow execution before a user-defined deadline. Meanwhile, the works that aimed to minimize the makespan by fully utilizing the available budget to lease as much as possible the faster resources are limited. We identified algorithms that considered budget in their scheduling such as \cite{7371366}, \cite{ghasemzadeh_et_al:LIPIcs:2017:7088}, \cite{ARABNEJAD2017211}, and \cite{Zhou2018} that exploited the workflow budget as a complementary constraint to the deadline. Therefore, none of them aims to fully utilize the available budget to get a faster execution time.

Furthermore, the works in multiple workflows scheduling that specifically aim to achieve multi-objective (i.e., time and cost minimization) is also very limited. While the metaheuristics and evolutionary programming have been used to accomplish this multi-objective such as work by Fard et al. \cite{6217435}, the implementation for multiple workflows scheduling is limited by its absolute pre-processing requirement. However, a more lightweight list-based heuristic approach such as MOHEFT \cite{6427573} and DPDS \cite{MALAWSKI20151} can be considered for their adaptation in multiple workflows scheduling.

\subsection{Energy-efficient Computing}
Beloglazov et al. \cite{BELOGLAZOV201147} have extensively explored the issue of green computing in clouds. There are several works of multiple workflows scheduling in our study that addressed this energy-efficient issue. A work discussed energy-efficient strategy at the infrastructure level by implementing a live migration technique for scheduling \cite{7276993}, while another work tackled the problem at workload level by allocating the load to specific physical machines \cite{7576489}. For the providers that rely on IaaS clouds for their source to lease the computational resources, adopting workload level strategies for energy-aware scheduling is the possible way as they do not have the control over the raw computational infrastructures as IaaS cloud providers do.

\subsection{Privacy-aware Scheduling}
The privacy constraint is an essential aspect that has been tackled in the OPHC-TR \cite{7037702} by separating the execution in a private and public cloud based on their level of privacy represented in the processed data. However, it is crucial to consider the security aspect in managing privacy, since both issues are highly inter-related. One of the workflow scheduling algorithms that consider security is the SABA algorithm \cite{ZENG2015141}. However, it is designed for a single workflow scheduling and intended to explore the relationship between cost and security aspects in the scheduling, instead of focusing on the privacy aspect.

Further exploration of privacy and security in the multiple workflows scheduling has to be done as it resembles the real world workflow application problems in multi-tenant distributed systems. Another way to deal with privacy is by adopting a reliable security protocol for data processing in cloud environments, such as a homomorphic encryption \cite{6779008}. However, the increase in security must have influenced the computational time, and this becomes a scheduling challenge to address.

\subsection{Internet of Things (IoT) Workflows}
A vision paper by Gubbi et al. \cite{GUBBI20131645} mentions an essential and future use of the Internet of Things (IoT) in a workflow form. The idea, then, has been implemented in several works, including a disaster warning system \cite{6716026}, a smart city system \cite{6962168}, and big data framework \cite{7912282}. This type of workflow increases in numbers and its broad adoption is predicted to be widely seen shortly. In the meantime, therefore, the need for a multi-tenant platform that can handle such workflows may arise. The computational characteristics of IoT workflows are different from regular workflows. Moreover, IoT applications are highly demanding network resources to handle end-to-end services from sensors at one point, to users at the other end. Therefore, a specific problem related to network-intensive requirements such as bandwidth-aware and latency-aware must be considered in the scheduling algorithms for IoT workflow application.

\section{Summary}

This paper presents a study on algorithms for multiple workflows scheduling in multi-tenant distributed systems. In particular, the research focuses on the heterogeneity of workloads, the model for deploying multiple workflows, the priority assignment model for multiple users, the scheduling techniques for multiple workflows, and the resource provisioning strategies in multi-tenant distributed systems. It presents a taxonomy covering the focus of study based on a comprehensive review of multiple workflows scheduling algorithms. The taxonomy is accompanied by classification from surveyed algorithms to show the existing solutions for multiple workflows scheduling in various aspects. The current algorithms within the scope of the study are reviewed and classified to open up the problems in this area and provide the readers with a helicopter view on multiple workflows scheduling. Some descriptions and discussions of various solutions are also covered in this paper to give a more detailed and comprehensive understanding of the state-of-the-art techniques and even to get an insight on further research and development in this area.

\begin{acks}
	
	This research is partially supported by LPDP (Indonesia Endowment Fund for Education) and ARC (Australia Research Council) research grant.
	
\end{acks}

\bibliographystyle{ACM-Reference-Format}
\bibliography{references}


\begin{thebibliography}{114}


\ifx \showCODEN    \undefined \def \showCODEN     #1{\unskip}     \fi
\ifx \showDOI      \undefined \def \showDOI       #1{#1}\fi
\ifx \showISBNx    \undefined \def \showISBNx     #1{\unskip}     \fi
\ifx \showISBNxiii \undefined \def \showISBNxiii  #1{\unskip}     \fi
\ifx \showISSN     \undefined \def \showISSN      #1{\unskip}     \fi
\ifx \showLCCN     \undefined \def \showLCCN      #1{\unskip}     \fi
\ifx \shownote     \undefined \def \shownote      #1{#1}          \fi
\ifx \showarticletitle \undefined \def \showarticletitle #1{#1}   \fi
\ifx \showURL      \undefined \def \showURL       {\relax}        \fi
\providecommand\bibfield[2]{#2}
\providecommand\bibinfo[2]{#2}
\providecommand\natexlab[1]{#1}
\providecommand\showeprint[2][]{arXiv:#2}

\bibitem[\protect\citeauthoryear{Ahmad, Liew, Rafique, Munir, and Khan}{Ahmad
  et~al\mbox{.}}{2014}]%
        {7034777}
\bibfield{author}{\bibinfo{person}{Saima~G. Ahmad}, \bibinfo{person}{Chee~S.
  Liew}, \bibinfo{person}{Muhammad~M. Rafique}, \bibinfo{person}{Ehsan~U.
  Munir}, {and} \bibinfo{person}{Samee~U. Khan}.}
  \bibinfo{year}{2014}\natexlab{}.
\newblock \showarticletitle{Data-Intensive Workflow Optimization Based on
  Application Task Graph Partitioning in Heterogeneous Computing Systems}. In
  \bibinfo{booktitle}{\emph{Proceedings of The 4th IEEE International
  Conference on Big Data and Cloud Computing}}. \bibinfo{pages}{129--136}.
\newblock
\urldef\tempurl%
\url{https://doi.org/10.1109/BDCloud.2014.63}
\showDOI{\tempurl}


\bibitem[\protect\citeauthoryear{Alkhanak, Lee, and Khan}{Alkhanak
  et~al\mbox{.}}{2015}]%
        {ALKHANAK20153}
\bibfield{author}{\bibinfo{person}{Ehab~N. Alkhanak}, \bibinfo{person}{Sai~P.
  Lee}, {and} \bibinfo{person}{Saif U.~R. Khan}.}
  \bibinfo{year}{2015}\natexlab{}.
\newblock \showarticletitle{Cost-aware {Challenges} for {Workflow} {Scheduling}
  {Approaches} in {Cloud} {Computing} {Environments}: {Taxonomy} and
  {Opportunities}}.
\newblock \bibinfo{journal}{\emph{Future Generation Computer Systems}}
  \bibinfo{volume}{50}, \bibinfo{number}{Supplement C} (\bibinfo{year}{2015}),
  \bibinfo{pages}{3--21}.
\newblock
\showISSN{0167-739X}
\urldef\tempurl%
\url{https://doi.org/10.1016/j.future.2015.01.007}
\showDOI{\tempurl}


\bibitem[\protect\citeauthoryear{Alzahrani, Tari, Lee, Alsadie, and
  Zomaya}{Alzahrani et~al\mbox{.}}{2017}]%
        {8171362}
\bibfield{author}{\bibinfo{person}{Eidah~J. Alzahrani}, \bibinfo{person}{Zahir
  Tari}, \bibinfo{person}{Young~C. Lee}, \bibinfo{person}{Deafallah Alsadie},
  {and} \bibinfo{person}{Albert~Y. Zomaya}.} \bibinfo{year}{2017}\natexlab{}.
\newblock \showarticletitle{ad{C}{F}{S}: {Adaptive} {Completely} {Fair}
  {Scheduling} {Policy} for {Containerised} {Workflows} {Systems}}. In
  \bibinfo{booktitle}{\emph{Proceedings of The 16th IEEE International
  Symposium on Network Computing and Applications}}. \bibinfo{pages}{1--8}.
\newblock
\urldef\tempurl%
\url{https://doi.org/10.1109/NCA.2017.8171362}
\showDOI{\tempurl}


\bibitem[\protect\citeauthoryear{Arabnejad and Barbosa}{Arabnejad and
  Barbosa}{2015}]%
        {7371366}
\bibfield{author}{\bibinfo{person}{Hamid Arabnejad} {and}
  \bibinfo{person}{Jorge~G. Barbosa}.} \bibinfo{year}{2015}\natexlab{}.
\newblock \showarticletitle{Multi-workflow {Q}o{S}-constrained {Scheduling} for
  {Utility} {Computing}}. In \bibinfo{booktitle}{\emph{Proceedings of The 18th
  IEEE International Conference on Computational Science and Engineering}}.
  \bibinfo{pages}{137--144}.
\newblock
\urldef\tempurl%
\url{https://doi.org/10.1109/CSE.2015.29}
\showDOI{\tempurl}


\bibitem[\protect\citeauthoryear{Arabnejad and Barbosa}{Arabnejad and
  Barbosa}{2017a}]%
        {ARABNEJAD2017120}
\bibfield{author}{\bibinfo{person}{Hamid Arabnejad} {and}
  \bibinfo{person}{Jorge~G. Barbosa}.} \bibinfo{year}{2017}\natexlab{a}.
\newblock \showarticletitle{Maximizing {The} {Completion} {Rate} of
  {Concurrent} {Scientific} {Applications} {Under} {Time} and {Budget}
  {Constraints}}.
\newblock \bibinfo{journal}{\emph{Journal of Computational Science}}
  \bibinfo{volume}{23}, \bibinfo{number}{Supplement C} (\bibinfo{year}{2017}),
  \bibinfo{pages}{120--129}.
\newblock
\showISSN{1877-7503}
\urldef\tempurl%
\url{https://doi.org/10.1016/j.jocs.2016.10.013}
\showDOI{\tempurl}


\bibitem[\protect\citeauthoryear{Arabnejad and Barbosa}{Arabnejad and
  Barbosa}{2017b}]%
        {ARABNEJAD2017211}
\bibfield{author}{\bibinfo{person}{Hamid Arabnejad} {and}
  \bibinfo{person}{Jorge~G. Barbosa}.} \bibinfo{year}{2017}\natexlab{b}.
\newblock \showarticletitle{{Multi}-{Q}o{S} {Constrained} and {Profit-aware}
  {Scheduling} {Approach} for {Concurrent} {Workflows} on {Heterogeneous}
  {Systems}}.
\newblock \bibinfo{journal}{\emph{Future Generation Computer Systems}}
  \bibinfo{volume}{68}, \bibinfo{number}{Supplement C} (\bibinfo{year}{2017}),
  \bibinfo{pages}{211--221}.
\newblock
\showISSN{0167-739X}
\urldef\tempurl%
\url{https://doi.org/10.1016/j.future.2016.10.003}
\showDOI{\tempurl}


\bibitem[\protect\citeauthoryear{Arabnejad, Barbosa, and Suter}{Arabnejad
  et~al\mbox{.}}{2014}]%
        {arabnejad:hal-00926460}
\bibfield{author}{\bibinfo{person}{Hamid Arabnejad}, \bibinfo{person}{Jorge~G.
  Barbosa}, {and} \bibinfo{person}{Fr{\'e}d{\'e}ric Suter}.}
  \bibinfo{year}{2014}\natexlab{}.
\newblock \showarticletitle{{Fair {Resource} {Sharing} for {Dynamic}
  {Scheduling} of {Workflows} on {Heterogeneous} {Systems}}}.
\newblock In \bibinfo{booktitle}{\emph{{High-Performance Computing on Complex
  Environments}}}.
\newblock


\bibitem[\protect\citeauthoryear{Barbosa, Morais, Nobrega, and
  Monteiro}{Barbosa et~al\mbox{.}}{2005}]%
        {4154152}
\bibfield{author}{\bibinfo{person}{Jorge~G. Barbosa}, \bibinfo{person}{Celeste
  Morais}, \bibinfo{person}{Ruben Nobrega}, {and} \bibinfo{person}{Ant\`{o}nio
  Monteiro}.} \bibinfo{year}{2005}\natexlab{}.
\newblock \showarticletitle{Static {S}cheduling of {D}ependent {P}arallel
  {T}asks on {H}eterogeneous {C}lusters}. In
  \bibinfo{booktitle}{\emph{Proceedings of The IEEE International Conference on
  Cluster Computing}}. \bibinfo{pages}{1--8}.
\newblock
\showISSN{1552-5244}
\urldef\tempurl%
\url{https://doi.org/10.1109/CLUSTR.2005.347024}
\showDOI{\tempurl}


\bibitem[\protect\citeauthoryear{Barbosa and Moreira}{Barbosa and
  Moreira}{2011}]%
        {BARBOSA2011428}
\bibfield{author}{\bibinfo{person}{Jorge~G. Barbosa} {and}
  \bibinfo{person}{Belmiro Moreira}.} \bibinfo{year}{2011}\natexlab{}.
\newblock \showarticletitle{Dynamic Scheduling of A Batch of Parallel Task Jobs
  on Heterogeneous Clusters}.
\newblock \bibinfo{journal}{\emph{Parallel Comput.}} \bibinfo{volume}{37},
  \bibinfo{number}{8} (\bibinfo{year}{2011}), \bibinfo{pages}{428--438}.
\newblock
\showISSN{0167-8191}
\urldef\tempurl%
\url{https://doi.org/10.1016/j.parco.2010.12.004}
\showDOI{\tempurl}


\bibitem[\protect\citeauthoryear{Belkin, Haas, Arnold, Leong, Huerta, Lesny,
  and Neubauer}{Belkin et~al\mbox{.}}{2018}]%
        {belkin2018container}
\bibfield{author}{\bibinfo{person}{Maxim Belkin}, \bibinfo{person}{Roland
  Haas}, \bibinfo{person}{Galen~W. Arnold}, \bibinfo{person}{Hon~W. Leong},
  \bibinfo{person}{Eliu~A. Huerta}, \bibinfo{person}{David Lesny}, {and}
  \bibinfo{person}{Mark Neubauer}.} \bibinfo{year}{2018}\natexlab{}.
\newblock \bibinfo{title}{Container {S}olutions for {H}{P}{C} {S}ystems: {A}
  {C}ase {S}tudy of {U}sing {S}hifter on {B}lue {W}aters}.
\newblock   (\bibinfo{year}{2018}).
\newblock
\showeprint[arxiv]{cs.DC/1808.00556}


\bibitem[\protect\citeauthoryear{Beloglazov, Buyya, Lee, and Zomaya}{Beloglazov
  et~al\mbox{.}}{2011}]%
        {BELOGLAZOV201147}
\bibfield{author}{\bibinfo{person}{Anton Beloglazov}, \bibinfo{person}{Rajkumar
  Buyya}, \bibinfo{person}{Young~C. Lee}, {and} \bibinfo{person}{Albert~Y.
  Zomaya}.} \bibinfo{year}{2011}\natexlab{}.
\newblock \showarticletitle{Chapter 3 - A Taxonomy and Survey of
  Energy-Efficient Data Centers and Cloud Computing Systems}.
\newblock \bibinfo{series}{Advances in Computers}, Vol.~\bibinfo{volume}{82}.
  \bibinfo{publisher}{Elsevier}, \bibinfo{pages}{47--111}.
\newblock
\showISSN{0065-2458}
\urldef\tempurl%
\url{https://doi.org/h10.1016/B978-0-12-385512-1.00003-7}
\showDOI{\tempurl}


\bibitem[\protect\citeauthoryear{Bryant, Van, Riedel, Gardner, Bejar, Hover,
  Tovar, Hurtado, and Thain}{Bryant et~al\mbox{.}}{2018}]%
        {Bryant:2018:VVC:3219104.3219125}
\bibfield{author}{\bibinfo{person}{Lincoln Bryant}, \bibinfo{person}{Jeremy
  Van}, \bibinfo{person}{Benedikt Riedel}, \bibinfo{person}{Robert~W. Gardner},
  \bibinfo{person}{Jose~C. Bejar}, \bibinfo{person}{John Hover},
  \bibinfo{person}{Ben Tovar}, \bibinfo{person}{Kenyi Hurtado}, {and}
  \bibinfo{person}{Douglas Thain}.} \bibinfo{year}{2018}\natexlab{}.
\newblock \showarticletitle{V{C}3: {A} {V}irtual {C}luster {S}ervice for
  {C}ommunity {C}omputation}. In \bibinfo{booktitle}{\emph{Proceedings of The
  Practice and Experience on Advanced Research Computing}}.
  \bibinfo{publisher}{ACM}, \bibinfo{pages}{30:1--30:8}.
\newblock
\showISBNx{978-1-4503-6446-1}
\urldef\tempurl%
\url{https://doi.org/10.1145/3219104.3219125}
\showDOI{\tempurl}


\bibitem[\protect\citeauthoryear{Chen, Zhu, Wu, and Huo}{Chen
  et~al\mbox{.}}{2018b}]%
        {Chen2018}
\bibfield{author}{\bibinfo{person}{Huangke Chen}, \bibinfo{person}{Jianghan
  Zhu}, \bibinfo{person}{Guohua Wu}, {and} \bibinfo{person}{Lisu Huo}.}
  \bibinfo{year}{2018}\natexlab{b}.
\newblock \showarticletitle{Cost-efficient {R}eactive {S}cheduling for
  {R}eal-time {W}orkflows in {C}louds}.
\newblock \bibinfo{journal}{\emph{The Journal of Supercomputing}}
  \bibinfo{volume}{74}, \bibinfo{number}{11} (\bibinfo{year}{2018}),
  \bibinfo{pages}{6291--6309}.
\newblock
\showISSN{1573-0484}
\urldef\tempurl%
\url{https://doi.org/10.1007/s11227-018-2561-9}
\showDOI{\tempurl}


\bibitem[\protect\citeauthoryear{Chen, Zhu, Zhang, Ma, and Shen}{Chen
  et~al\mbox{.}}{2017}]%
        {Chen2017}
\bibfield{author}{\bibinfo{person}{Huangke Chen}, \bibinfo{person}{Jianghan
  Zhu}, \bibinfo{person}{Zhenshi Zhang}, \bibinfo{person}{Manhao Ma}, {and}
  \bibinfo{person}{Xin Shen}.} \bibinfo{year}{2017}\natexlab{}.
\newblock \showarticletitle{Real-time {Workflows} {Oriented} {Online}
  {Scheduling} in {Uncertain} {Cloud} {Environment}}.
\newblock \bibinfo{journal}{\emph{The Journal of Supercomputing}}
  \bibinfo{volume}{73}, \bibinfo{number}{11} (\bibinfo{year}{2017}),
  \bibinfo{pages}{4906--4922}.
\newblock
\showISSN{1573-0484}
\urldef\tempurl%
\url{https://doi.org/10.1007/s11227-017-2060-4}
\showDOI{\tempurl}


\bibitem[\protect\citeauthoryear{Chen, Zhu, Liu, and Pedrycz}{Chen
  et~al\mbox{.}}{2018a}]%
        {8443134}
\bibfield{author}{\bibinfo{person}{Huangke Chen}, \bibinfo{person}{Xiaomin
  Zhu}, \bibinfo{person}{Guipeng Liu}, {and} \bibinfo{person}{Witold Pedrycz}.}
  \bibinfo{year}{2018}\natexlab{a}.
\newblock \showarticletitle{Uncertainty-{A}ware {O}nline {S}cheduling for
  {R}eal-{T}ime {W}orkflows in {C}loud {S}ervice {E}nvironment}.
\newblock \bibinfo{journal}{\emph{IEEE Transactions on Services Computing}}
  (\bibinfo{year}{2018}), \bibinfo{pages}{1--1}.
\newblock
\showISSN{1939-1374}
\urldef\tempurl%
\url{https://doi.org/10.1109/TSC.2018.2866421}
\showDOI{\tempurl}


\bibitem[\protect\citeauthoryear{Chen, Zhu, Qiu, Guo, Yang, and Lu}{Chen
  et~al\mbox{.}}{2016b}]%
        {7576489}
\bibfield{author}{\bibinfo{person}{Huangke Chen}, \bibinfo{person}{Xiaomin
  Zhu}, \bibinfo{person}{Dishan Qiu}, \bibinfo{person}{Hui Guo},
  \bibinfo{person}{Laurence~T. Yang}, {and} \bibinfo{person}{Peizhong Lu}.}
  \bibinfo{year}{2016}\natexlab{b}.
\newblock \showarticletitle{EONS: Minimizing Energy Consumption for Executing
  Real-Time Workflows in Virtualized Cloud Data Centers}. In
  \bibinfo{booktitle}{\emph{Proceedings of The 45th International Conference on
  Parallel Processing Workshops}}. \bibinfo{pages}{385--392}.
\newblock
\urldef\tempurl%
\url{https://doi.org/10.1109/ICPPW.2016.60}
\showDOI{\tempurl}


\bibitem[\protect\citeauthoryear{Chen, Zhu, Qiu, and Liu}{Chen
  et~al\mbox{.}}{2016a}]%
        {7820319}
\bibfield{author}{\bibinfo{person}{Huangke Chen}, \bibinfo{person}{Xiaomin
  Zhu}, \bibinfo{person}{Dishan Qiu}, {and} \bibinfo{person}{Ling Liu}.}
  \bibinfo{year}{2016}\natexlab{a}.
\newblock \showarticletitle{Uncertainty-aware {Real-time} {Workflow}
  {Scheduling} in {The} {Cloud}}. In \bibinfo{booktitle}{\emph{Proceedings of
  The 9th IEEE International Conference on Cloud Computing}}.
  \bibinfo{pages}{577--584}.
\newblock
\urldef\tempurl%
\url{https://doi.org/10.1109/CLOUD.2016.0082}
\showDOI{\tempurl}


\bibitem[\protect\citeauthoryear{Chen, Lee, Fekete, and Zomaya}{Chen
  et~al\mbox{.}}{2015}]%
        {Chen2015}
\bibfield{author}{\bibinfo{person}{Wei Chen}, \bibinfo{person}{Young~C. Lee},
  \bibinfo{person}{Alan Fekete}, {and} \bibinfo{person}{Albert~Y. Zomaya}.}
  \bibinfo{year}{2015}\natexlab{}.
\newblock \showarticletitle{Adaptive {Multiple}-workflow {Scheduling} with
  {Task} {Rearrangement}}.
\newblock \bibinfo{journal}{\emph{The Journal of Supercomputing}}
  \bibinfo{volume}{71}, \bibinfo{number}{4} (\bibinfo{year}{2015}),
  \bibinfo{pages}{1297--1317}.
\newblock
\showISSN{1573-0484}
\urldef\tempurl%
\url{https://doi.org/10.1007/s11227-014-1361-0}
\showDOI{\tempurl}


\bibitem[\protect\citeauthoryear{Combe, Martin, and D.~Pietro}{Combe
  et~al\mbox{.}}{2016}]%
        {7742298}
\bibfield{author}{\bibinfo{person}{Theo Combe}, \bibinfo{person}{Antony
  Martin}, {and} \bibinfo{person}{Roberto D.~Pietro}.}
  \bibinfo{year}{2016}\natexlab{}.
\newblock \showarticletitle{To {D}ocker or {N}ot to {D}ocker: {A} {S}ecurity
  {P}erspective}.
\newblock \bibinfo{journal}{\emph{IEEE Cloud Computing}} \bibinfo{volume}{3},
  \bibinfo{number}{5} (\bibinfo{year}{2016}), \bibinfo{pages}{54--62}.
\newblock
\showISSN{2325-6095}
\urldef\tempurl%
\url{https://doi.org/10.1109/MCC.2016.100}
\showDOI{\tempurl}


\bibitem[\protect\citeauthoryear{da~Silva, Juve, Rynge, Deelman, and
  Livny}{da~Silva et~al\mbox{.}}{2015}]%
        {doi:10.1142/S0129626415410030}
\bibfield{author}{\bibinfo{person}{Rafael~F. da Silva}, \bibinfo{person}{Gideon
  Juve}, \bibinfo{person}{Mats Rynge}, \bibinfo{person}{Ewa Deelman}, {and}
  \bibinfo{person}{Miron Livny}.} \bibinfo{year}{2015}\natexlab{}.
\newblock \showarticletitle{Online {Task} {Resource} {Consumption} {Prediction}
  for {Scientific} {Workflows}}.
\newblock \bibinfo{journal}{\emph{Parallel Processing Letters}}
  \bibinfo{volume}{25}, \bibinfo{number}{03} (\bibinfo{year}{2015}),
  \bibinfo{pages}{1541003}.
\newblock
\urldef\tempurl%
\url{https://doi.org/10.1142/S0129626415410030}
\showDOI{\tempurl}


\bibitem[\protect\citeauthoryear{Deelman, Peterka, Altintas, Carothers, van
  Dam, Moreland, Parashar, Ramakrishnan, Taufer, and Vetter}{Deelman
  et~al\mbox{.}}{2018}]%
        {doi:10.1177/1094342017704893}
\bibfield{author}{\bibinfo{person}{Ewa Deelman}, \bibinfo{person}{Tom Peterka},
  \bibinfo{person}{Ilkay Altintas}, \bibinfo{person}{Christopher~D. Carothers},
  \bibinfo{person}{Kerstin~K. van Dam}, \bibinfo{person}{Kenneth Moreland},
  \bibinfo{person}{Manish Parashar}, \bibinfo{person}{Lavanya Ramakrishnan},
  \bibinfo{person}{Michela Taufer}, {and} \bibinfo{person}{Jeffrey Vetter}.}
  \bibinfo{year}{2018}\natexlab{}.
\newblock \showarticletitle{The {Future} of {Scientific} {Workflows}}.
\newblock \bibinfo{journal}{\emph{The International Journal of High Performance
  Computing Applications}} \bibinfo{volume}{32}, \bibinfo{number}{1}
  (\bibinfo{year}{2018}), \bibinfo{pages}{159--175}.
\newblock
\urldef\tempurl%
\url{https://doi.org/10.1177/1094342017704893}
\showDOI{\tempurl}


\bibitem[\protect\citeauthoryear{Deelman, Singh, Livny, Berriman, and
  Good}{Deelman et~al\mbox{.}}{2008}]%
        {Deelman:2008:CDS:1413370.1413421}
\bibfield{author}{\bibinfo{person}{Ewa Deelman}, \bibinfo{person}{Gurmeet
  Singh}, \bibinfo{person}{Miron Livny}, \bibinfo{person}{Bruce Berriman},
  {and} \bibinfo{person}{John Good}.} \bibinfo{year}{2008}\natexlab{}.
\newblock \showarticletitle{The Cost of Doing Science on The Cloud: The Montage
  Example}. In \bibinfo{booktitle}{\emph{Proceedings of The ACM/IEEE Conference
  on Supercomputing}}. \bibinfo{pages}{1--12}.
\newblock
\showISBNx{978-1-4244-2835-9}
\urldef\tempurl%
\url{https://doi.org/10.1109/SC.2008.5217932}
\showDOI{\tempurl}


\bibitem[\protect\citeauthoryear{Deelman, Vahi, Juve, Rynge, Callaghan,
  Maechling, Mayani, Chen, da~Silva, Livny, and Wenger}{Deelman
  et~al\mbox{.}}{2015}]%
        {DEELMAN201517}
\bibfield{author}{\bibinfo{person}{Ewa Deelman}, \bibinfo{person}{Karan Vahi},
  \bibinfo{person}{Gideon Juve}, \bibinfo{person}{Mats Rynge},
  \bibinfo{person}{Scott Callaghan}, \bibinfo{person}{Philip~J. Maechling},
  \bibinfo{person}{Rajiv Mayani}, \bibinfo{person}{Weiwei Chen},
  \bibinfo{person}{Rafael~Ferreira da Silva}, \bibinfo{person}{Miron Livny},
  {and} \bibinfo{person}{Kent Wenger}.} \bibinfo{year}{2015}\natexlab{}.
\newblock \showarticletitle{Pegasus, {A} {Workflow} {Management} {System} for
  {Science} {Automation}}.
\newblock \bibinfo{journal}{\emph{Future Generation Computer Systems}}
  \bibinfo{volume}{46}, \bibinfo{number}{Supplement C} (\bibinfo{year}{2015}),
  \bibinfo{pages}{17--35}.
\newblock
\showISSN{0167-739X}
\urldef\tempurl%
\url{https://doi.org/10.1016/j.future.2014.10.008}
\showDOI{\tempurl}


\bibitem[\protect\citeauthoryear{Doukas and Antonelli}{Doukas and
  Antonelli}{2014}]%
        {6962168}
\bibfield{author}{\bibinfo{person}{Charalampos Doukas} {and}
  \bibinfo{person}{Fabio Antonelli}.} \bibinfo{year}{2014}\natexlab{}.
\newblock \showarticletitle{A Full End-to-end Platform as a Service for Smart
  City Applications}. In \bibinfo{booktitle}{\emph{Proceedings of The 10th IEEE
  International Conference on Wireless and Mobile Computing, Networking and
  Communications}}. \bibinfo{pages}{181--186}.
\newblock
\showISSN{2160-4886}
\urldef\tempurl%
\url{https://doi.org/10.1109/WiMOB.2014.6962168}
\showDOI{\tempurl}


\bibitem[\protect\citeauthoryear{Durillo, Fard, and Prodan}{Durillo
  et~al\mbox{.}}{2012}]%
        {6427573}
\bibfield{author}{\bibinfo{person}{Juan~J. Durillo}, \bibinfo{person}{Hamid~M.
  Fard}, {and} \bibinfo{person}{Radu Prodan}.} \bibinfo{year}{2012}\natexlab{}.
\newblock \showarticletitle{MOHEFT: A Multi-objective List-based Method for
  Workflow Scheduling}. In \bibinfo{booktitle}{\emph{Proceedings of The 4th
  IEEE International Conference on Cloud Computing Technology and Science}}.
  \bibinfo{pages}{185--192}.
\newblock
\urldef\tempurl%
\url{https://doi.org/10.1109/CloudCom.2012.6427573}
\showDOI{\tempurl}


\bibitem[\protect\citeauthoryear{Duro, Blas, and Carretero}{Duro
  et~al\mbox{.}}{2013}]%
        {Duro:2013:HPS:2488551.2488598}
\bibfield{author}{\bibinfo{person}{Francisco~R. Duro},
  \bibinfo{person}{Javier~G. Blas}, {and} \bibinfo{person}{Jesus Carretero}.}
  \bibinfo{year}{2013}\natexlab{}.
\newblock \showarticletitle{A {H}ierarchical {P}arallel {S}torage {S}ystem
  {B}ased on {D}istributed {M}emory for {L}arge {S}cale {S}ystems}. In
  \bibinfo{booktitle}{\emph{Proceedings of The 20th European MPI Users' Group
  Meeting}}. \bibinfo{publisher}{ACM}, \bibinfo{pages}{139--140}.
\newblock
\showISBNx{978-1-4503-1903-4}
\urldef\tempurl%
\url{https://doi.org/10.1145/2488551.2488598}
\showDOI{\tempurl}


\bibitem[\protect\citeauthoryear{Ebrahimirad, Goudarzi, and Rajabi}{Ebrahimirad
  et~al\mbox{.}}{2015}]%
        {Ebrahimirad2015}
\bibfield{author}{\bibinfo{person}{Vahid Ebrahimirad}, \bibinfo{person}{Maziar
  Goudarzi}, {and} \bibinfo{person}{Aboozar Rajabi}.}
  \bibinfo{year}{2015}\natexlab{}.
\newblock \showarticletitle{Energy-{A}ware {S}cheduling for
  {P}recedence-{C}onstrained {P}arallel {V}irtual {M}achines in {V}irtualized
  {D}ata {C}enters}.
\newblock \bibinfo{journal}{\emph{Journal of Grid Computing}}
  \bibinfo{volume}{13}, \bibinfo{number}{2} (\bibinfo{year}{2015}),
  \bibinfo{pages}{233--253}.
\newblock
\showISSN{1572-9184}
\urldef\tempurl%
\url{https://doi.org/10.1007/s10723-015-9327-x}
\showDOI{\tempurl}


\bibitem[\protect\citeauthoryear{Esteves and Veiga}{Esteves and Veiga}{2016}]%
        {doi:10.1093/comjnl/bxu158}
\bibfield{author}{\bibinfo{person}{S{\'e}rgio Esteves} {and}
  \bibinfo{person}{Lu{\'i}s Veiga}.} \bibinfo{year}{2016}\natexlab{}.
\newblock \showarticletitle{WaaS: {Workflow}-as-a-{Service} for {The} {Cloud}
  with {Scheduling} of {Continuous} and {Data}-intensive {Workflows}}.
\newblock \bibinfo{journal}{\emph{Comput. J.}} \bibinfo{volume}{59},
  \bibinfo{number}{3} (\bibinfo{year}{2016}), \bibinfo{pages}{371--383}.
\newblock
\urldef\tempurl%
\url{https://doi.org/10.1093/comjnl/bxu158}
\showDOI{\tempurl}


\bibitem[\protect\citeauthoryear{Fard, Prodan, Durillo, and Fahringer}{Fard
  et~al\mbox{.}}{2012}]%
        {6217435}
\bibfield{author}{\bibinfo{person}{Hamid~M. Fard}, \bibinfo{person}{Radu
  Prodan}, \bibinfo{person}{Juan~J. Durillo}, {and} \bibinfo{person}{Thomas
  Fahringer}.} \bibinfo{year}{2012}\natexlab{}.
\newblock \showarticletitle{A {M}ulti-objective {A}pproach for {W}orkflow
  {S}cheduling in {H}eterogeneous {E}nvironments}. In
  \bibinfo{booktitle}{\emph{Proceedings of The 12th IEEE/ACM International
  Symposium on Cluster, Cloud and Grid Computing}}. \bibinfo{pages}{300--309}.
\newblock
\urldef\tempurl%
\url{https://doi.org/10.1109/CCGrid.2012.114}
\showDOI{\tempurl}


\bibitem[\protect\citeauthoryear{Fazio, Celesti, Ranjan, Liu, Chen, and
  Villari}{Fazio et~al\mbox{.}}{2016}]%
        {7742215}
\bibfield{author}{\bibinfo{person}{Maria Fazio}, \bibinfo{person}{Antonio
  Celesti}, \bibinfo{person}{Rajiv Ranjan}, \bibinfo{person}{Chang Liu},
  \bibinfo{person}{Lydia Chen}, {and} \bibinfo{person}{Massimo Villari}.}
  \bibinfo{year}{2016}\natexlab{}.
\newblock \showarticletitle{Open {I}ssues in {S}cheduling {M}icroservices in
  {T}he {C}loud}.
\newblock \bibinfo{journal}{\emph{IEEE Cloud Computing}} \bibinfo{volume}{3},
  \bibinfo{number}{5} (\bibinfo{year}{2016}), \bibinfo{pages}{81--88}.
\newblock
\showISSN{2325-6095}
\urldef\tempurl%
\url{https://doi.org/10.1109/MCC.2016.112}
\showDOI{\tempurl}


\bibitem[\protect\citeauthoryear{Gaikwad, Mandal, Ruth, Juve, Kr\'ol, and
  Deelman}{Gaikwad et~al\mbox{.}}{2016}]%
        {7568396}
\bibfield{author}{\bibinfo{person}{Prathamesh Gaikwad},
  \bibinfo{person}{Anirban Mandal}, \bibinfo{person}{Paul Ruth},
  \bibinfo{person}{Gideon Juve}, \bibinfo{person}{Dariusz Kr\'ol}, {and}
  \bibinfo{person}{Ewa Deelman}.} \bibinfo{year}{2016}\natexlab{}.
\newblock \showarticletitle{Anomaly {Detection} for {Scientific} {Workflow}
  {Applications} on {Networked} {Clouds}}. In
  \bibinfo{booktitle}{\emph{Proceedings of The International Conference on High
  Performance Computing Simulation}}. \bibinfo{pages}{645--652}.
\newblock
\urldef\tempurl%
\url{https://doi.org/10.1109/HPCSim.2016.7568396}
\showDOI{\tempurl}


\bibitem[\protect\citeauthoryear{Gerlach, Tang, Wilke, Olson, and
  Meyer}{Gerlach et~al\mbox{.}}{2015}]%
        {7092947}
\bibfield{author}{\bibinfo{person}{Wolfgang Gerlach}, \bibinfo{person}{Wei
  Tang}, \bibinfo{person}{Andreas Wilke}, \bibinfo{person}{Dan Olson}, {and}
  \bibinfo{person}{Folker Meyer}.} \bibinfo{year}{2015}\natexlab{}.
\newblock \showarticletitle{Container {Orchestration} for {Scientific}
  {Workflows}}. In \bibinfo{booktitle}{\emph{Proceedings of The IEEE
  International Conference on Cloud Engineering}}. \bibinfo{pages}{377--378}.
\newblock
\urldef\tempurl%
\url{https://doi.org/10.1109/IC2E.2015.87}
\showDOI{\tempurl}


\bibitem[\protect\citeauthoryear{Ghasemzadeh, Arabnejad, and
  Barbosa}{Ghasemzadeh et~al\mbox{.}}{2017}]%
        {ghasemzadeh_et_al:LIPIcs:2017:7088}
\bibfield{author}{\bibinfo{person}{Mozhgan Ghasemzadeh}, \bibinfo{person}{Hamid
  Arabnejad}, {and} \bibinfo{person}{Jorge~G. Barbosa}.}
  \bibinfo{year}{2017}\natexlab{}.
\newblock \showarticletitle{{Deadline-{Budget} {Constrained} {Scheduling}
  {Algorithm} for {Scientific} {Workflows} in a {Cloud} {Environment}}}. In
  \bibinfo{booktitle}{\emph{Proceedings of The 20th International Conference on
  Principles of Distributed Systems}}, Vol.~\bibinfo{volume}{70}.
  \bibinfo{pages}{19:1--19:16}.
\newblock
\showISBNx{978-3-95977-031-6}
\showISSN{1868-8969}
\urldef\tempurl%
\url{https://doi.org/10.4230/LIPIcs.OPODIS.2016.19}
\showDOI{\tempurl}


\bibitem[\protect\citeauthoryear{Goncalves, Assuncao, and Cunha}{Goncalves
  et~al\mbox{.}}{2012}]%
        {6427527}
\bibfield{author}{\bibinfo{person}{Carlos Goncalves}, \bibinfo{person}{Luis
  Assuncao}, {and} \bibinfo{person}{Jose~C. Cunha}.}
  \bibinfo{year}{2012}\natexlab{}.
\newblock \showarticletitle{Data {A}nalytics in {T}he {C}loud with {F}lexible
  {M}ap{R}educe {W}orkflows}. In \bibinfo{booktitle}{\emph{Proceedings of The
  4th IEEE International Conference on Cloud Computing Technology and Science
  Proceedings}}. \bibinfo{pages}{427--434}.
\newblock
\urldef\tempurl%
\url{https://doi.org/10.1109/CloudCom.2012.6427527}
\showDOI{\tempurl}


\bibitem[\protect\citeauthoryear{Gubbi, Buyya, Marusic, and Palaniswami}{Gubbi
  et~al\mbox{.}}{2013}]%
        {GUBBI20131645}
\bibfield{author}{\bibinfo{person}{Jayavardhana Gubbi},
  \bibinfo{person}{Rajkumar Buyya}, \bibinfo{person}{Slaven Marusic}, {and}
  \bibinfo{person}{Marimuthu Palaniswami}.} \bibinfo{year}{2013}\natexlab{}.
\newblock \showarticletitle{Internet of Things (IoT): A Vision, Architectural
  Elements, and Future Directions}.
\newblock \bibinfo{journal}{\emph{Future Generation Computer Systems}}
  \bibinfo{volume}{29}, \bibinfo{number}{7} (\bibinfo{year}{2013}),
  \bibinfo{pages}{1645--1660}.
\newblock
\showISSN{0167-739X}
\urldef\tempurl%
\url{https://doi.org/10.1016/j.future.2013.01.010}
\showDOI{\tempurl}


\bibitem[\protect\citeauthoryear{Hilman, Rodr{\'i}guez, and Buyya}{Hilman
  et~al\mbox{.}}{2018}]%
        {8603156}
\bibfield{author}{\bibinfo{person}{Muhammad~H. Hilman},
  \bibinfo{person}{Maria~A. Rodr{\'i}guez}, {and} \bibinfo{person}{Rajkumar
  Buyya}.} \bibinfo{year}{2018}\natexlab{}.
\newblock \showarticletitle{Task {R}untime {P}rediction in {S}cientific
  {W}orkflows {U}sing an {O}nline {I}ncremental {L}earning {A}pproach}. In
  \bibinfo{booktitle}{\emph{Proceedings of The 11th IEEE/ACM International
  Conference on Utility and Cloud Computing}}. \bibinfo{pages}{93--102}.
\newblock
\urldef\tempurl%
\url{https://doi.org/10.1109/UCC.2018.00018}
\showDOI{\tempurl}


\bibitem[\protect\citeauthoryear{Howe, Cole, Souroush, Koutris, Key,
  Khoussainova, and Battle}{Howe et~al\mbox{.}}{2011}]%
        {10.1007/978-3-642-22351-8_31}
\bibfield{author}{\bibinfo{person}{Bill Howe}, \bibinfo{person}{Garret Cole},
  \bibinfo{person}{Emad Souroush}, \bibinfo{person}{Paraschos Koutris},
  \bibinfo{person}{Alicia Key}, \bibinfo{person}{Nodira Khoussainova}, {and}
  \bibinfo{person}{Leilani Battle}.} \bibinfo{year}{2011}\natexlab{}.
\newblock \showarticletitle{Database-as-a-{S}ervice for {L}ong-{T}ail
  {S}cience}. In \bibinfo{booktitle}{\emph{Scientific and Statistical Database
  Management}}. \bibinfo{publisher}{Springer Berlin Heidelberg},
  \bibinfo{address}{Berlin, Heidelberg}, \bibinfo{pages}{480--489}.
\newblock
\showISBNx{978-3-642-22351-8}


\bibitem[\protect\citeauthoryear{Hsu, Huang, and Wang}{Hsu
  et~al\mbox{.}}{2011}]%
        {HSU2011860}
\bibfield{author}{\bibinfo{person}{Chihchiang Hsu}, \bibinfo{person}{Kuochan
  Huang}, {and} \bibinfo{person}{Fengjian Wang}.}
  \bibinfo{year}{2011}\natexlab{}.
\newblock \showarticletitle{Online Scheduling of Workflow Applications in Grid
  Environments}.
\newblock \bibinfo{journal}{\emph{Future Generation Computer Systems}}
  \bibinfo{volume}{27}, \bibinfo{number}{6} (\bibinfo{year}{2011}),
  \bibinfo{pages}{860--870}.
\newblock
\showISSN{0167-739X}
\urldef\tempurl%
\url{https://doi.org/10.1016/j.future.2010.10.015}
\showDOI{\tempurl}


\bibitem[\protect\citeauthoryear{Hung, Meiss, Keswani, Xiong, Sobie, and
  Yeung}{Hung et~al\mbox{.}}{2017}]%
        {Hung099010}
\bibfield{author}{\bibinfo{person}{Linghong Hung}, \bibinfo{person}{Trevor
  Meiss}, \bibinfo{person}{Jayant Keswani}, \bibinfo{person}{Yuguang Xiong},
  \bibinfo{person}{Eric Sobie}, {and} \bibinfo{person}{Kayee Yeung}.}
  \bibinfo{year}{2017}\natexlab{}.
\newblock \showarticletitle{Building {Containerized} {Workflows} for
  {R}{N}{A}-seq {Data} {Using} {The} {Bio}{Depot}-workflow-Builder ({B}w{B})}.
\newblock \bibinfo{journal}{\emph{bioRxiv}} (\bibinfo{year}{2017}).
\newblock
\urldef\tempurl%
\url{https://doi.org/10.1101/099010}
\showDOI{\tempurl}


\bibitem[\protect\citeauthoryear{Jackson, Ramakrishnan, Muriki, Canon, Cholia,
  Shalf, Wasserman, and Wright}{Jackson et~al\mbox{.}}{2010}]%
        {jackson2010performance}
\bibfield{author}{\bibinfo{person}{Keith~R. Jackson}, \bibinfo{person}{Lavanya
  Ramakrishnan}, \bibinfo{person}{Krishna Muriki}, \bibinfo{person}{Shane
  Canon}, \bibinfo{person}{Shreyas Cholia}, \bibinfo{person}{John Shalf},
  \bibinfo{person}{Harvey~J. Wasserman}, {and} \bibinfo{person}{Nicholas~J.
  Wright}.} \bibinfo{year}{2010}\natexlab{}.
\newblock \showarticletitle{Performance {Analysis} of {High} {Performance}
  {Computing} {Applications} on {The} {Amazon} {Web} {Services} {Cloud}}. In
  \bibinfo{booktitle}{\emph{Proceedings of The 2nd IEEE International
  Conference on Cloud Computing Technology and Science}}.
  \bibinfo{pages}{159--168}.
\newblock
\urldef\tempurl%
\url{https://doi.org/10.1109/CloudCom.2010.69}
\showDOI{\tempurl}


\bibitem[\protect\citeauthoryear{Jiang, Lee, and Zomaya}{Jiang
  et~al\mbox{.}}{2017}]%
        {Jiang2017b}
\bibfield{author}{\bibinfo{person}{Qingye Jiang}, \bibinfo{person}{Young~C.
  Lee}, {and} \bibinfo{person}{Albert~Y. Zomaya}.}
  \bibinfo{year}{2017}\natexlab{}.
\newblock \showarticletitle{Serverless {Execution} of {Scientific}
  {Workflows}}. In \bibinfo{booktitle}{\emph{Proceedings of The 15th
  International Conference Service-Oriented Computing}}.
  \bibinfo{pages}{706--721}.
\newblock
\showISBNx{978-3-319-69035-3}
\urldef\tempurl%
\url{https://doi.org/10.1007/978-3-319-69035-3_51}
\showDOI{\tempurl}


\bibitem[\protect\citeauthoryear{Jones, Arcand, Bergeron, Bestor, Byun,
  Milechin, Gadepally, Hubbell, Kepner, Michaleas, Mullen, Prout, Rosa, Samsi,
  Yee, and Reuther}{Jones et~al\mbox{.}}{2016}]%
        {7761629}
\bibfield{author}{\bibinfo{person}{Mike Jones}, \bibinfo{person}{Bill Arcand},
  \bibinfo{person}{Bill Bergeron}, \bibinfo{person}{David Bestor},
  \bibinfo{person}{Chansup Byun}, \bibinfo{person}{Lauren Milechin},
  \bibinfo{person}{Vijay Gadepally}, \bibinfo{person}{Matt Hubbell},
  \bibinfo{person}{Jeremy Kepner}, \bibinfo{person}{Pete Michaleas},
  \bibinfo{person}{Julie Mullen}, \bibinfo{person}{Andy Prout},
  \bibinfo{person}{Tony Rosa}, \bibinfo{person}{Siddharth Samsi},
  \bibinfo{person}{Charles Yee}, {and} \bibinfo{person}{Albert Reuther}.}
  \bibinfo{year}{2016}\natexlab{}.
\newblock \showarticletitle{Scalability of VM Provisioning Systems}. In
  \bibinfo{booktitle}{\emph{Proceedings of The IEEE High Performance Extreme
  Computing Conference}}. \bibinfo{pages}{1--5}.
\newblock
\urldef\tempurl%
\url{https://doi.org/10.1109/HPEC.2016.7761629}
\showDOI{\tempurl}


\bibitem[\protect\citeauthoryear{Jrad, Tao, and Streit}{Jrad
  et~al\mbox{.}}{2013}]%
        {Jrad:2013:BFM:2462326.2462339}
\bibfield{author}{\bibinfo{person}{Foued Jrad}, \bibinfo{person}{Jie Tao},
  {and} \bibinfo{person}{Achim Streit}.} \bibinfo{year}{2013}\natexlab{}.
\newblock \showarticletitle{A Broker-based Framework for Multi-cloud
  Workflows}. In \bibinfo{booktitle}{\emph{Proceedings of The International
  Workshop on Multi-cloud Applications and Federated Clouds}}.
  \bibinfo{pages}{61--68}.
\newblock
\showISBNx{978-1-4503-2050-4}
\urldef\tempurl%
\url{https://doi.org/10.1145/2462326.2462339}
\showDOI{\tempurl}


\bibitem[\protect\citeauthoryear{Juve, Chervenak, Deelman, Bharathi, Mehta, and
  Vahi}{Juve et~al\mbox{.}}{2013}]%
        {JUVE2013682}
\bibfield{author}{\bibinfo{person}{Gideon Juve}, \bibinfo{person}{Ann
  Chervenak}, \bibinfo{person}{Ewa Deelman}, \bibinfo{person}{Shishir
  Bharathi}, \bibinfo{person}{Gaurang Mehta}, {and} \bibinfo{person}{Karan
  Vahi}.} \bibinfo{year}{2013}\natexlab{}.
\newblock \showarticletitle{Characterizing and {Profiling} {Scientific}
  {Workflows}}.
\newblock \bibinfo{journal}{\emph{Future Generation Computer Systems}}
  \bibinfo{volume}{29}, \bibinfo{number}{3} (\bibinfo{year}{2013}),
  \bibinfo{pages}{682--692}.
\newblock
\showISSN{0167-739X}
\urldef\tempurl%
\url{https://doi.org/10.1016/j.future.2012.08.015}
\showDOI{\tempurl}


\bibitem[\protect\citeauthoryear{Kozhirbayev and Sinnott}{Kozhirbayev and
  Sinnott}{2017}]%
        {KOZHIRBAYEV2017175}
\bibfield{author}{\bibinfo{person}{Zhanibek Kozhirbayev} {and}
  \bibinfo{person}{Richard~O. Sinnott}.} \bibinfo{year}{2017}\natexlab{}.
\newblock \showarticletitle{A {Performance} {Comparison} of {Container-based}
  {Technologies} for {The} {Cloud}}.
\newblock \bibinfo{journal}{\emph{Future Generation Computer Systems}}
  \bibinfo{volume}{68}, \bibinfo{number}{Supplement C} (\bibinfo{year}{2017}),
  \bibinfo{pages}{175--182}.
\newblock
\showISSN{0167-739X}
\urldef\tempurl%
\url{https://doi.org/10.1016/j.future.2016.08.025}
\showDOI{\tempurl}


\bibitem[\protect\citeauthoryear{Kurtzer, Sochat, and Bauer}{Kurtzer
  et~al\mbox{.}}{2017}]%
        {10.1371/journal.pone.0177459}
\bibfield{author}{\bibinfo{person}{Gregory~M. Kurtzer},
  \bibinfo{person}{Vanessa Sochat}, {and} \bibinfo{person}{Michael~W. Bauer}.}
  \bibinfo{year}{2017}\natexlab{}.
\newblock \showarticletitle{Singularity: {S}cientific {C}ontainers for
  {M}obility of {C}ompute}.
\newblock \bibinfo{journal}{\emph{PLOS ONE}} \bibinfo{volume}{12},
  \bibinfo{number}{5} (\bibinfo{year}{2017}), \bibinfo{pages}{1--20}.
\newblock
\urldef\tempurl%
\url{https://doi.org/10.1371/journal.pone.0177459}
\showDOI{\tempurl}


\bibitem[\protect\citeauthoryear{Le and Paz}{Le and Paz}{2017}]%
        {Le:2017:PAA:3093338.3106737}
\bibfield{author}{\bibinfo{person}{Emily Le} {and} \bibinfo{person}{David
  Paz}.} \bibinfo{year}{2017}\natexlab{}.
\newblock \showarticletitle{Performance {A}nalysis of {A}pplications {U}sing
  {S}ingularity {C}ontainer on {S}{D}{S}{C} {C}omet}. In
  \bibinfo{booktitle}{\emph{Proceedings of The Practice and Experience in
  Advanced Research Computing 2017 on Sustainability, Success and Impact}}.
  \bibinfo{publisher}{ACM}, \bibinfo{pages}{66:1--66:4}.
\newblock
\showISBNx{978-1-4503-5272-7}
\urldef\tempurl%
\url{https://doi.org/10.1145/3093338.3106737}
\showDOI{\tempurl}


\bibitem[\protect\citeauthoryear{Leitner and Cito}{Leitner and Cito}{2016}]%
        {leitner2016performance}
\bibfield{author}{\bibinfo{person}{Philipp Leitner} {and}
  \bibinfo{person}{J\"{u}rgen Cito}.} \bibinfo{year}{2016}\natexlab{}.
\newblock \showarticletitle{Patterns in {The} {Chaos}: {A} {Study} of
  {Performance} {Variation} and {Predictability} in {Public} {I}aa{S}
  {Clouds}}.
\newblock \bibinfo{journal}{\emph{ACM Transaction of Internet Technology}}
  \bibinfo{volume}{16}, \bibinfo{number}{3} (\bibinfo{year}{2016}),
  \bibinfo{pages}{1--23}.
\newblock
\showISSN{1533-5399}
\urldef\tempurl%
\url{https://doi.org/10.1145/2885497}
\showDOI{\tempurl}


\bibitem[\protect\citeauthoryear{Lin, Guo, and Lin}{Lin et~al\mbox{.}}{2016}]%
        {CPE:CPE3582}
\bibfield{author}{\bibinfo{person}{Bing Lin}, \bibinfo{person}{Wenzhong Guo},
  {and} \bibinfo{person}{Xiuyan Lin}.} \bibinfo{year}{2016}\natexlab{}.
\newblock \showarticletitle{Online {Optimization} {Scheduling} for {Scientific}
  {Workflows} with {Deadline} {Constraint} on {Hybrid} {Clouds}}.
\newblock \bibinfo{journal}{\emph{Concurrency and Computation: Practice and
  Experience}} \bibinfo{volume}{28}, \bibinfo{number}{11}
  (\bibinfo{year}{2016}), \bibinfo{pages}{3079--3095}.
\newblock
\showISSN{1532-0634}
\urldef\tempurl%
\url{https://doi.org/10.1002/cpe.3582}
\showDOI{\tempurl}


\bibitem[\protect\citeauthoryear{Liu, Ren, Dai, Zhang, Zhou, Zhang, Min, and
  Najjari}{Liu et~al\mbox{.}}{2019}]%
        {8669862}
\bibfield{author}{\bibinfo{person}{Jiagang Liu}, \bibinfo{person}{Ju Ren},
  \bibinfo{person}{Wei Dai}, \bibinfo{person}{Deyu Zhang},
  \bibinfo{person}{Pude Zhou}, \bibinfo{person}{Yaoxue Zhang},
  \bibinfo{person}{Geyong Min}, {and} \bibinfo{person}{Noushin Najjari}.}
  \bibinfo{year}{2019}\natexlab{}.
\newblock \showarticletitle{Online {M}ulti-{W}orkflow {S}cheduling under
  {U}ncertain {T}ask {E}xecution {T}ime in {I}aa{S} {C}louds}.
\newblock \bibinfo{journal}{\emph{IEEE Transactions on Cloud Computing}}
  (\bibinfo{year}{2019}), \bibinfo{pages}{1--1}.
\newblock
\showISSN{2168-7161}
\urldef\tempurl%
\url{https://doi.org/10.1109/TCC.2019.2906300}
\showDOI{\tempurl}


\bibitem[\protect\citeauthoryear{Liu, Aida, Yokoyama, and Masatani}{Liu
  et~al\mbox{.}}{2016}]%
        {7600178}
\bibfield{author}{\bibinfo{person}{Kai Liu}, \bibinfo{person}{Kento Aida},
  \bibinfo{person}{Shigetoshi Yokoyama}, {and} \bibinfo{person}{Yoshinobu
  Masatani}.} \bibinfo{year}{2016}\natexlab{}.
\newblock \showarticletitle{Flexible {Container-based} {Computing} {Platform}
  on {Cloud} for {Scientific} {Workflows}}. In
  \bibinfo{booktitle}{\emph{Proceedings of The International Conference on
  Cloud Computing Research and Innovations}}. \bibinfo{pages}{56--63}.
\newblock
\urldef\tempurl%
\url{https://doi.org/10.1109/ICCCRI.2016.17}
\showDOI{\tempurl}


\bibitem[\protect\citeauthoryear{Lizhen, Meng, and Bi}{Lizhen
  et~al\mbox{.}}{2009}]%
        {5420119}
\bibfield{author}{\bibinfo{person}{Cui Lizhen}, \bibinfo{person}{Xu Meng},
  {and} \bibinfo{person}{Yanbing Bi}.} \bibinfo{year}{2009}\natexlab{}.
\newblock \showarticletitle{A {S}cheduling {S}trategy for {M}ultiple {Q}o{S}
  {C}onstrained {G}rid {W}orkflows}. In \bibinfo{booktitle}{\emph{Proceedings
  of The Joint Conferences on Pervasive Computing}}. \bibinfo{pages}{561--566}.
\newblock
\urldef\tempurl%
\url{https://doi.org/10.1109/JCPC.2009.5420119}
\showDOI{\tempurl}


\bibitem[\protect\citeauthoryear{Lud{\"a}scher, Altintas, Berkley, Higgins,
  Jaeger, Jones, Lee, Tao, and Zhao}{Lud{\"a}scher et~al\mbox{.}}{2006}]%
        {CPE:CPE994}
\bibfield{author}{\bibinfo{person}{Bertram Lud{\"a}scher},
  \bibinfo{person}{Ilkay Altintas}, \bibinfo{person}{Chad Berkley},
  \bibinfo{person}{Dan Higgins}, \bibinfo{person}{Efrat Jaeger},
  \bibinfo{person}{Matthew Jones}, \bibinfo{person}{Edward~A. Lee},
  \bibinfo{person}{Jing Tao}, {and} \bibinfo{person}{Yang Zhao}.}
  \bibinfo{year}{2006}\natexlab{}.
\newblock \showarticletitle{Scientific {Workflow} {Management} and {The}
  {Kepler} {System}}.
\newblock \bibinfo{journal}{\emph{Concurrency and Computation: Practice and
  Experience}} \bibinfo{volume}{18}, \bibinfo{number}{10}
  (\bibinfo{year}{2006}), \bibinfo{pages}{1039--1065}.
\newblock
\showISSN{1532-0634}
\urldef\tempurl%
\url{https://doi.org/10.1002/cpe.994}
\showDOI{\tempurl}


\bibitem[\protect\citeauthoryear{Madhavapeddy, Mortier, Rotsos, Scott, Singh,
  Gazagnaire, Smith, Hand, and Crowcroft}{Madhavapeddy et~al\mbox{.}}{2013}]%
        {Madhavapeddy:2013:ULO:2451116.2451167}
\bibfield{author}{\bibinfo{person}{Anil Madhavapeddy}, \bibinfo{person}{Richard
  Mortier}, \bibinfo{person}{Charalampos Rotsos}, \bibinfo{person}{David
  Scott}, \bibinfo{person}{Balraj Singh}, \bibinfo{person}{Thomas Gazagnaire},
  \bibinfo{person}{Steven Smith}, \bibinfo{person}{Steven Hand}, {and}
  \bibinfo{person}{Jon Crowcroft}.} \bibinfo{year}{2013}\natexlab{}.
\newblock \showarticletitle{Unikernels: {L}ibrary {O}perating {S}ystems for the
  {C}loud}. In \bibinfo{booktitle}{\emph{Proceedings of The 18th International
  Conference on Architectural Support for Programming Languages and Operating
  Systems}}. \bibinfo{publisher}{ACM}, \bibinfo{pages}{461--472}.
\newblock
\showISBNx{978-1-4503-1870-9}
\urldef\tempurl%
\url{https://doi.org/10.1145/2451116.2451167}
\showDOI{\tempurl}


\bibitem[\protect\citeauthoryear{Maechling, Deelman, Zhao, Graves, Mehta,
  Gupta, Mehringer, Kesselman, Callaghan, Okaya, Francoeur, Gupta, Cui, Vahi,
  Jordan, and Field}{Maechling et~al\mbox{.}}{2007}]%
        {Maechling2007}
\bibfield{author}{\bibinfo{person}{Philip Maechling}, \bibinfo{person}{Ewa
  Deelman}, \bibinfo{person}{Li Zhao}, \bibinfo{person}{Robert Graves},
  \bibinfo{person}{Gaurang Mehta}, \bibinfo{person}{Nitin Gupta},
  \bibinfo{person}{John Mehringer}, \bibinfo{person}{Carl Kesselman},
  \bibinfo{person}{Scott Callaghan}, \bibinfo{person}{David Okaya},
  \bibinfo{person}{Hunter Francoeur}, \bibinfo{person}{Vipin Gupta},
  \bibinfo{person}{Yifeng Cui}, \bibinfo{person}{Karan Vahi},
  \bibinfo{person}{Thomas Jordan}, {and} \bibinfo{person}{Edward Field}.}
  \bibinfo{year}{2007}\natexlab{}.
\newblock \bibinfo{booktitle}{\emph{SCEC CyberShake Workflows---Automating
  Probabilistic Seismic Hazard Analysis Calculations}}.
\newblock \bibinfo{publisher}{Springer London}, \bibinfo{pages}{143--163}.
\newblock
\showISBNx{978-1-84628-757-2}
\urldef\tempurl%
\url{https://doi.org/10.1007/978-1-84628-757-2_10}
\showDOI{\tempurl}


\bibitem[\protect\citeauthoryear{Malawski}{Malawski}{2016}]%
        {malawski2016towards}
\bibfield{author}{\bibinfo{person}{Maciej Malawski}.}
  \bibinfo{year}{2016}\natexlab{}.
\newblock \showarticletitle{Towards {S}erverless {E}xecution of {S}cientific
  {W}orkflows-{H}yper{F}low {C}ase {S}tudy}. In
  \bibinfo{booktitle}{\emph{Proceedings of The Workshop of Workflows in Support
  of Large-Scale Sciences}}. \bibinfo{pages}{25--33}.
\newblock


\bibitem[\protect\citeauthoryear{Malawski, Gajek, Zima, Balis, and
  Figiela}{Malawski et~al\mbox{.}}{2017}]%
        {MALAWSKI2017}
\bibfield{author}{\bibinfo{person}{Maciej Malawski}, \bibinfo{person}{Adam
  Gajek}, \bibinfo{person}{Adam Zima}, \bibinfo{person}{Bartosz Balis}, {and}
  \bibinfo{person}{Kamil Figiela}.} \bibinfo{year}{2017}\natexlab{}.
\newblock \showarticletitle{Serverless {Execution} of {Scientific} {Workflows}:
  {Experiments} with {HyperFlow}, {A}{W}{S} {Lambda} and {Google} {Cloud}
  {Functions}}.
\newblock \bibinfo{journal}{\emph{Future Generation Computer Systems.}}
  (\bibinfo{year}{2017}).
\newblock
\showISSN{0167-739X}
\urldef\tempurl%
\url{https://doi.org/10.1016/j.future.2017.10.029}
\showDOI{\tempurl}


\bibitem[\protect\citeauthoryear{Malawski, Juve, Deelman, and
  Nabrzyski}{Malawski et~al\mbox{.}}{2015}]%
        {MALAWSKI20151}
\bibfield{author}{\bibinfo{person}{Maciej Malawski}, \bibinfo{person}{Gideon
  Juve}, \bibinfo{person}{Ewa Deelman}, {and} \bibinfo{person}{Jarek
  Nabrzyski}.} \bibinfo{year}{2015}\natexlab{}.
\newblock \showarticletitle{Algorithms for {Cost-} and {Deadline-constrained}
  {Provisioning} for {Scientific} {Workflow} {Ensembles} in {I}aa{S} {Clouds}}.
\newblock \bibinfo{journal}{\emph{Future Generation Computer Systems}}
  \bibinfo{volume}{48}, \bibinfo{number}{Supplement C} (\bibinfo{year}{2015}),
  \bibinfo{pages}{1--18}.
\newblock
\showISSN{0167-739X}
\urldef\tempurl%
\url{https://doi.org/10.1016/j.future.2015.01.004}
\showDOI{\tempurl}


\bibitem[\protect\citeauthoryear{Mao and Humphrey}{Mao and Humphrey}{2012}]%
        {6253534}
\bibfield{author}{\bibinfo{person}{Ming Mao} {and} \bibinfo{person}{Marty
  Humphrey}.} \bibinfo{year}{2012}\natexlab{}.
\newblock \showarticletitle{A {Performance} {Study} on {The} {V}{M} {Startup}
  {Time} in {The} {Cloud}}. In \bibinfo{booktitle}{\emph{Proceedings of The 5th
  IEEE International Conference on Cloud Computing}}.
  \bibinfo{pages}{423--430}.
\newblock
\showISSN{2159-6182}
\urldef\tempurl%
\url{https://doi.org/10.1109/CLOUD.2012.103}
\showDOI{\tempurl}


\bibitem[\protect\citeauthoryear{Montes, Zou, Singh, Tao, and Parashar}{Montes
  et~al\mbox{.}}{2014}]%
        {6973738}
\bibfield{author}{\bibinfo{person}{Javier~D. Montes}, \bibinfo{person}{Mengsong
  Zou}, \bibinfo{person}{Rahul Singh}, \bibinfo{person}{Shu Tao}, {and}
  \bibinfo{person}{Manish Parashar}.} \bibinfo{year}{2014}\natexlab{}.
\newblock \showarticletitle{Data-Driven Workflows in Multi-cloud Marketplaces}.
  In \bibinfo{booktitle}{\emph{Proceedings of The 7th IEEE International
  Conference on Cloud Computing}}. \bibinfo{pages}{168--175}.
\newblock
\showISSN{2159-6182}
\urldef\tempurl%
\url{https://doi.org/10.1109/CLOUD.2014.32}
\showDOI{\tempurl}


\bibitem[\protect\citeauthoryear{Murphy, Kagey, Fenn, and Goasguen}{Murphy
  et~al\mbox{.}}{2009}]%
        {Murphy:2009:DPV:1577849.1577925}
\bibfield{author}{\bibinfo{person}{Michael~A. Murphy}, \bibinfo{person}{Brandon
  Kagey}, \bibinfo{person}{Michael Fenn}, {and} \bibinfo{person}{Sebastien
  Goasguen}.} \bibinfo{year}{2009}\natexlab{}.
\newblock \showarticletitle{Dynamic Provisioning of Virtual Organization
  Clusters}. In \bibinfo{booktitle}{\emph{Proceedings of The 9th IEEE/ACM
  International Symposium on Cluster Computing and The Grid}}.
  \bibinfo{pages}{364--371}.
\newblock
\urldef\tempurl%
\url{https://doi.org/10.1109/CCGRID.2009.37}
\showDOI{\tempurl}


\bibitem[\protect\citeauthoryear{Nadeem, Alghazzawi, Mashat, Fakeeh, Almalaise,
  and Hagras}{Nadeem et~al\mbox{.}}{2017}]%
        {Nadeem2017}
\bibfield{author}{\bibinfo{person}{Farrukh Nadeem}, \bibinfo{person}{Daniyal
  Alghazzawi}, \bibinfo{person}{Abdulfattah Mashat}, \bibinfo{person}{Khalid
  Fakeeh}, \bibinfo{person}{Abdullah Almalaise}, {and} \bibinfo{person}{Hani
  Hagras}.} \bibinfo{year}{2017}\natexlab{}.
\newblock \showarticletitle{Modeling and {Predicting} {Execution} {Time} of
  {Scientific} {Workflows} in The {Grid} {Using} {Radial} {Basis} {Function}
  {Neural} {Network}}.
\newblock \bibinfo{journal}{\emph{Cluster Computing}} \bibinfo{volume}{20},
  \bibinfo{number}{3} (\bibinfo{year}{2017}), \bibinfo{pages}{2805--2819}.
\newblock
\showISSN{1573-7543}
\urldef\tempurl%
\url{https://doi.org/10.1007/s10586-017-1018-x}
\showDOI{\tempurl}


\bibitem[\protect\citeauthoryear{Nadeem and Fahringer}{Nadeem and
  Fahringer}{2009}]%
        {5071887}
\bibfield{author}{\bibinfo{person}{Farrukh Nadeem} {and}
  \bibinfo{person}{Thomas Fahringer}.} \bibinfo{year}{2009}\natexlab{}.
\newblock \showarticletitle{Using {T}emplates to {P}redict {E}xecution {T}ime
  of {S}cientific {W}orkflow {A}pplications in {T}he {G}rid}. In
  \bibinfo{booktitle}{\emph{Proceedings of The 9th IEEE/ACM International
  Symposium on Cluster Computing and The Grid}}. \bibinfo{pages}{316--323}.
\newblock
\urldef\tempurl%
\url{https://doi.org/10.1109/CCGRID.2009.77}
\showDOI{\tempurl}


\bibitem[\protect\citeauthoryear{Nardelli, Nastic, Dustdar, Villari, and
  Ranjan}{Nardelli et~al\mbox{.}}{2017}]%
        {7912282}
\bibfield{author}{\bibinfo{person}{Matteo Nardelli}, \bibinfo{person}{Stefan
  Nastic}, \bibinfo{person}{Schahram Dustdar}, \bibinfo{person}{Massimo
  Villari}, {and} \bibinfo{person}{Rajiv Ranjan}.}
  \bibinfo{year}{2017}\natexlab{}.
\newblock \showarticletitle{Osmotic Flow: Osmotic Computing + IoT Workflow}.
\newblock \bibinfo{journal}{\emph{IEEE Cloud Computing}} \bibinfo{volume}{4},
  \bibinfo{number}{2} (\bibinfo{year}{2017}), \bibinfo{pages}{68--75}.
\newblock
\showISSN{2325-6095}
\urldef\tempurl%
\url{https://doi.org/10.1109/MCC.2017.22}
\showDOI{\tempurl}


\bibitem[\protect\citeauthoryear{Oinn, Addis, Ferris, Marvin, Senger,
  Greenwood, Carver, Glover, Pocock, Wipat, and Li}{Oinn et~al\mbox{.}}{2004}]%
        {doi:10.1093/bioinformatics/bth361}
\bibfield{author}{\bibinfo{person}{Tom Oinn}, \bibinfo{person}{Matthew Addis},
  \bibinfo{person}{Justin Ferris}, \bibinfo{person}{Darren Marvin},
  \bibinfo{person}{Martin Senger}, \bibinfo{person}{Mark Greenwood},
  \bibinfo{person}{Tim Carver}, \bibinfo{person}{Kevin Glover},
  \bibinfo{person}{Matthew~R. Pocock}, \bibinfo{person}{Anil Wipat}, {and}
  \bibinfo{person}{Peter Li}.} \bibinfo{year}{2004}\natexlab{}.
\newblock \showarticletitle{Taverna: {A} {Tool} for {The} {Composition} and
  {Enactment} of {Bioinformatics} {Workflows}}.
\newblock \bibinfo{journal}{\emph{Bioinformatics}} \bibinfo{volume}{20},
  \bibinfo{number}{17} (\bibinfo{year}{2004}), \bibinfo{pages}{3045--3054}.
\newblock
\urldef\tempurl%
\url{https://doi.org/10.1093/bioinformatics/bth361}
\showDOI{\tempurl}


\bibitem[\protect\citeauthoryear{Omote, Shinagawa, and Kato}{Omote
  et~al\mbox{.}}{2015}]%
        {Omote:2015:IAE:2694344.2694349}
\bibfield{author}{\bibinfo{person}{Yushi Omote}, \bibinfo{person}{Takahiro
  Shinagawa}, {and} \bibinfo{person}{Kazuhiko Kato}.}
  \bibinfo{year}{2015}\natexlab{}.
\newblock \showarticletitle{Improving {A}gility and {E}lasticity in
  {B}are-metal {C}louds}. In \bibinfo{booktitle}{\emph{Proceedings of The 20th
  International Conference on Architectural Support for Programming Languages
  and Operating Systems}}. \bibinfo{publisher}{ACM}, \bibinfo{pages}{145--159}.
\newblock
\showISBNx{978-1-4503-2835-7}
\urldef\tempurl%
\url{https://doi.org/10.1145/2694344.2694349}
\showDOI{\tempurl}


\bibitem[\protect\citeauthoryear{Pham, Durillo, and Fahringer}{Pham
  et~al\mbox{.}}{2017}]%
        {8013738}
\bibfield{author}{\bibinfo{person}{Thanh~P. Pham}, \bibinfo{person}{Juan~J.
  Durillo}, {and} \bibinfo{person}{Thomas Fahringer}.}
  \bibinfo{year}{2017}\natexlab{}.
\newblock \showarticletitle{Predicting {Workflow} {Task} {Execution} {Time} in
  {The} {Cloud} {Using} {A} {Two}-{Stage} {Machine} {Learning} {Approach}}.
\newblock \bibinfo{journal}{\emph{IEEE Transactions on Cloud Computing}}
  \bibinfo{number}{99} (\bibinfo{year}{2017}), \bibinfo{pages}{1--1}.
\newblock
\urldef\tempurl%
\url{https://doi.org/10.1109/TCC.2017.2732344}
\showDOI{\tempurl}


\bibitem[\protect\citeauthoryear{Pietri and Sakellariou}{Pietri and
  Sakellariou}{2014}]%
        {7103444}
\bibfield{author}{\bibinfo{person}{Ilia Pietri} {and} \bibinfo{person}{Rizos
  Sakellariou}.} \bibinfo{year}{2014}\natexlab{}.
\newblock \showarticletitle{Energy-{A}ware {W}orkflow {S}cheduling {U}sing
  {F}requency {S}caling}. In \bibinfo{booktitle}{\emph{Proceedings of The 43rd
  International Conference on Parallel Processing Workshops}}.
  \bibinfo{pages}{104--113}.
\newblock
\showISSN{0190-3918}
\urldef\tempurl%
\url{https://doi.org/10.1109/ICPPW.2014.26}
\showDOI{\tempurl}


\bibitem[\protect\citeauthoryear{Poola, Garg, Buyya, Yang, and
  Ramamohanarao}{Poola et~al\mbox{.}}{2014}]%
        {6838754}
\bibfield{author}{\bibinfo{person}{Deepak Poola}, \bibinfo{person}{Saurab
  Garg}, \bibinfo{person}{Rajkumar Buyya}, \bibinfo{person}{Yun Yang}, {and}
  \bibinfo{person}{Kotagiri Ramamohanarao}.} \bibinfo{year}{2014}\natexlab{}.
\newblock \showarticletitle{Robust {S}cheduling of {S}cientific {W}orkflows
  with {D}eadline and {B}udget {C}onstraints in {C}louds}. In
  \bibinfo{booktitle}{\emph{Proceedings of The 28th IEEE International
  Conference on Advanced Information Networking and Applications}}.
  \bibinfo{pages}{858--865}.
\newblock
\showISSN{1550-445X}
\urldef\tempurl%
\url{https://doi.org/10.1109/AINA.2014.105}
\showDOI{\tempurl}


\bibitem[\protect\citeauthoryear{Qasha, Cala, and Watson}{Qasha
  et~al\mbox{.}}{2016}]%
        {7830693}
\bibfield{author}{\bibinfo{person}{Rawaa Qasha}, \bibinfo{person}{Jacek Cala},
  {and} \bibinfo{person}{Paul Watson}.} \bibinfo{year}{2016}\natexlab{}.
\newblock \showarticletitle{Dynamic {Deployment} of {Scientific} {Workflows} in
  {The} {Cloud} {Using} {Container} {Virtualization}}. In
  \bibinfo{booktitle}{\emph{Proceedings of The IEEE International Conference on
  Cloud Computing Technology and Science}}. \bibinfo{pages}{269--276}.
\newblock
\urldef\tempurl%
\url{https://doi.org/10.1109/CloudCom.2016.0052}
\showDOI{\tempurl}


\bibitem[\protect\citeauthoryear{Qin and Jiang}{Qin and Jiang}{2006}]%
        {QIN2006331}
\bibfield{author}{\bibinfo{person}{Xiao Qin} {and} \bibinfo{person}{Hong
  Jiang}.} \bibinfo{year}{2006}\natexlab{}.
\newblock \showarticletitle{A {N}ovel {F}ault-tolerant {S}cheduling {A}lgorithm
  for {P}recedence {C}onstrained {T}asks in {R}eal-time {H}eterogeneous
  {S}ystems}.
\newblock \bibinfo{journal}{\emph{Parallel Comput.}} \bibinfo{volume}{32},
  \bibinfo{number}{5} (\bibinfo{year}{2006}), \bibinfo{pages}{331--356}.
\newblock
\showISSN{0167-8191}
\urldef\tempurl%
\url{https://doi.org/10.1016/j.parco.2006.06.006}
\showDOI{\tempurl}


\bibitem[\protect\citeauthoryear{Rimal and El-Refaey}{Rimal and
  El-Refaey}{2010}]%
        {5541997}
\bibfield{author}{\bibinfo{person}{Bhaskar~P. Rimal} {and}
  \bibinfo{person}{Mohamed~A. El-Refaey}.} \bibinfo{year}{2010}\natexlab{}.
\newblock \showarticletitle{A Framework of Scientific Workflow Management
  Systems for Multi-tenant Cloud Orchestration Environment}. In
  \bibinfo{booktitle}{\emph{Proceedings of The 19th IEEE International
  Workshops on Enabling Technologies: Infrastructures for Collaborative
  Enterprises}}. \bibinfo{pages}{88--93}.
\newblock
\showISSN{1524-4547}
\urldef\tempurl%
\url{https://doi.org/10.1109/WETICE.2010.20}
\showDOI{\tempurl}


\bibitem[\protect\citeauthoryear{Rimal and Maier}{Rimal and Maier}{2017}]%
        {7457258}
\bibfield{author}{\bibinfo{person}{Bhaskar~P. Rimal} {and}
  \bibinfo{person}{Martin Maier}.} \bibinfo{year}{2017}\natexlab{}.
\newblock \showarticletitle{Workflow {Scheduling} in {Multi-tenant} {Cloud}
  {Computing} {Environments}}.
\newblock \bibinfo{journal}{\emph{IEEE Transactions on Parallel and Distributed
  Systems}} \bibinfo{volume}{28}, \bibinfo{number}{1} (\bibinfo{year}{2017}),
  \bibinfo{pages}{290--304}.
\newblock
\showISSN{1045-9219}
\urldef\tempurl%
\url{https://doi.org/10.1109/TPDS.2016.2556668}
\showDOI{\tempurl}


\bibitem[\protect\citeauthoryear{Rodriguez and Buyya}{Rodriguez and
  Buyya}{2017}]%
        {CPE:CPE4041}
\bibfield{author}{\bibinfo{person}{Maria~A. Rodriguez} {and}
  \bibinfo{person}{Rajkumar Buyya}.} \bibinfo{year}{2017}\natexlab{}.
\newblock \showarticletitle{A {Taxonomy} and {Survey} on {Scheduling}
  {Algorithms} for {Scientific} {Workflows} in {I}aa{S} {Cloud} {Computing}
  {Environments}}.
\newblock \bibinfo{journal}{\emph{Concurrency and Computation: Practice and
  Experience}} \bibinfo{volume}{29}, \bibinfo{number}{8}
  (\bibinfo{year}{2017}), \bibinfo{pages}{e4041--n/a}.
\newblock
\showISSN{1532-0634}
\urldef\tempurl%
\url{https://doi.org/10.1002/cpe.4041}
\showDOI{\tempurl}


\bibitem[\protect\citeauthoryear{Rodriguez and Buyya}{Rodriguez and
  Buyya}{2018}]%
        {RODRIGUEZ2018739}
\bibfield{author}{\bibinfo{person}{Maria~A. Rodriguez} {and}
  \bibinfo{person}{Rajkumar Buyya}.} \bibinfo{year}{2018}\natexlab{}.
\newblock \showarticletitle{Scheduling {Dynamic} {Workloads} in {Multi-tenant}
  {Scientific} {Workflow} as a {Service} {Platforms}}.
\newblock \bibinfo{journal}{\emph{Future Generation Computer Systems}}
  \bibinfo{volume}{79}, \bibinfo{number}{Part 2} (\bibinfo{year}{2018}),
  \bibinfo{pages}{739--750}.
\newblock
\showISSN{0167-739X}
\urldef\tempurl%
\url{https://doi.org/10.1016/j.future.2017.05.009}
\showDOI{\tempurl}


\bibitem[\protect\citeauthoryear{Rodriguez, Kotagiri, and Buyya}{Rodriguez
  et~al\mbox{.}}{2018}]%
        {RODRIGUEZ2018}
\bibfield{author}{\bibinfo{person}{Maria~A. Rodriguez},
  \bibinfo{person}{Ramamohanarao Kotagiri}, {and} \bibinfo{person}{Rajkumar
  Buyya}.} \bibinfo{year}{2018}\natexlab{}.
\newblock \showarticletitle{Detecting Performance Anomalies in Scientific
  Workflows Using Hierarchical Temporal Memory}.
\newblock \bibinfo{journal}{\emph{Future Generation Computer Systems}}
  (\bibinfo{year}{2018}).
\newblock
\showISSN{0167-739X}
\urldef\tempurl%
\url{https://doi.org/10.1016/j.future.2018.05.014}
\showDOI{\tempurl}


\bibitem[\protect\citeauthoryear{Sahoo, Hoi, and Li}{Sahoo
  et~al\mbox{.}}{2019}]%
        {Sahoo:2019:LSO:3301280.3299875}
\bibfield{author}{\bibinfo{person}{Doyen Sahoo}, \bibinfo{person}{Steven C.~H.
  Hoi}, {and} \bibinfo{person}{Bin Li}.} \bibinfo{year}{2019}\natexlab{}.
\newblock \showarticletitle{Large {S}cale {O}nline {M}ultiple {K}ernel
  {R}egression with {A}pplication to {T}ime-{S}eries {P}rediction}.
\newblock \bibinfo{journal}{\emph{ACM Transaction on Knowledge Discovery from
  Data}} \bibinfo{volume}{13}, \bibinfo{number}{1} (\bibinfo{year}{2019}),
  \bibinfo{pages}{9:1--9:33}.
\newblock
\showISSN{1556-4681}
\urldef\tempurl%
\url{https://doi.org/10.1145/3299875}
\showDOI{\tempurl}


\bibitem[\protect\citeauthoryear{Samak, Gunter, Goode, Deelman, Juve, Mehta,
  Silva, and Vahi}{Samak et~al\mbox{.}}{2011}]%
        {6063014}
\bibfield{author}{\bibinfo{person}{Taghrid Samak}, \bibinfo{person}{Dan
  Gunter}, \bibinfo{person}{Monte Goode}, \bibinfo{person}{Ewa Deelman},
  \bibinfo{person}{Gideon Juve}, \bibinfo{person}{Gaurang Mehta},
  \bibinfo{person}{Fabio Silva}, {and} \bibinfo{person}{Karan Vahi}.}
  \bibinfo{year}{2011}\natexlab{}.
\newblock \showarticletitle{Online {Fault} and {Anomaly} {Detection} for
  {Large}-{Scale} {Scientific} {Workflows}}. In
  \bibinfo{booktitle}{\emph{Proceedings of The IEEE International Conference on
  High Performance Computing and Communications}}. \bibinfo{pages}{373--381}.
\newblock
\urldef\tempurl%
\url{https://doi.org/10.1109/HPCC.2011.55}
\showDOI{\tempurl}


\bibitem[\protect\citeauthoryear{Sharif, Taheri, Zomaya, and Nepal}{Sharif
  et~al\mbox{.}}{2014}]%
        {7037702}
\bibfield{author}{\bibinfo{person}{Shaghayegh Sharif}, \bibinfo{person}{Javid
  Taheri}, \bibinfo{person}{Albert~Y. Zomaya}, {and} \bibinfo{person}{Surya
  Nepal}.} \bibinfo{year}{2014}\natexlab{}.
\newblock \showarticletitle{Online {Multiple} {Workflow} {Scheduling} under
  {Privacy} and {Deadline} in {Hybrid} {Cloud} {Environment}}. In
  \bibinfo{booktitle}{\emph{Proceedings of The 6th IEEE International
  Conference on Cloud Computing Technology and Science}}.
  \bibinfo{pages}{455--462}.
\newblock
\urldef\tempurl%
\url{https://doi.org/10.1109/CloudCom.2014.128}
\showDOI{\tempurl}


\bibitem[\protect\citeauthoryear{Shea, Wang, Wang, and Liu}{Shea
  et~al\mbox{.}}{2014}]%
        {6848061}
\bibfield{author}{\bibinfo{person}{Ryan Shea}, \bibinfo{person}{Feng Wang},
  \bibinfo{person}{Haiyang Wang}, {and} \bibinfo{person}{Jiangchuan Liu}.}
  \bibinfo{year}{2014}\natexlab{}.
\newblock \showarticletitle{A {Deep} {Investigation} {Into} {Network}
  {Performance} in {Virtual} {Machine} {Based} {Cloud} {Environments}}. In
  \bibinfo{booktitle}{\emph{Proceeding of The IEEE Conference on Computer
  Communications}}. \bibinfo{pages}{1285--1293}.
\newblock
\showISSN{0743-166X}
\urldef\tempurl%
\url{https://doi.org/10.1109/INFOCOM.2014.6848061}
\showDOI{\tempurl}


\bibitem[\protect\citeauthoryear{Singh and Chana}{Singh and Chana}{2016}]%
        {Singh2016}
\bibfield{author}{\bibinfo{person}{Sukhpal Singh} {and}
  \bibinfo{person}{Inderveer Chana}.} \bibinfo{year}{2016}\natexlab{}.
\newblock \showarticletitle{A {S}urvey on {R}esource {S}cheduling in {C}loud
  {C}omputing: {I}ssues and {C}hallenges}.
\newblock \bibinfo{journal}{\emph{Journal of Grid Computing}}
  \bibinfo{volume}{14}, \bibinfo{number}{2} (\bibinfo{year}{2016}),
  \bibinfo{pages}{217--264}.
\newblock
\showISSN{1572-9184}
\urldef\tempurl%
\url{https://doi.org/10.1007/s10723-015-9359-2}
\showDOI{\tempurl}


\bibitem[\protect\citeauthoryear{Smanchat and Viriyapant}{Smanchat and
  Viriyapant}{2015}]%
        {SMANCHAT20151}
\bibfield{author}{\bibinfo{person}{Sucha Smanchat} {and}
  \bibinfo{person}{Kanchana Viriyapant}.} \bibinfo{year}{2015}\natexlab{}.
\newblock \showarticletitle{Taxonomies of {Workflow} {Scheduling} {Problem} and
  {Techniques} in {The} {Cloud}}.
\newblock \bibinfo{journal}{\emph{Future Generation Computer Systems}}
  \bibinfo{volume}{52}, \bibinfo{number}{Supplement C} (\bibinfo{year}{2015}),
  \bibinfo{pages}{1--12}.
\newblock
\showISSN{0167-739X}
\urldef\tempurl%
\url{https://doi.org/10.1016/j.future.2015.04.019}
\showDOI{\tempurl}


\bibitem[\protect\citeauthoryear{Stavrinides, Duro, Karatza, Blas, and
  Carretero}{Stavrinides et~al\mbox{.}}{2017}]%
        {STAVRINIDES2017120}
\bibfield{author}{\bibinfo{person}{Georgios~L. Stavrinides},
  \bibinfo{person}{Francisco~R. Duro}, \bibinfo{person}{Helen~D. Karatza},
  \bibinfo{person}{Javier~G. Blas}, {and} \bibinfo{person}{Jesus Carretero}.}
  \bibinfo{year}{2017}\natexlab{}.
\newblock \showarticletitle{Different {Aspects} of {Workflow} {Scheduling} in
  {Large-scale} {Distributed} {Systems}}.
\newblock \bibinfo{journal}{\emph{Simulation Modelling Practice and Theory}}
  \bibinfo{volume}{70}, \bibinfo{number}{Supplement C} (\bibinfo{year}{2017}),
  \bibinfo{pages}{120--134}.
\newblock
\showISSN{1569-190X}
\urldef\tempurl%
\url{https://doi.org/10.1016/j.simpat.2016.10.009}
\showDOI{\tempurl}


\bibitem[\protect\citeauthoryear{Stavrinides and Karatza}{Stavrinides and
  Karatza}{2010}]%
        {STAVRINIDES20101004}
\bibfield{author}{\bibinfo{person}{Georgios~L. Stavrinides} {and}
  \bibinfo{person}{Helen~D. Karatza}.} \bibinfo{year}{2010}\natexlab{}.
\newblock \showarticletitle{Scheduling Multiple Task Graphs with End-to-end
  Deadlines in Distributed Real-time Systems Utilizing Imprecise Computations}.
\newblock \bibinfo{journal}{\emph{Journal of Systems and Software}}
  \bibinfo{volume}{83}, \bibinfo{number}{6} (\bibinfo{year}{2010}),
  \bibinfo{pages}{1004--1014}.
\newblock
\showISSN{0164-1212}
\urldef\tempurl%
\url{https://doi.org/10.1016/j.jss.2009.12.025}
\showDOI{\tempurl}


\bibitem[\protect\citeauthoryear{Stavrinides and Karatza}{Stavrinides and
  Karatza}{2011}]%
        {STAVRINIDES2011540}
\bibfield{author}{\bibinfo{person}{Georgios~L. Stavrinides} {and}
  \bibinfo{person}{Helen~D. Karatza}.} \bibinfo{year}{2011}\natexlab{}.
\newblock \showarticletitle{Scheduling Multiple Task Graphs in Heterogeneous
  Distributed Real-time Systems by Exploiting Schedule Holes with Bin Packing
  Techniques}.
\newblock \bibinfo{journal}{\emph{Simulation Modelling Practice and Theory}}
  \bibinfo{volume}{19}, \bibinfo{number}{1} (\bibinfo{year}{2011}),
  \bibinfo{pages}{540--552}.
\newblock
\showISSN{1569-190X}
\urldef\tempurl%
\url{https://doi.org/10.1016/j.simpat.2010.08.010}
\showDOI{\tempurl}


\bibitem[\protect\citeauthoryear{Stavrinides and Karatza}{Stavrinides and
  Karatza}{2015}]%
        {7300823}
\bibfield{author}{\bibinfo{person}{Georgios~L. Stavrinides} {and}
  \bibinfo{person}{Helen~D. Karatza}.} \bibinfo{year}{2015}\natexlab{}.
\newblock \showarticletitle{A {Cost-effective} and {Q}o{S}-aware {Approach} to
  {Scheduling} {Real-time} {Workflow} {Applications} in {P}aa{S} and {S}aa{S}
  {Clouds}}. In \bibinfo{booktitle}{\emph{Proceedings of The 3rd International
  Conference on Future Internet of Things and Cloud}}.
  \bibinfo{pages}{231--239}.
\newblock
\urldef\tempurl%
\url{https://doi.org/10.1109/FiCloud.2015.93}
\showDOI{\tempurl}


\bibitem[\protect\citeauthoryear{Svitil}{Svitil}{2016}]%
        {ras2016gravitational}
\bibfield{author}{\bibinfo{person}{Kathy Svitil}.}
  \bibinfo{year}{2016}\natexlab{}.
\newblock \showarticletitle{Gravitational Waves Detected 100 Years After
  Einstein's Prediction}.
\newblock  (\bibinfo{year}{2016}).
\newblock
\urldef\tempurl%
\url{http://www.caltech.edu/news/gravitational-waves-detected-100-years-after-einstein-s-prediction-49777}
\showURL{%
\tempurl}


\bibitem[\protect\citeauthoryear{Tang, Li, Liao, Fang, and Wu}{Tang
  et~al\mbox{.}}{2011}]%
        {TANG20111083}
\bibfield{author}{\bibinfo{person}{Xiaoyong Tang}, \bibinfo{person}{Kenli Li},
  \bibinfo{person}{Guiping Liao}, \bibinfo{person}{Kui Fang}, {and}
  \bibinfo{person}{Fan Wu}.} \bibinfo{year}{2011}\natexlab{}.
\newblock \showarticletitle{A {S}tochastic {S}cheduling {A}lgorithm for
  {P}recedence {C}onstrained {T}asks on {G}rid}.
\newblock \bibinfo{journal}{\emph{Future Generation Computer Systems}}
  \bibinfo{volume}{27}, \bibinfo{number}{8} (\bibinfo{year}{2011}),
  \bibinfo{pages}{1083--1091}.
\newblock
\showISSN{0167-739X}
\urldef\tempurl%
\url{https://doi.org/10.1016/j.future.2011.04.007}
\showDOI{\tempurl}


\bibitem[\protect\citeauthoryear{Tang, Qi, Cheng, Li, Khan, and Li}{Tang
  et~al\mbox{.}}{2016}]%
        {Tang2016}
\bibfield{author}{\bibinfo{person}{Zhuo Tang}, \bibinfo{person}{Ling Qi},
  \bibinfo{person}{Zhenzhen Cheng}, \bibinfo{person}{Kenli Li},
  \bibinfo{person}{Samee~U. Khan}, {and} \bibinfo{person}{Keqin Li}.}
  \bibinfo{year}{2016}\natexlab{}.
\newblock \showarticletitle{An {E}nergy-{E}fficient {T}ask {S}cheduling
  {A}lgorithm in {D}{V}{F}{S}-enabled {C}loud {E}nvironment}.
\newblock \bibinfo{journal}{\emph{Journal of Grid Computing}}
  \bibinfo{volume}{14}, \bibinfo{number}{1} (\bibinfo{year}{2016}),
  \bibinfo{pages}{55--74}.
\newblock
\showISSN{1572-9184}
\urldef\tempurl%
\url{https://doi.org/10.1007/s10723-015-9334-y}
\showDOI{\tempurl}


\bibitem[\protect\citeauthoryear{Taylor, Deelman, Gannon, and Shields}{Taylor
  et~al\mbox{.}}{2014}]%
        {Taylor:2014:WES:2655383}
\bibfield{author}{\bibinfo{person}{Ian~J. Taylor}, \bibinfo{person}{Ewa
  Deelman}, \bibinfo{person}{Dennis~B. Gannon}, {and} \bibinfo{person}{Matthew
  Shields}.} \bibinfo{year}{2014}\natexlab{}.
\newblock \bibinfo{booktitle}{\emph{Workflows for e-{S}cience: {S}cientific
  {W}orkflows for {G}rids}}.
\newblock \bibinfo{publisher}{Springer Publishing Company, Incorporated}.
\newblock
\showISBNx{1849966192, 9781849966191}


\bibitem[\protect\citeauthoryear{Tian, Xiao, Xu, and Xiao}{Tian
  et~al\mbox{.}}{2012}]%
        {tian2012hybrid}
\bibfield{author}{\bibinfo{person}{Guo-Zhong Tian}, \bibinfo{person}{Chuang-Bai
  Xiao}, \bibinfo{person}{Zhu-Sheng Xu}, {and} \bibinfo{person}{Xia Xiao}.}
  \bibinfo{year}{2012}\natexlab{}.
\newblock \showarticletitle{Hybrid {S}cheduling {S}trategy for {M}ultiple
  {D}{A}{G}s {W}orkflow in {H}eterogeneous {S}ystem}.
\newblock \bibinfo{journal}{\emph{Ruanjian Xuebao/Journal of Software}}
  \bibinfo{volume}{23}, \bibinfo{number}{10} (\bibinfo{year}{2012}),
  \bibinfo{pages}{2720--2734}.
\newblock


\bibitem[\protect\citeauthoryear{Topcuoglu, Hariri, and Wu}{Topcuoglu
  et~al\mbox{.}}{2002}]%
        {993206}
\bibfield{author}{\bibinfo{person}{Haluk Topcuoglu}, \bibinfo{person}{Salim
  Hariri}, {and} \bibinfo{person}{Min~You Wu}.}
  \bibinfo{year}{2002}\natexlab{}.
\newblock \showarticletitle{Performance-Effective and Low-Complexity Task
  Scheduling for Heterogeneous Computing}.
\newblock \bibinfo{journal}{\emph{IEEE Transactions on Parallel and Distributed
  Systems}} \bibinfo{volume}{13}, \bibinfo{number}{3} (\bibinfo{year}{2002}),
  \bibinfo{pages}{260--274}.
\newblock
\showISSN{1045-9219}
\urldef\tempurl%
\url{https://doi.org/10.1109/71.993206}
\showDOI{\tempurl}


\bibitem[\protect\citeauthoryear{Tsai, Liu, and Huang}{Tsai
  et~al\mbox{.}}{2015}]%
        {Tsai2015}
\bibfield{author}{\bibinfo{person}{Yinglin Tsai}, \bibinfo{person}{Hsiaoching
  Liu}, {and} \bibinfo{person}{Kuochan Huang}.}
  \bibinfo{year}{2015}\natexlab{}.
\newblock \showarticletitle{Adaptive {Dual-criteria} {Task} {Group}
  {Allocation} for {Clustering-based} {Multi-workflow} {Scheduling} on
  {Parallel} {Computing} {Platform}}.
\newblock \bibinfo{journal}{\emph{The Journal of Supercomputing}}
  \bibinfo{volume}{71}, \bibinfo{number}{10} (\bibinfo{year}{2015}),
  \bibinfo{pages}{3811--3831}.
\newblock
\showISSN{1573-0484}
\urldef\tempurl%
\url{https://doi.org/10.1007/s11227-015-1469-x}
\showDOI{\tempurl}


\bibitem[\protect\citeauthoryear{Viriyasitavat, Xu, and
  Viriyasitavat}{Viriyasitavat et~al\mbox{.}}{2014}]%
        {6716026}
\bibfield{author}{\bibinfo{person}{Wattana Viriyasitavat},
  \bibinfo{person}{Lida Xu}, {and} \bibinfo{person}{Wantanee Viriyasitavat}.}
  \bibinfo{year}{2014}\natexlab{}.
\newblock \showarticletitle{Compliance Checking for Requirement-Oriented
  Service Workflow Interoperations}.
\newblock \bibinfo{journal}{\emph{IEEE Transactions on Industrial Informatics}}
  \bibinfo{volume}{10}, \bibinfo{number}{2} (\bibinfo{year}{2014}),
  \bibinfo{pages}{1469--1477}.
\newblock
\showISSN{1551-3203}
\urldef\tempurl%
\url{https://doi.org/10.1109/TII.2014.2301132}
\showDOI{\tempurl}


\bibitem[\protect\citeauthoryear{Wang, Korambath, Altintas, Davis, and
  Crawl}{Wang et~al\mbox{.}}{2014}]%
        {WANG2014546}
\bibfield{author}{\bibinfo{person}{Jianwu Wang}, \bibinfo{person}{Prakashan
  Korambath}, \bibinfo{person}{Ilkay Altintas}, \bibinfo{person}{Jim Davis},
  {and} \bibinfo{person}{Daniel Crawl}.} \bibinfo{year}{2014}\natexlab{}.
\newblock \showarticletitle{Workflow as a {Service} in {The} {Cloud}:
  {Architecture} and {Scheduling} {Algorithms}}.
\newblock \bibinfo{journal}{\emph{Procedia Computer Science}}
  \bibinfo{volume}{29}, \bibinfo{number}{Supplement C} (\bibinfo{year}{2014}),
  \bibinfo{pages}{546--556}.
\newblock
\showISSN{1877-0509}
\urldef\tempurl%
\url{https://doi.org/10.1016/j.procs.2014.05.049}
\showDOI{\tempurl}


\bibitem[\protect\citeauthoryear{Wang, Cao, Wang, Feng, Zhang, and Guo}{Wang
  et~al\mbox{.}}{2017}]%
        {7951947}
\bibfield{author}{\bibinfo{person}{Yuxin Wang}, \bibinfo{person}{Shijie Cao},
  \bibinfo{person}{Guan Wang}, \bibinfo{person}{Zhen Feng},
  \bibinfo{person}{Chi Zhang}, {and} \bibinfo{person}{He Guo}.}
  \bibinfo{year}{2017}\natexlab{}.
\newblock \showarticletitle{Fairness {Scheduling} with {Dynamic} {Priority} for
  {Multi} {Workflow} on {Heterogeneous} {Systems}}. In
  \bibinfo{booktitle}{\emph{Proceedings of The 2nd IEEE International
  Conference on Cloud Computing and Big Data Analysis}}.
  \bibinfo{pages}{404--409}.
\newblock
\urldef\tempurl%
\url{https://doi.org/10.1109/ICCCBDA.2017.7951947}
\showDOI{\tempurl}


\bibitem[\protect\citeauthoryear{Wang, Huang, and Wang}{Wang
  et~al\mbox{.}}{2016}]%
        {WANG201635}
\bibfield{author}{\bibinfo{person}{Yirong Wang}, \bibinfo{person}{Kuochan
  Huang}, {and} \bibinfo{person}{Fengjian Wang}.}
  \bibinfo{year}{2016}\natexlab{}.
\newblock \showarticletitle{Scheduling {Online} {Mixed-parallel} {Workflows} of
  {Rigid} {Tasks} in {Heterogeneous} {Multi-cluster} {Environments}}.
\newblock \bibinfo{journal}{\emph{Future Generation Computer Systems}}
  \bibinfo{volume}{60}, \bibinfo{number}{Supplement C} (\bibinfo{year}{2016}),
  \bibinfo{pages}{35--47}.
\newblock
\showISSN{0167-739X}
\urldef\tempurl%
\url{https://doi.org/10.1016/j.future.2016.01.013}
\showDOI{\tempurl}


\bibitem[\protect\citeauthoryear{Wieczorek, Hoheisel, and Prodan}{Wieczorek
  et~al\mbox{.}}{2008}]%
        {Wieczorek2008}
\bibfield{author}{\bibinfo{person}{Marek Wieczorek}, \bibinfo{person}{Andreas
  Hoheisel}, {and} \bibinfo{person}{Radu Prodan}.}
  \bibinfo{year}{2008}\natexlab{}.
\newblock \bibinfo{booktitle}{\emph{Taxonomies of the {M}ulti-{C}riteria {G}rid
  {W}orkflow {S}cheduling {P}roblem}}.
\newblock \bibinfo{publisher}{Springer US}, \bibinfo{pages}{237--264}.
\newblock
\showISBNx{978-0-387-78446-5}
\urldef\tempurl%
\url{https://doi.org/10.1007/978-0-387-78446-5_16}
\showDOI{\tempurl}


\bibitem[\protect\citeauthoryear{Williams, Koller, Lucina, and
  Prakash}{Williams et~al\mbox{.}}{2018}]%
        {Williams:2018:UP:3267809.3267845}
\bibfield{author}{\bibinfo{person}{Dan Williams}, \bibinfo{person}{Ricardo
  Koller}, \bibinfo{person}{Martin Lucina}, {and} \bibinfo{person}{Nikhil
  Prakash}.} \bibinfo{year}{2018}\natexlab{}.
\newblock \showarticletitle{Unikernels {A}s {P}rocesses}. In
  \bibinfo{booktitle}{\emph{Proceedings of The ACM Symposium on Cloud
  Computing}}. \bibinfo{publisher}{ACM}, \bibinfo{pages}{199--211}.
\newblock
\showISBNx{978-1-4503-6011-1}
\urldef\tempurl%
\url{https://doi.org/10.1145/3267809.3267845}
\showDOI{\tempurl}


\bibitem[\protect\citeauthoryear{Witt, Bux, Gusew, and Leser}{Witt
  et~al\mbox{.}}{2019}]%
        {WITT201933}
\bibfield{author}{\bibinfo{person}{Carl Witt}, \bibinfo{person}{Marc Bux},
  \bibinfo{person}{Wladislaw Gusew}, {and} \bibinfo{person}{Ulf Leser}.}
  \bibinfo{year}{2019}\natexlab{}.
\newblock \showarticletitle{Predictive {P}erformance {M}odeling for
  {D}istributed {B}atch {P}rocessing {U}sing {B}lack {B}ox {M}onitoring and
  {M}achine {L}earning}.
\newblock \bibinfo{journal}{\emph{Information Systems}}  \bibinfo{volume}{82}
  (\bibinfo{year}{2019}), \bibinfo{pages}{33--52}.
\newblock
\showISSN{0306-4379}
\urldef\tempurl%
\url{https://doi.org/10.1016/j.is.2019.01.006}
\showDOI{\tempurl}


\bibitem[\protect\citeauthoryear{Wu, Wu, and Tan}{Wu et~al\mbox{.}}{2015}]%
        {Wu2015}
\bibfield{author}{\bibinfo{person}{Fuhui Wu}, \bibinfo{person}{Qingbo Wu},
  {and} \bibinfo{person}{Yusong Tan}.} \bibinfo{year}{2015}\natexlab{}.
\newblock \showarticletitle{Workflow {Scheduling} in {Cloud}: {A} {Survey}}.
\newblock \bibinfo{journal}{\emph{The Journal of Supercomputing}}
  \bibinfo{volume}{71}, \bibinfo{number}{9} (\bibinfo{year}{2015}),
  \bibinfo{pages}{3373--3418}.
\newblock
\showISSN{1573-0484}
\urldef\tempurl%
\url{https://doi.org/10.1007/s11227-015-1438-4}
\showDOI{\tempurl}


\bibitem[\protect\citeauthoryear{Xie, Li, Xiao, and Chen}{Xie
  et~al\mbox{.}}{2014}]%
        {6838775}
\bibfield{author}{\bibinfo{person}{Guoqi Xie}, \bibinfo{person}{Renfa Li},
  \bibinfo{person}{Xiongren Xiao}, {and} \bibinfo{person}{Yuekun Chen}.}
  \bibinfo{year}{2014}\natexlab{}.
\newblock \showarticletitle{A High-Performance DAG Task Scheduling Algorithm
  for Heterogeneous Networked Embedded Systems}. In
  \bibinfo{booktitle}{\emph{Proceedings of The 28th IEEE International
  Conference on Advanced Information Networking and Applications}}.
  \bibinfo{pages}{1011--1016}.
\newblock
\showISSN{1550-445X}
\urldef\tempurl%
\url{https://doi.org/10.1109/AINA.2014.123}
\showDOI{\tempurl}


\bibitem[\protect\citeauthoryear{Xie, Liu, Yang, and Li}{Xie
  et~al\mbox{.}}{2017a}]%
        {CPE:CPE3782}
\bibfield{author}{\bibinfo{person}{Guoqi Xie}, \bibinfo{person}{Liangjiao Liu},
  \bibinfo{person}{Liu Yang}, {and} \bibinfo{person}{Renfa Li}.}
  \bibinfo{year}{2017}\natexlab{a}.
\newblock \showarticletitle{Scheduling {Trade-off} of {Dynamic} {Multiple}
  {Parallel} {Workflows} on {Heterogeneous} {Distributed} {Computing}
  {Systems}}.
\newblock \bibinfo{journal}{\emph{Concurrency and Computation: Practice and
  Experience}} \bibinfo{volume}{29}, \bibinfo{number}{2}
  (\bibinfo{year}{2017}), \bibinfo{pages}{e3782--n/a}.
\newblock
\showISSN{1532-0634}
\urldef\tempurl%
\url{https://doi.org/10.1002/cpe.3782}
\showDOI{\tempurl}


\bibitem[\protect\citeauthoryear{Xie, Zeng, Jiang, Fan, Li, and Li}{Xie
  et~al\mbox{.}}{2017b}]%
        {XIE2017}
\bibfield{author}{\bibinfo{person}{Guoqi Xie}, \bibinfo{person}{Gang Zeng},
  \bibinfo{person}{Junqiang Jiang}, \bibinfo{person}{Chunnian Fan},
  \bibinfo{person}{Renfa Li}, {and} \bibinfo{person}{Keqin Li}.}
  \bibinfo{year}{2017}\natexlab{b}.
\newblock \showarticletitle{Energy {Management} for {Multiple} {Real-time}
  {Workflows} on {Cyber–physical} {Cloud} {Systems}}.
\newblock \bibinfo{journal}{\emph{Future Generation Computer Systems.}}
  (\bibinfo{year}{2017}).
\newblock
\showISSN{0167-739X}
\urldef\tempurl%
\url{https://doi.org/10.1016/j.future.2017.05.033}
\showDOI{\tempurl}


\bibitem[\protect\citeauthoryear{Xu, Cui, Wang, and Bi}{Xu
  et~al\mbox{.}}{2009}]%
        {5207867}
\bibfield{author}{\bibinfo{person}{Meng Xu}, \bibinfo{person}{Lizhen Cui},
  \bibinfo{person}{Haiyang Wang}, {and} \bibinfo{person}{Yanbing Bi}.}
  \bibinfo{year}{2009}\natexlab{}.
\newblock \showarticletitle{A {Multiple} {Q}o{S} {Constrained} {Scheduling}
  {Strategy} of {Multiple} {Workflows} for {Cloud} {Computing}}. In
  \bibinfo{booktitle}{\emph{Proceedings of The IEEE International Symposium on
  Parallel and Distributed Processing with Applications}}.
  \bibinfo{pages}{629--634}.
\newblock
\showISSN{2158-9178}
\urldef\tempurl%
\url{https://doi.org/10.1109/ISPA.2009.95}
\showDOI{\tempurl}


\bibitem[\protect\citeauthoryear{Xu, Dou, Zhang, and Chen}{Xu
  et~al\mbox{.}}{2016}]%
        {7276993}
\bibfield{author}{\bibinfo{person}{Xiaolong Xu}, \bibinfo{person}{Wanchun Dou},
  \bibinfo{person}{Xuyun Zhang}, {and} \bibinfo{person}{Jinjun Chen}.}
  \bibinfo{year}{2016}\natexlab{}.
\newblock \showarticletitle{EnReal: {An} {Energy-aware} {Resource} {Allocation}
  {Method} for {Scientific} {Workflow} {Executions} in {Cloud} {Environment}}.
\newblock \bibinfo{journal}{\emph{IEEE Transactions on Cloud Computing}}
  \bibinfo{volume}{4}, \bibinfo{number}{2} (\bibinfo{year}{2016}),
  \bibinfo{pages}{166--179}.
\newblock
\showISSN{2168-7161}
\urldef\tempurl%
\url{https://doi.org/10.1109/TCC.2015.2453966}
\showDOI{\tempurl}


\bibitem[\protect\citeauthoryear{Yu and Buyya}{Yu and Buyya}{2005}]%
        {Yu2005}
\bibfield{author}{\bibinfo{person}{Jia Yu} {and} \bibinfo{person}{Rajkumar
  Buyya}.} \bibinfo{year}{2005}\natexlab{}.
\newblock \showarticletitle{A {Taxonomy} of {Workflow} {Management} {Systems}
  for {Grid} {Computing}}.
\newblock \bibinfo{journal}{\emph{Journal of Grid Computing}}
  \bibinfo{volume}{3}, \bibinfo{number}{3} (\bibinfo{year}{2005}),
  \bibinfo{pages}{171--200}.
\newblock
\showISSN{1572-9184}
\urldef\tempurl%
\url{https://doi.org/10.1007/s10723-005-9010-8}
\showDOI{\tempurl}


\bibitem[\protect\citeauthoryear{Yu and Shi}{Yu and Shi}{2008}]%
        {4626773}
\bibfield{author}{\bibinfo{person}{Zhifeng Yu} {and} \bibinfo{person}{Weisong
  Shi}.} \bibinfo{year}{2008}\natexlab{}.
\newblock \showarticletitle{A Planner-Guided Scheduling Strategy for Multiple
  Workflow Applications}. In \bibinfo{booktitle}{\emph{Proceedings of The
  International Conference on Parallel Processing}}. \bibinfo{pages}{1--8}.
\newblock
\showISSN{0190-3918}
\urldef\tempurl%
\url{https://doi.org/10.1109/ICPP-W.2008.10}
\showDOI{\tempurl}


\bibitem[\protect\citeauthoryear{Zeng, Veeravalli, and Li}{Zeng
  et~al\mbox{.}}{2015}]%
        {ZENG2015141}
\bibfield{author}{\bibinfo{person}{Lingfang Zeng}, \bibinfo{person}{Bharadwaj
  Veeravalli}, {and} \bibinfo{person}{Xiaorong Li}.}
  \bibinfo{year}{2015}\natexlab{}.
\newblock \showarticletitle{SABA: A Security-aware and Budget-aware Workflow
  Scheduling Strategy in Clouds}.
\newblock \bibinfo{journal}{\emph{J. Parallel and Distrib. Comput.}}
  \bibinfo{volume}{75} (\bibinfo{year}{2015}), \bibinfo{pages}{141--151}.
\newblock
\showISSN{0743-7315}
\urldef\tempurl%
\url{https://doi.org/10.1016/j.jpdc.2014.09.002}
\showDOI{\tempurl}


\bibitem[\protect\citeauthoryear{Zhang, Wang, Zhao, and Tao}{Zhang
  et~al\mbox{.}}{2013}]%
        {6681002}
\bibfield{author}{\bibinfo{person}{Shuo Zhang}, \bibinfo{person}{Baosheng
  Wang}, \bibinfo{person}{Baokang Zhao}, {and} \bibinfo{person}{Jing Tao}.}
  \bibinfo{year}{2013}\natexlab{}.
\newblock \showarticletitle{An {E}nergy-{A}ware {T}ask {S}cheduling {A}lgorithm
  for a {H}eterogeneous {D}ata {C}enter}. In
  \bibinfo{booktitle}{\emph{Proceedings of The 12th IEEE International
  Conference on Trust, Security and Privacy in Computing and Communications}}.
  \bibinfo{pages}{1471--1477}.
\newblock
\showISSN{2324-898X}
\urldef\tempurl%
\url{https://doi.org/10.1109/TrustCom.2013.178}
\showDOI{\tempurl}


\bibitem[\protect\citeauthoryear{Zhao, Li, and Liu}{Zhao et~al\mbox{.}}{2014}]%
        {6779008}
\bibfield{author}{\bibinfo{person}{Feng Zhao}, \bibinfo{person}{Chao Li}, {and}
  \bibinfo{person}{Chunfeng Liu}.} \bibinfo{year}{2014}\natexlab{}.
\newblock \showarticletitle{A Cloud Computing Security Solution based on Fully
  Homomorphic Encryption}. In \bibinfo{booktitle}{\emph{Proceedings of The 16th
  International Conference on Advanced Communication Technology}}.
  \bibinfo{pages}{485--488}.
\newblock
\showISSN{1738-9445}
\urldef\tempurl%
\url{https://doi.org/10.1109/ICACT.2014.6779008}
\showDOI{\tempurl}


\bibitem[\protect\citeauthoryear{Zhou, He, and Liu}{Zhou et~al\mbox{.}}{2016}]%
        {7044594}
\bibfield{author}{\bibinfo{person}{Amelia~C. Zhou}, \bibinfo{person}{Bingsheng
  He}, {and} \bibinfo{person}{Cheng Liu}.} \bibinfo{year}{2016}\natexlab{}.
\newblock \showarticletitle{Monetary {Cost} {Optimizations} for {Hosting}
  {Workflow}-as-a-{Service} in {IaaS} {Clouds}}.
\newblock \bibinfo{journal}{\emph{IEEE Transactions on Cloud Computing}}
  \bibinfo{volume}{4}, \bibinfo{number}{1} (\bibinfo{year}{2016}),
  \bibinfo{pages}{34--48}.
\newblock
\showISSN{2168-7161}
\urldef\tempurl%
\url{https://doi.org/10.1109/TCC.2015.2404807}
\showDOI{\tempurl}


\bibitem[\protect\citeauthoryear{Zhou, Li, Xu, and Qi}{Zhou
  et~al\mbox{.}}{2018}]%
        {Zhou2018}
\bibfield{author}{\bibinfo{person}{Naqin Zhou}, \bibinfo{person}{FuFang Li},
  \bibinfo{person}{Kefu Xu}, {and} \bibinfo{person}{Deyu Qi}.}
  \bibinfo{year}{2018}\natexlab{}.
\newblock \showarticletitle{Concurrent Workflow Budget- and
  Deadline-constrained Scheduling in Heterogeneous Distributed Environments}.
\newblock \bibinfo{journal}{\emph{Soft Computing}} (\bibinfo{year}{2018}).
\newblock
\showISSN{1433--7479}
\urldef\tempurl%
\url{https://doi.org/10.1007/s00500-018-3229-3}
\showDOI{\tempurl}


\bibitem[\protect\citeauthoryear{Zhu, Wang, Guo, Zhu, Yang, and Liu}{Zhu
  et~al\mbox{.}}{2016}]%
        {7435325}
\bibfield{author}{\bibinfo{person}{Xiaomin Zhu}, \bibinfo{person}{Ji Wang},
  \bibinfo{person}{Hui Guo}, \bibinfo{person}{Dakai Zhu},
  \bibinfo{person}{Laurence~T. Yang}, {and} \bibinfo{person}{Ling Liu}.}
  \bibinfo{year}{2016}\natexlab{}.
\newblock \showarticletitle{Fault-Tolerant Scheduling for Real-Time Scientific
  Workflows with Elastic Resource Provisioning in Virtualized Clouds}.
\newblock \bibinfo{journal}{\emph{IEEE Transactions on Parallel and Distributed
  Systems}} \bibinfo{volume}{27}, \bibinfo{number}{12} (\bibinfo{year}{2016}),
  \bibinfo{pages}{3501--3517}.
\newblock
\showISSN{1045-9219}
\urldef\tempurl%
\url{https://doi.org/10.1109/TPDS.2016.2543731}
\showDOI{\tempurl}


\end{thebibliography}

\end{document}